\documentclass[12pt]{article}

\usepackage{array} 
\usepackage{epsfig}
\usepackage{amssymb}
\usepackage{graphics,graphpap}

\renewcommand{\thefootnote}{\fnsymbol{footnote}}

\newcommand{\eqref}[1]{(\ref{#1})}

\newcommand{\al}{\alpha}
\newcommand{\be}{\beta}
\newcommand{\ga}{\gamma}
\newcommand{\de}{\delta}

\newcommand{\GeV}{\,\mbox{GeV}}

\newcommand{\matel}[3]{\langle #1|#2|#3\rangle}
\newcommand{\state}[1]{|#1\rangle}

\begin{document}

\begin{titlepage}
\begin{flushright}\begin{tabular}{l}
IPPP/06/79\\
DCPT/06/158
\end{tabular}
\end{flushright}
\vskip1.5cm
\begin{center}
   {\Large \bf \boldmath $B \to V\gamma$ Beyond QCD Factorisation}
    \vskip1.3cm {\sc
Patricia Ball\footnote{Patricia.Ball@durham.ac.uk}, Gareth
    W. Jones\footnote{G.W.Jones@durham.ac.uk} and Roman 
Zwicky\footnote{Roman.Zwicky@durham.ac.uk}
  \vskip0.5cm
        {\em IPPP, Department of Physics, 
University of Durham, Durham DH1 3LE, UK}} \\
\vskip2.5cm 


\vskip5cm

{\large\bf Abstract:\\[10pt]} \parbox[t]{\textwidth}{
We calculate the main observables in $B_{u,d}\to
(\rho,\omega,K^*)\gamma$ and $B_s\to (\bar
K^*,\phi)\gamma$ decays, i.e.\ branching ratios and CP and
isospin asymmetries. We include QCD factorisation results
and also the dominant contributions beyond QCD factorisation,
namely long-distance photon emission and soft-gluon emission from
quark loops. All contributions beyond QCD factorisation 
are estimated from light-cone sum rules.
We devise in particular a method for calculating
soft-gluon emission, building on earlier ideas 
developed for analogous contributions in non-leptonic decays.
Our results are relevant for new-physics searches at the $B$
factories, the LHC and a future super-flavour factory. Using current
experimental data, we also extract
$|V_{td}/V_{ts}|$ and the angle $\gamma$ of the unitarity triangle. We
give detailed tables of theoretical uncertainties of the relevant
quantities which facilitates future determinations of these CKM
parameters from updated experimental results.
}

\end{center}

\vfill

\end{titlepage}

\setcounter{footnote}{0}
\renewcommand{\thefootnote}{\arabic{footnote}}

\newpage

\section{Introduction}\label{sec:1}

The flavour-changing neutral-current (FCNC) 
transitions $b\to s\gamma$ and $b\to d\gamma$ are among
the most valuable probes of flavour physics. Assuming the Standard
Model (SM) to be valid, these processes offer the possibility to extract the
CKM matrix elements $|V_{t(d,s)}|$, in complementarity to, on the one
hand, the determination from $B$ mixing and, on the other hand, the SM 
unitarity triangle (UT) based on the tree-level observables
$|V_{ub}/V_{cb}|$ and the angle $\gamma$. 
These decays are also characterized by their high
sensitivity to new physics (NP) contributions 
and by the particularly large impact
of short-distance QCD corrections, see Ref.~\cite{Hurth} for a
review. Considerable time and effort has gone into the calculation
of these corrections which are now approaching next-to-next-to-leading
order accuracy \cite{misiak,NNLObsgamma}. On the experimental side,  
both exclusive and inclusive $b\to s\gamma$ branching ratios are known
with good accuracy, $5\%$ for $B\to K^*\gamma$ and $7\%$ for
$B\to X_s\gamma$, while the situation is less favourable
for $b\to d\gamma$ transitions: measurements are only available for
exclusive channels. In Tab.~\ref{tab1} we give the branching ratios of
all established exclusive $b\to(d,s)\gamma$ channels.

Whereas the inclusive modes can be computed perturbatively, using
fixed-order heavy
quark expansion or soft-collinear effective theory (SCET)
 \cite{misiak,neubert}, the treatment of
exclusive channels is more complicated. With presently available
methods it is impossible to simulate the full amplitude on the
lattice, the reason being the occurence of non-local correlation
functions associated with the insertion of the electromagnetic
interaction operator into the effective Hamiltonian for $b\to
(s,d)\gamma$. Instead, one has to resort to effective field theory
methods, which yield an expansion in inverse powers of the $b$ quark
mass, $m_b$. It was shown, in SCET, 
that the relevant hadronic matrix elements factorise to all
orders in $\alpha_s$ and leading order in $1/m_b$ 
and can be written as \cite{BHN}
\begin{equation}\label{1}
\langle V\gamma|Q_i| B\rangle =
e^* \cdot \left[ T_1^{B\to V}(0)\, T^I_{i} +
\int^1_0 d\xi\, du\, T^{II}_i(\xi,u)\, \phi_B(\xi)\, 
\phi_{2;V}^\perp(u)\right]
\times \left\{1 + O(1/m_b)\vphantom{\int^1_0}\right\}.
\end{equation}
This formula coincides with that obtained earlier in QCD factorisation (QCDF)
to next-to-leading order in $\alpha_s$ 
\cite{BVga1,AP,BVga2,BoBu1,BoschThesis,BoBu2}. 
In (\ref{1}), $e_\mu$ is the photon's polarisation four-vector, $Q_i$ is one of
the operators in the  effective Hamiltonian for $b\to (s,d)$ transitions,
$T_1^{B\to V}$ is a $B\to V$ transition form factor,
and $\phi_B$, $\phi_{2;V}^\perp$ 
are leading-twist light-cone distribution amplitudes (DAs)
of the $B$ meson and the vector meson $V$, respectively.
These quantities are universal non-perturbative objects and
describe the long-distance dynamics of  matrix elements, which
is factorised from the perturbative short-distance interactions
included in the hard-scattering kernels $T^I_{i}$ and $T^{II}_i$.
$B\to V\gamma$ decays have also been investigated in the alternative
approach of perturbative QCD factorisation (pQCD) \cite{pQCD}.

Eq.~\eqref{1} is sufficient to calculate observables that are dominated
by the leading-order term in the heavy-quark expansion, like 
${\cal B}(B\to K^*\gamma)$. For ${\cal B}(B\to
(\rho,\omega)\gamma)$, however,
power-suppressed corrections play an important
r\^ole, for instance weak annihilation (WA) which is mediated by a
tree-level diagram. In this case, the parametric suppression by one
power of $1/m_b$ is alleviated by an enhancement factor $2\pi^2$
relative to the loop-suppressed contributions at leading order in
$1/m_b$. Power-suppressed contributions also determine the
time-dependent CP asymmetry in $B\to V\gamma$, see
Refs.~\cite{alt,grin04,grin05,cpas}, 
as well as isospin asymmetries \cite{kagan} ---
all observables with a potentially large contribution from NP. 
The purpose of this paper is to calculate the  dominant 
power-suppressed contributions to (\ref{1}) 
and the resulting branching ratios and CP and isospin
asymmetries, using the most up-to-date hadronic input parameters for
form factors and light-cone DAs. Although $1/m_b$ effects are, in
principle, accessible in SCET, the vast majority of studies in this
framework only includes leading-order effects, the reason being a
proliferation of new effective operators at power-suppressed accuracy, 
see Ref.~\cite{daniel}, whose matrix elements induce subleading form
factors and are largely unknown. For this reason, in this paper 
we adopt a different
approach not based on SCET and calculate power-suppressed corrections
using the method of QCD sum rules on the light-cone (LCSRs).
The present paper is an extension of our previous
work Ref.~\cite{BZ06b}, where we calculated the ratio of branching
ratios ${\cal B}(B\to (\rho,\omega)\gamma)/{\cal B}(B\to K^*\gamma)$
in order to determine the ratio of CKM matrix elements
$|V_{td}/V_{ts}|$ from data. Some of these power corrections, namely those
related to WA contributions and the isospin asymmetry
in $B\to K^*\gamma$, have already been calculated in QCDF
\cite{BoBu1,BoschThesis,BoBu2,kagan}. Other power corrections cannot be
calculated in the framework of QCDF. The most relevant of these
come from soft-gluon emission from quark loops and long-distance
photon emission from soft quarks. We have already calculated some of these
contributions before, using LCSRs: 
long-distance photon emission in
Ref.~\cite{emi} and soft-gluon emission from charm loops
in Ref.~\cite{cpas}, using heavy-quark expansion in powers of
$1/m_c$. In this paper, we complete these calculations and develop
a method to also calculate soft-gluon emission
from light-quark loops, thus allowing us to predict branching ratios
and isospin and CP asymmetries for exclusive $B\to V\gamma$
transitions with increased precision. We also include
the $B_s$ decays $B_s\to\phi\gamma$, which is a $b\to s\gamma$
transition, and $B_s\to K^*\gamma$, which is $b\to d\gamma$. All 
these decays
will be studied in detail at the LHC, and those of $B_{u,d}$ at future
super-flavour factories \cite{superB}.
\begin{table}[t]
\renewcommand{\arraystretch}{1.4}\addtolength{\arraycolsep}{3pt}
$$
\begin{array}{l||c|c||l|c}
{\cal B} \times 10^6 & \mbox{BaBar \cite{Babar}} & \mbox{Belle
  \cite{Belle}} & {\cal B} \times 10^6 & \mbox{HFAG \cite{HFAG}}
\\\hline
B\to (\rho,\omega)\gamma & 1.25^{+0.25}_{-0.24}\pm 0.09 &
                           1.32^{+0.34}_{-0.31}{}^{+0.10}_{-0.09} 
                         &
B^+\to K^{*+}\gamma & 40.3\pm 2.6
\\
B^+\to \rho^+\gamma &      1.10^{+0.37}_{-0.33}\pm 0.09 &
                           0.55^{+0.42}_{-0.36}{}^{+0.09}_{-0.08} 
                    &
B^0\to K^{*0}\gamma &  40.1\pm 2.0
\\
B^0\to\rho^0\gamma  &      0.79^{+0.22}_{-0.20}\pm0.06 &
                           1.25^{+0.37}_{-0.33}{}^{+0.07}_{-0.06} 
\\
B^0\to\omega\gamma  &      <0.78 &
                           0.96^{+0.34}_{-0.27}{}^{+0.05}_{-0.10}
\end{array}
$$
\caption[]{\small Experimental branching ratios of exclusive $b\to
  (d,s)\gamma$
  transitions. All entries are CP averaged.
The first error is statistical, the second
  systematic. $B\to (\rho,\omega)\gamma$ is the CP
  average of the
  isospin average over $\rho$ and $\omega$ channels:\\ 
$\overline{\cal B}(B\to
  (\rho,\omega)\gamma) = \frac{1}{2}
  \left\{ \overline{\cal B}(B^\pm\to \rho^\pm\gamma) +
  \frac{\tau_{B^\pm}}{\tau_{B^0}} \left[ \overline{\cal B}(B^0\to 
  \rho^0\gamma) +
  \overline{\cal B}(B^0\to \omega \gamma)\right]\right\}$.}\label{tab1}
\end{table}

Our paper is organized as follows: in Sec.~\ref{sec:2} we introduce
notations and recall QCDF formulae. In Sec.~\ref{sec:3} we calculate the
WA contributions and in Sec.~\ref{sec:4} the
long-distance contributions to the $B\to V\gamma$ amplitude from
heavy- and light-quark loops.
In Sec.~\ref{sec:5} we present results
for branching ratios and asymmetries; we summarize and conclude in
Sec.~\ref{sec:6}. The appendix contains a discussion of the
longitudinal and transverse decay constants of vector mesons.

\section{Framework and Basic Formulae}\label{sec:2}

The effective Hamiltonian for $b\to D\gamma$ transitions, with
$D=s,d$, reads:
\begin{equation}\label{heff}
H_{\rm eff}=\frac{G_F}{\sqrt{2}}\sum_{U=u,c}\lambda_U^{(D)} 
\left[ C_1 Q^U_1 + C_2 Q^U_2 +\sum_{i=3\ldots 8} C_i Q_i\right],
\end{equation}
where $\lambda_U^{(D)}=V^*_{UD}V_{Ub}$. This Hamiltonian implicitly
relies on the SM unitarity relation, also referred to as the GIM mechanism,
\begin{equation}\label{SMunitarity}
\lambda_t^{(D)} + \lambda_c^{(D)} + \lambda_u^{(D)}=0\,,
\end{equation}
which enters the calculation of the penguin contributions.
The operators are given by
\begin{eqnarray}
  Q^U_1 &=& (\bar D_i U_j)_{V-A}(\bar U_j b_i)_{V-A}\,,
\qquad Q^U_2 = (\bar DU)_{V-A}(\bar Ub)_{V-A}\,,\nonumber\\
  Q_3 &=& (\bar Db)_{V-A} \sum_q (\bar qq)_{V-A} \,,
\qquad Q_4 = (\bar D_i b_j)_{V-A} \sum_q (\bar q_j q_i)_{V-A} \,,\nonumber\\
  Q_5 &=& (\bar Db)_{V-A} \sum_q (\bar qq)_{V+A}\,, 
\qquad Q_6 = (\bar D_i b_j)_{V-A} \sum_q (\bar q_j q_i)_{V+A}\,,\nonumber\\
Q_7 &=& \frac{e}{8\pi^2}m_b\, 
        \bar D\sigma^{\mu\nu}(1+\gamma_5)F_{\mu\nu}\,b
         + \frac{e}{8\pi^2}m_D\, 
        \bar D\sigma^{\mu\nu}(1-\gamma_5)F_{\mu\nu}\,b \,,\nonumber\\
Q_8 &=& \frac{g}{8\pi^2}m_b\, 
        \bar D\sigma^{\mu\nu}(1+\gamma_5)G_{\mu\nu}\, b
        + \frac{g}{8\pi^2}m_D\, 
        \bar D\sigma^{\mu\nu}(1-\gamma_5)G_{\mu\nu}\, b\, , \label{eq:WC}
\end{eqnarray}
where $(\bar q  Q)_{V-A}(\bar r  R)_{V\pm A} \equiv 
(\bar q \gamma_\mu(1-\gamma_5) Q)
(\bar r \gamma^\mu(1\pm\gamma_5) R)$. The sign conventions for the 
electromagnetic and strong couplings correspond to the covariant derivative
$D_\mu=\partial_\mu +ie Q_f A_\mu + i g T^a A^a_\mu$. With these
definitions the coefficients $C_{7,8}$ are negative in the
SM, which is the choice generally adopted
in the literature. The above operator basis, which we
shall label BBL after the authors of Ref.~\cite{beyondLL}, is the same
as that of Ref.~\cite{BoBu1}, except that $Q_1$ and $Q_2$ are
exchanged. For some applications, in particular calculations of 
inclusive $b\to D\gamma$ transitions, a different operator basis
proves more suitable: the basis adopted for instance 
in Refs.~\cite{munz,buras}, labelled CMM in the following, has
$Q_{7(8)}^{\rm CMM} = Q_{7(8)}^{\rm BBL}$, but differs in $Q_{1\dots 6}$.
It turns out that we need the Wilson coefficients in both bases:
\begin{itemize}
\item $C_{1\dots 6}^{\rm BBL}(\mu\sim m_b)$, calculated according to 
  Ref.~\cite{beyondLL}, for power-suppressed (WA and soft-gluon emission)
  contributions; 
\item $C_{1\dots 8}^{\rm CMM}(\mu\sim m_b)$, calculated according to 
 Ref.~\cite{munz}, for
   hard-vertex corrections in QCDF which are given in terms
  of two-loop matrix elements for $b\to D\gamma$ transitions obtained in
  Ref.~\cite{buras}, in the CMM basis;
\item $C_{1\dots 6}^{\rm BBL}(\mu\sim 2\,{\rm GeV})$ and 
  $C_{8}^{\rm CMM}(\mu\sim 2\,{\rm GeV})$ for hard-spectator
  corrections in QCDF; although these coefficients refer to a
  different basis,  it is correct to use them together as the 
  corresponding operators $Q_{7,8}$ are identical in both bases and 
  independent of the  basis chosen for the four-quark operators.
\end{itemize}
Numerical values of all $C_i$ are given in Tab.~\ref{tab2}.
\begin{table}[bt]
\renewcommand{\arraystretch}{1.3}
\addtolength{\arraycolsep}{2pt}
$$
\begin{array}{c|c|c|c|c|c|c}
C^{\rm CMM}_1(m_b) & C^{\rm CMM}_2(m_b) & C^{\rm CMM}_3(m_b) & 
C^{\rm CMM}_4(m_b) & C^{\rm CMM}_5(m_b) & C^{\rm CMM}_6(m_b) &
C^{\rm CMM}_7(m_b)
\\\hline
-0.322 & 1.009 & -0.005 & -0.874 & 0.0004 & -0.001 & -0.309 
\\\hline\hline
C^{\rm BBL}_1(m_b) & C^{\rm BBL}_2(m_b) & C^{\rm BBL}_3(m_b) & 
C^{\rm BBL}_4(m_b) & C^{\rm BBL}_5(m_b) & C^{\rm BBL}_6(m_b) 
& C^{\rm CMM}_8(m_b) 
\\\hline
-0.189 & 1.081  & 0.014 & -0.036 & 0.009  & -0.042 & -0.170
\\\hline\hline
C^{\rm BBL}_1(\mu_h) & C^{\rm BBL}_2(\mu_h) & C^{\rm BBL}_3(\mu_h) & 
C^{\rm BBL}_4(\mu_h) & C^{\rm BBL}_5(\mu_h) & C^{\rm BBL}_6(\mu_h) &
C^{\rm CMM}_8(\mu_h)
\\\hline
-0.288 & 1.133 & 0.021 & -0.051 & 0.010 & -0.065 & -0.191
\end{array}
$$
\vspace*{-10pt}
\caption[]{\small NLO Wilson coefficients to be used in this
  paper, at the scales $m_b=4.2\,$GeV and $\mu_h=2.2\,$GeV. The 
  coefficients labelled BBL correspond to the operator
  basis of Ref.~\cite{beyondLL} and given in Eq.~(\ref{eq:WC}), 
  whereas CMM denotes the basis of Ref.~\cite{munz}. We use 
  $\alpha_s(m_Z) = 0.1176$ \cite{PDG} and ${m}_t({m}_t) = 
  163.6\,$GeV \cite{mt}. Note that $C_1^{\rm BBL}$ and $C_2^{\rm BBL}$ 
  are exchanged with respect to the basis of Ref.~\cite{BoBu1} and
  that $C_{7(8)}^{\rm BBL}=C_{7(8)}^{\rm CMM}$, see text. Following
  \cite{BoschThesis}, the CMM set is used for calculating hard-vertex
  corrections to the QCDF formulae and the BBL set at the lower scale
  $\mu_h$ is used to calculate hard-spectator corrections. The BBL set at
  scale $m_b$ is used for the calculation of power corrections.
}\label{tab2}
\end{table}
Note that the question whether to use Wilson coefficients (and other
scale-dependent hadronic quantities) at LO or NLO accuracy is actually
non-trivial. Strictly speaking, NLO accuracy is mandatory only for
$C_7$, as the hadronic matrix element for this term only is also known
to NLO accuracy, see below. We will evaluate all $O(\alpha_s)$ and
power-suppressed corrections using both LO and NLO scaling for Wilson
coefficients and hadronic matrix elements and include the resulting
discrepancies in the theoretical uncertainty.

\begin{figure}[tb]
\vskip-1cm
$$\epsfxsize=0.9\textwidth\epsffile{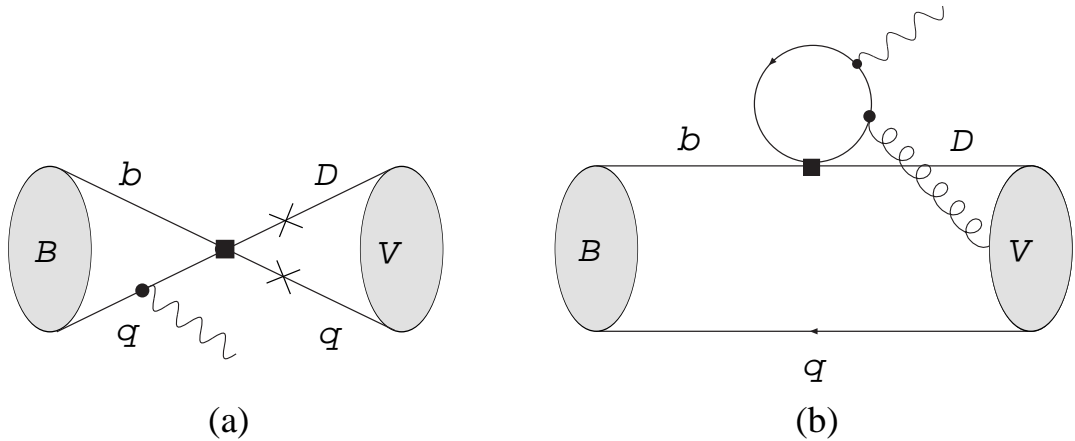}$$
\vspace*{-30pt}
\caption[]{\small (a): WA diagram. The square
  denotes insertion of the operator $Q_i$. Photon emission from 
  lines other than the $B$ spectator is power-suppressed, except for 
emission from the final-state quark lines for the operators $Q_{5,6}$,
  denoted by crosses.
(b): soft-gluon emission from a quark loop. Again the square dot
  denotes the insertion of the operator $Q_i$. There is also a second
  diagram where the soft gluon is picked up by the $B$ meson.}\label{fig1}
\end{figure}

The calculation of the decay amplitudes of exclusive $B\to V\gamma$
decays also requires the knowledge of hadronic matrix elements of type
$\langle V\gamma|Q_i| B\rangle$. A complete calculation of these
quantities is not possible to date, but the leading term in an
expansion in $1/m_b$ is obtained from QCDF, see Eq.~(\ref{1}).
The factorisation formula is valid in the heavy-quark limit
$m_b\to\infty$ and is subject to
corrections of order $\Lambda_{\rm QCD}/m_b$. Some of these
corrections are numerically very relevant: for instance, the
contributions from all operators but $Q_2$ are loop suppressed; hence,
the tree-level WA diagram in Fig.~\ref{fig1}, which is
suppressed by one power of $m_b$, comes with a
relative enhancement factor $\sim 4\pi^2$. 
This contribution, with the operator $Q_2$, is doubly
Cabibbo-suppressed for $b\to s\gamma$ transitions,
but carries no CKM suppression factor in $b\to d\gamma$
transitions. WA can also be induced by the penguin
operators $Q_{3\ldots 6}$ and in this case carries no CKM suppression
in $b\to s\gamma$, but comes with small (loop-suppressed)
Wilson-coefficients. Other examples for relevant power-suppressed
corrections are CP and isospin asymmetries,
Refs.~\cite{alt,cpas, kagan}, which actually vanish in the heavy quark
limit. This indicates that in $B\to V\gamma$ transitions
simple $1/m_b$ counting is, in general, not sufficient to determine the
numerical relevance of a particular contribution, but that all
relevant factors,
\begin{itemize}
\item order of power-suppression in $1/m_b$;
\item loop suppression or tree enhancement;
\item CKM suppression;
\item size of hadronic matrix elements;
\end{itemize}
have to be taken into account. This is a consequence of the fact that
in radiative transitions the ``naively'' leading term in $Q_7$ is loop 
suppressed, which is qualitatively different from other applications
of QCDF, for instance in $B^-\to \pi^-\pi^0$, where the 
leading hadronic matrix element describes a tree-level process.

The exclusive $B\to V\gamma$ process is actually described by two
physical amplitudes, one for each polarisation of the photon:
\begin{equation}\label{4}
\bar{\cal A}_{L(R)} = {\cal A}(\bar B\to V
\gamma_{L(R)})\,,  \qquad
{\cal A}_{L(R)} = {\cal A}(B\to \bar V \gamma_{L(R)})\,,
\end{equation}
where $\bar B$ denotes a $(b\bar q)$ and $V$ a $(D\bar q)$ bound
state.\footnote{Note that in this paper $K^*$ is a $(s\bar q)$
  bound state, in contrast to the standard labelling, according to
  which $K^{*0}=(d\bar s)$ and $\bar K^{*0}=(s\bar d)$. This is
  because the calculation of form factors and other matrix
  elements involves light-cone DAs of the vector
  meson $V$ and in the standard notation used in that context,
  $K^*$ always contains an $s$ quark, and $\bar K^*$ an $\bar s$
  quark. This distinction is relevant because of
  a sign change of G-odd matrix elements under $(s\bar
  q)\leftrightarrow (q\bar s)$, see Tabs.~\ref{tab3},
  \ref{tab:kappas}, \ref{twist4}.} 
In the notation introduced in Ref.~\cite{BoBu1} in the context
of QCDF, the
decay amplitudes can be written as
\begin{eqnarray}
\bar{\cal A}_{L(R)} &=& \frac{G_F}{\sqrt{2}}\,\left( \lambda_u^D
a_{7}^u( V\gamma_{L(R)}) +
\lambda_c^D a_{7}^c( V\gamma_{L(R)})\right) \langle V
\gamma_{L(R)} | Q_7^{L(R)} | \bar B\rangle
\nonumber\\
&\equiv& \frac{G_F}{\sqrt{2}}\,\left( \lambda_u^D
a_{7L(R)}^u( V) + \lambda_c^D a_{7L(R)}^c( V)\right) \langle V
\gamma_{L(R)} | Q_7^{L(R)} | \bar B\rangle\,,\nonumber\\
{\cal A}_{L(R)}
& = & \frac{G_F}{\sqrt{2}}\,\left( (\lambda_u^D)^*
a_{7R(L)}^u( V) + (\lambda_c^D)^* a_{7R(L)}^c( V)\right) \langle \bar V
\gamma_{L(R)} | (Q_7^{R(L)})^\dagger | B\rangle\,.\label{ME}
\end{eqnarray}
The
  $a_7^{c,u}$ calculated in Refs.~\cite{BoBu1,BoschThesis}, coincide,
to leading order in  $1/m_b$, with our
  $a_{7L}^U$, whereas $a_{7R}^U$ are set zero in
  \cite{BoBu1,BoschThesis}. Our
expression (\ref{ME}) is purely formal and does not imply that the
$a_{7R(L)}^{U}$ factorise at order $1/m_b$. As a matter of fact,
they don't.
The operators $Q_7^{L(R)}$ are given by
$$
Q_7^{L(R)} = \frac{e}{8\pi^2}\, m_b \bar D \sigma_{\mu\nu}
             \left(1 \pm \gamma_5\right)b F^{\mu\nu}
$$
and generate left- (right-) handed photons in the decay
$b\to D\gamma$. The matrix elements in (\ref{ME}) can be expressed in 
terms of the form factor $T_1^{B\to V}$  as
\begin{eqnarray}
\lefteqn{\langle V(p,\eta) \gamma_{L(R)}(q,e) | Q_7^{L(R)} | \bar
B \rangle }\hspace*{1cm}\nonumber\\
&=& -\frac{e}{2\pi^2}\, m_b T_1^{B\to V}(0) \left[
\epsilon^{\mu\nu\rho\sigma} e_\mu^* \eta_\nu^* p_\rho q_\sigma \pm i
\{ (e^* \eta^*) (pq) - (e^*p)(\eta^* q)\}\right]
\nonumber\\
&\equiv& -\frac{e}{2\pi^2}\, m_b T_1^{B\to V}(0) S_{L(R)}\,,\nonumber\\
\lefteqn{\langle \bar V(p,\eta) \gamma_{L(R)}(q,e) |
  (Q_7^{R(L)})^\dagger | B \rangle  = -\frac{e}{2\pi^2}\, m_b
  T_1^{B\to V}(0) S_{L(R)}\,,}\hspace*{1cm}\label{6}
\end{eqnarray}
where $S_{L,R}$ are the helicity amplitudes corresponding to left- and
right-handed photons, respectively, and $e_\mu(\eta_\mu)$ is the polarisation
four-vector of the photon (vector meson). The definition of $T_1^{B\to V}$
can be found in Ref.~\cite{BZ04b}; our convention for the
epsilon tensor follows that of  Bjorken \& Drell:
${\rm Tr}[\gamma^\al\gamma^\be\gamma^\ga\gamma^\de\gamma_5] = 4i 
\epsilon^{\al\be\ga\de}$.
Up-to-date values for  all
decays studied in this paper are given in Tab.~\ref{tab3}. The
non-perturbative parameters are taken from experiment, where
available, from lattice ($f_B$), from QCD sum rules ($a_i^\perp$)
and from QCD sum rules on the light-cone ($T_1$); for the decay
constants $f^\perp$ results are available from both lattice and QCD
sum rules; they are discussed in the appendix. No lattice results
are available for $a_i^\perp$ and only partial results for $T_1$
\cite{damir}. The numbers in Tab.~\ref{tab3}
differ slightly
from those given in Ref.~\cite{BZ04b} because we include updates
of the hadronic input parameters.
We do not include isospin breaking in the form factors since it is
caused by the difference of quark masses and electric charges and 
expected to be of the order of $1\%$ only.
This is indeed the size of isospin-breaking in the form factor
indicated by recent 
measurements of $D^0 \to (K^-,\pi^-) e^+ \nu$ and 
$D^+ \to  (\bar K^0,\pi^0) e^+ \nu$  at CLEO \cite{CLEO}. At this
point we would also like to comment on the UT angle
$\gamma$. The value given in Tab.~\ref{tab3} comes from Belle's 
Dalitz-plot analysis of the CP asymmetry in $B^-\to (K_S^0 \pi^+\pi^-)
K^-$, with $K_S^0 \pi^+\pi^-$ being a three-body final state common
to both $D^0$ and $\bar D^0$. This method to measure $\gamma$ from a
new-physics free tree-level process was suggested in Ref.~\cite{GGSZ}
and has been implemented by both BaBar \cite{babargamma} and Belle 
\cite{Bellegamma}, but the BaBar result currently suffers from huge
errors. Other determinations of $\gamma$ from QCDF or SCET analyses, 
or SU(3) or U-spin fits of non-leptonic $B$ decays, or global UT fits,
all come with theoretical uncertainties and/or possible contamination by
unresolved new physics, so we decide to stick, as a reference point, to
the tree-level result of Belle. For all observables with a pronounced
dependence on $\gamma$, i.e.\ $b\to d\gamma$ branching ratios and
isospin asymmetries, we will present results as a function of $\gamma$.
\begin{table}[btp]
\addtolength{\arraycolsep}{3pt}
\renewcommand{\arraystretch}{1.3}
$$
\begin{array}{c|c|c|c|c|c}\hline\hline
\multicolumn{6}{c}{\mbox{CKM parameters and couplings}}\\\hline
\lambda \mbox{~\cite{PDG}} & |V_{cb}| \mbox{~\cite{inclmoments}} & 
|V_{ub}| & \gamma \mbox{~\cite{Bellegamma}} & \alpha_s(m_Z)
\mbox{~\cite{PDG}} & \alpha\\\hline
0.227(1) & 42.0(7)\times 10^{-3} & 4.0(7)\times
10^{-3} & (53\pm 20)^\circ & 0.1176(20) & 1/137\\\hline\hline
\multicolumn{6}{c}{\mbox{B parameters}}\\\hline
f_{B_q}\mbox{~\cite{Onogi}} & f_{B_s}\mbox{~\cite{Onogi}} &
    \lambda_{B_q}(\mu_h) \mbox{~\cite{BZ06b}} & \lambda_{B_s}(\mu_h)
& \mu_h     \\\hline
200(25)\,{\rm MeV} & 240(30)\,{\rm MeV} & 0.51(12)\,{\rm GeV} &
0.6(2)\,{\rm GeV} & 2.2\,{\rm GeV}\\\hline\hline
\multicolumn{6}{c}{\mbox{$\rho$ parameters}}\\\hline
f_{\rho} & f_{\rho}^\perp & a_1^\perp({\rho}) &
a_2^\perp({\rho}) & T_1^{B\to\rho}(0)\\\hline
216(3)\,{\rm MeV} & 165(9)\,{\rm MeV} & 0 & 0.15(7) & 0.27(4)
\\\hline\hline
\multicolumn{6}{c}{\mbox{$\omega$  parameters}}\\\hline
f_{\omega} & f_{\omega}^\perp & a_1^\perp({\omega}) &
a_2^\perp({\omega}) & T_1^{B\to\omega}(0)\\\hline
187(5)\,{\rm MeV} & 151(9)\,{\rm MeV} & 0 & 0.15(7) & 0.25(4)
\\\hline\hline
\multicolumn{6}{c}{\mbox{$K^*$ parameters}}\\\hline
f_{K^*} & f_{K^*}^\perp & a_1^\perp({K^*})\mbox{~\cite{BZ05b}} & 
a_2^\perp({K^*}) & T_1^{B_q\to K^*}(0) & 
T_1^{B_s\to \bar  K^*}(0)\\\hline
220(5)\,{\rm MeV} & 185(10)\,{\rm MeV} & 0.04(3) & 0.15(10) &
0.31(4) & 0.29(4)\\\hline\hline
\multicolumn{6}{c}{\mbox{$\phi$ parameters}}\\\hline
f_{\phi} & f_{\phi}^\perp & a_1^\perp({\phi}) &
a_2^\perp({\phi}) & T_1^{B_s\to\phi}(0)\\\hline
215(5)\,{\rm MeV} & 186(9)\,{\rm MeV} & 0 & 0.2(2) & 0.31(4) & \\\hline\hline
\multicolumn{6}{c}{\mbox{quark masses}}\\\hline
\multicolumn{2}{c|}{m_s(2\,{\rm GeV})\mbox{~\cite{ms}}} 
& m_b(m_b)\mbox{~\cite{inclmoments}} & 
m_c(m_c)\mbox{~\cite{czakon}} & \multicolumn{2}{c}{
m_t(m_t)\mbox{~\cite{mt}}}\\\hline
\multicolumn{2}{c|}{100(20)\,{\rm MeV}} 
& 4.20(4)\,{\rm GeV} & 1.30(2)\,{\rm GeV} & 
\multicolumn{2}{c}{163.6(2.0)\,{\rm GeV}} \\\hline\hline
\end{array}
$$
\caption[]{\small Summary of input parameters. The
 value of $|V_{ub}|$ is our own average over inclusive and exclusive 
determinations and the result from UTangles, see
 Refs.~\cite{HFAG,global,Vub}. None of our results is very sensitive
 to $|V_{ub}|$. For an explanation of our choice of the value of the UT angle
 $\gamma$, see text.
$\lambda_B$ is the first inverse moment
 of the $B$ meson's light-cone DA. 
$\lambda_{B_s}$ is obtained from
  $\lambda_{B_q}$ as described in the text. 
The vector meson
 decay constants $f_V$, $f_V^\perp$ are discussed in the appendix; the
 values of the Gegenbauer moments $a_i^\perp$ are compiled from
 various sources
 \cite{BZ06b,BB96,BBKT,elena} and include only small SU(3) breaking,
 in line with the findings for pseudoscalar mesons \cite{BBL06}.
The form factors $T_1$ are obtained from LCSRs and are 
updates of our previous results
 \cite{BZ04b}, including the updated values of the decay constants
 $f_{\rho,\omega,\phi}$ and of $a_1^\perp({K^*})$ \cite{BZ05b,BZ06a}. 
Note that
 $a_1^\perp({K^*})$ refers to a $(s\bar q)$ bound state; for a $(q\bar
 s)$ state it changes sign. All scale-dependent quantities
 are given at the scale $\mu=1\,$GeV unless stated otherwise.
}\label{tab3}
\end{table}

As for the $a_7$ coefficients, the $a_{7L}^{c,u}( V)$ are, in QCDF, 
of order 1 in a $1/m_b$ expansion,
\begin{equation}
a_{7L}^{c,u}( V) = C_7 + O(\alpha_s,1/m_b)\,,
\end{equation}
whereas $a_{7R}^{c,u}( V)$ are of order $1/m_b$ \cite{grin04,grin05}.
It proves convenient to split these coefficients into three
contributions which we will investigate separately:
\begin{eqnarray}
a_{7L}^U( V) &=& a_{7L}^{U,{\rm QCDF}}( V) + a_{7L}^{U,{\rm ann}}(
 V) + a_{7L}^{U,{\rm soft}}( V)+\dots\,,\nonumber\\
a_{7R}^U( V) &=& a_{7R}^{U,{\rm QCDF}}( V) + a_{7R}^{U,{\rm ann}}(V)+ 
a_{7R}^{U,{\rm soft}}( V)+\dots\,,\label{asplit}
\end{eqnarray} 
where $a_{7L}^{U,{\rm QCDF}}$ is the leading term in the $1/m_b$
expansion; all other terms are suppressed by at least one power of
$m_b$. We only include those
power-suppressed terms that are either numerically large or relevant
for isospin and CP asymmetries. The dots
denote terms of higher order in $\alpha_s$ and
further $1/m_b$ corrections to QCDF, most of which are uncalculable.
Explicit formulae for $a_{7L}^{U,{\rm QCDF}}$, complete to
$O(\alpha_s)$, can be found in Ref.~\cite{BoschThesis}. 
$a_{7L}^{U,{\rm ann}}$ encodes the $O(1/m_b)$ contribution of
the WA diagram of Fig.~\ref{fig1} which drives  the isospin
asymmetries and has been 
calculated, in QCDF, and to leading order in $\alpha_s$, in 
Refs.~\cite{BoschThesis,kagan} for $\rho$ and $K^*$, 
and in Ref.~\cite{BoBu2} for $\omega$. Preliminary results for the
$O(\alpha_s)$ corrections to WA in $B\to\rho \gamma$ were presented in
Ref.~\cite{chamonix}. This
contribution is also relevant for 
the branching ratio of $B\to(\rho,\omega)\gamma$; in this case, also
long-distance photon emission from the soft $B$ spectator quark, which is
$O(1/m_b^2)$, becomes relevant and
has been calculated in Refs.~\cite{emi,WA}. We discuss the WA
contributions in Sec.~\ref{sec:3}. The last terms in (\ref{asplit}), 
$a_{7L(R)}^{U,{\rm soft}}$, encode soft-gluon emission from a (light or
heavy quark) loop as shown in Fig.~\ref{fig1} 
and are particularly relevant for the CP asymmetry;
they will be discussed in Sec.~\ref{sec:4}.
In Ref.~\cite{kagan} also another class of $1/m_b$
corrections to $B\to K^*\gamma$ was calculated, namely $O(\alpha_s)$ 
corrections to the isospin asymmetry in this decay. As these
corrections break factorisation (require an infra-red cut-off in the 
momentum 
distribution of the valence quarks in the $K^*$ meson) and are
numerically small, we do not
include them in our analysis. 
As for $a_{7R}^U$, the dominant contributions to $a_{7R}^c(
K^*)$ were calculated in Ref.~\cite{cpas}. Here we extend this
analysis to other vector mesons and also develop a method to include
light-quark loops.


We conclude this section by providing explicit results for the QCDF
contributions to $a_7$, using the formulae of Ref.~\cite{BoschThesis} and the
Wilson coefficients and hadronic input parameters collected in
Tabs.~\ref{tab2} and \ref{tab3}. The decay constants $f_V$ and
$f_V^\perp$ are defined as
\begin{equation}
\langle 0 | \bar q \gamma_\mu D | V(p)\rangle = m_V f_V \eta_\mu,\qquad 
\langle 0 | \bar q \sigma_{\mu\nu} D | V(p)\rangle = i (\eta_\mu
p_\nu - p_\mu \eta_\nu) f_V^\perp\,;
\end{equation}
their numerical values are discussed in the appendix. The other
parameters in Tab.~\ref{tab3} pertaining to vector mesons are
$a_1^\perp$ and $a_2^\perp$, the first and second Gegenbauer moments of
their transversal light-cone DAs of leading
twist. In this paper we do not want to go into any detail about 
light-cone DAs, their conformal expansion in Gegenbauer polynomials
and the dependence of the Gegenbauer moments on the renormalisation
scale, but simply refer to the relevant
literature \cite{BB96,BBKT,elena}.

It turns out that, at the level of two
significant digits, all $a_{7L}^{c,{\rm QCDF}}$ are equal and so are the
$a_{7L}^{u,{\rm QCDF}}$. For central values of the input parameters of
Tab.~\ref{tab3} we obtain
\begin{eqnarray}
a^{c,{\rm QCDF}}_{7L}(V) & = & -(0.41+0.03i) - 
(0.01+0.01i)\,,\nonumber\\
a^{u,{\rm QCDF}}_{7L}(V) & = & -(0.45+0.07i) + (0.02-0i)\,.\label{10}
\end{eqnarray}
Here we have split the result into contributions from vertex corrections
(first term) and hard-spectator interactions (second term). The size
of the hard-spectator corrections is set by the factor
\begin{equation}
h_V = \frac{2 \pi^2}{9}\,\frac{f_B f_V^\perp}{m_B  
T_1^{B\to V}(0)\lambda_B}\,.
\end{equation}
Note that our value of
$\lambda_{B_s}$, the first inverse moment of the twist-2 $B$-meson
light-cone DA, is obtained from that of
$\lambda_{B_d}$ by a simple scaling argument:
$$
\frac{m_{B_s}}{\lambda_{B_s}}\,(\Lambda_{\rm QCD}+m_s) = 
\frac{m_{B_q}}{\lambda_{B_q}}\,\Lambda_{\rm QCD}\,,
$$
which follows from the assumption that the $B_q$ DA
peaks at the spectator momentum $k_+ = \Lambda_{\rm QCD}$, whereas
that of $B_s$ peaks at $\Lambda_{\rm QCD}+m_s$.

 The parameters
$a_{7R}^{U,{\rm QCDF}}$, at tree level, were
obtained in Ref.~\cite{cpas} and read:
\begin{equation}\label{11}
a_{7R}^{U,{\rm QCDF}}(K^*,\phi) = C_7\,\frac{m_s}{m_b}\,,\qquad
a_{7R}^{U,{\rm QCDF}}(\rho,\omega,\bar K^*) = C_7\,\frac{m_d}{m_b}\,.
\end{equation}

\section{Weak Annihilation Contributions}\label{sec:3}

The intrinsic WA diagram is shown in
Fig.~\ref{fig1}; the weak interaction operator is one of the
charged-current or QCD-penguin operators. All these contributions are
$O(1/m_b)$; photon emission from the $b$ quark and the quarks in the
vector meson is further suppressed and $O(1/m_b^2)$ -- unless the weak
interaction operator is $Q_{5,6}$, which can be Fierz transformed into
$(\bar D (1+\gamma_5) q) (\bar q (1-\gamma_5) b)$ and picks up an
additional factor $m_B$ from the projection onto the $B$ meson
DA which results in this contribution being
$O(1/m_b)$. In Tab.~\ref{tab4} we
show the relative weights of these diagrams in terms of CKM factors
and Wilson coefficients. The numerically largest contribution occurs
for $B^\pm\to \rho^\pm\gamma$: it comes with the large combination of 
Wilson coefficients $C_2+C_1/3=1.02$ and is not CKM suppressed. For
$B^0\to (\rho^0,\omega)\gamma$ it comes with the factor $C_1+C_2/3 = 0.17$
instead and an additional suppression factor $1/2$ from the electric
charge of the spectator quark ($d$ instead of $u$). For all other
decays, WA is suppressed by small (penguin) Wilson coefficients. We evaluate
the annihilation diagrams at the scale $\mu=m_b$. Apart from
$B\to(\rho,\omega)\gamma$, WA is not relevant so much for the
total values of $a_{7L}$, but rather for isospin breaking, which is
set by photon emission from the spectator quark. WA is the only
mechanism to contribute to isospin asymmetries at tree-level; see
Ref.~\cite{kagan} for $O(\alpha_s)$ contributions.

Formulae for $a_{7L}^{U,{\rm ann}}(\rho,K^*)$ in QCDF can be found in
Refs.~\cite{BoschThesis,BoBu2}; 
in this approximation, there is no contribution to 
$a_{7R}^{U,{\rm ann}}$.  In QCDF,  the $a_{7L}^{U,{\rm ann}}$ 
are expressed in terms of the hadronic quantities
\begin{equation}
b^V = \frac{2\pi^2}{T_1^{B\to V}(0)} \,\frac{f_B m_V f_V}{m_B m_b
  \lambda_B}\,, \qquad
d^V_{v} = -\frac{4\pi^2}{T_1^{B\to V}(0)} \,\frac{f_B f_V^\perp}{m_B
  m_b} \,\int_0^1 dv\,\frac{\phi_{2;V}^\perp(v)}{v}
\end{equation}
and $d^V_{\bar v}$, obtained by replacing $1/v\to 1/\bar v$ in the
integrand; $\phi_{2;V}^\perp$ is the twist-2 DA of
a transversely polarised vector meson. 
For $B_s$ decays one has to set $f_B\to f_{B_s}$ and
correspondingly for the other $B$ meson parameters. 
Numerically, one finds, for instance for the $\rho$, $b^\rho
= 0.22$ and $d^\rho = -0.59$, at the scale $\mu = 4.2\,$GeV. As
$T_1\sim 1/m_b^{3/2}$ and $f_B\sim m_b^{-1/2}$ in the heavy quark
limit, these terms are $O(1/m_b)$, but not numerically small because
of the tree-enhancement factors of $\pi^2$.

\begin{table}
\renewcommand{\arraystretch}{1.3}
\addtolength{\arraycolsep}{3pt}
$$
\begin{array}{l||c|c|c|c|c}
\mbox{WA} &  
B^-\to K^{*-} & \bar B^0\to K^{*0} & B\to (\rho,\omega) & B_s\to\phi & 
B_s\to \bar  K^*\\\hline
\mbox{induced by} & \mbox{C (and P)} & \mbox{P} & \mbox{C and P} &
\mbox{P} & \mbox{P}\\
\mbox{CKM} & \lambda^2 \mbox{~(and 1)} & 1 & 1 & 1 & 1
\end{array}
$$
\caption[]{\small Parametric size of WA  contributions to $B\to
  V\gamma$. C denotes the charged-current operators $Q_{1,2}$, P
  the penguin operators $Q_{3,4,5,6}$; their Wilson coefficients are 
  small, see Tab.~\ref{tab2}. CKM denotes the order in
  the Wolfenstein parameter 
$\lambda$ with respect to the dominant amplitude induced by $Q_7$.
}\label{tab4}
\end{table}

For $\omega$, $\bar K^*$ and $\phi$ we obtain
\begin{eqnarray}
\left. a_{7L}^{u,{\rm ann}}(\omega)\right|_{\rm QCDF}
 & = & Q_d b^\omega (a_1 + 2 (a_3+a_5)
+ a_4) + Q_d (d^\omega_v + d^\omega_{\bar v}) a_6\,,\nonumber\\
\left. a_{7L}^{c,{\rm ann}}(\omega)\right|_{\rm QCDF} 
& = & Q_d b^\omega (2 (a_3+a_5)
+ a_4) + Q_d (d^\omega_v + d^\omega_{\bar v}) a_6\,,\nonumber\\
\left. a_{7L}^{U,{\rm ann}}(\phi)\right|_{\rm QCDF} 
& = & Q_s b^\phi (a_3+a_5) + 
         Q_s (d^\phi_v + d^\phi_{\bar v}) a_6\,,\nonumber\\
\left. a_{7L}^{U,{\rm ann}}(\bar K^*)\right|_{\rm QCDF} 
& = & Q_s b^{\bar K^*} a_4 + 
         Q_s (d^{\bar K^*}_v Q_d/Q_s + d^{\bar K^*}_{\bar v}) a_6\,,
\label{15}
\end{eqnarray}
with $a_1 = C_1+C_2/3$, $a_3 = C_3+C_4/3$, $a_4 = C_4+C_3/3$, $a_5 =
 C_5+C_6/3$, $a_6 = C_6+C_5/3$; note that $a_1\leftrightarrow
 a_2$ as compared to \cite{BoschThesis} as in our operator basis
 (i.e.\ the BBL basis) $Q_1$
 and $Q_2$ are exchanged. 
The expressions for $\phi$ and $\bar K^*$ are new; for
$\omega$, we do not agree with \cite{BoBu2}. 
For completeness, we also give the annihilation coefficients for the
 $\rho$, $K^*$ and $\omega$, as obtained in Ref.~\cite{BoschThesis}:
\begin{eqnarray}
\left. a_{7L}^{u,{\rm ann}}(K^{*0})\right|_{\rm QCDF} & = & 
Q_d \left[ a_4 b ^{K^*} + a_6 (d_v^{K^*} + d_{\bar v}^{K^*})\right],
\nonumber\\
\left. a_{7L}^{u,{\rm ann}}(K^{*-})\right|_{\rm QCDF} & = & 
Q_u \left[ a_2 b ^{K^*} + a_4 b ^{K^*} + 
a_6 (Q_s/Q_u d_v^{K^*} + d_{\bar v}^{K^*})\right],
\nonumber\\
\left. a_{7L}^{u,{\rm ann}}(\rho^0)\right|_{\rm QCDF} & = & 
Q_d \left[ -a_1 b^\rho + a_4 b^\rho + a_6 (d_v^\rho + 
d_{\bar v}^\rho)\right],
\nonumber\\
\left. a_{7L}^{u,{\rm ann}}(\rho^-)\right|_{\rm QCDF} & = & 
Q_u \left[ a_2 b^\rho + a_4 b^\rho + a_6 (Q_d/Q_u 
d_v^\rho + d_{\bar v}^\rho)\right],
\nonumber\\
\left. a_{7L}^{c,{\rm ann}}(K^{*0})\right|_{\rm QCDF} & = & 
Q_d \left[ a_4 b ^{K^*} + a_6 (d_v^{K^*} + d_{\bar v}^{K^*})\right],
\nonumber\\
\left. a_{7L}^{c,{\rm ann}}(K^{*-})\right|_{\rm QCDF} & = & 
Q_u \left[ a_4 b ^{K^*} + 
a_6 (Q_s/Q_u d_v^{K^*} + d_{\bar v}^{K^*})\right],
\nonumber\\
\left. a_{7L}^{c,{\rm ann}}(\rho^0)\right|_{\rm QCDF} & = & 
Q_d \left[ a_4 b^\rho + a_6 (d_v^\rho + d_{\bar v}^\rho)\right],
\nonumber\\
\left. a_{7L}^{c,{\rm ann}}(\rho^-)\right|_{\rm QCDF} & = & 
Q_u \left[ a_4 b^\rho + a_6 (Q_d/Q_u 
d_v^\rho + d_{\bar v}^\rho)\right].
\end{eqnarray}
Apart from $\rho$ and $\omega$, all these coefficients are
numerically small and do not change the branching ratio significantly;
the terms in $a_6$, however, are relevant 
for the isospin asymmetries. 

\begin{figure}
$$\epsfxsize=\textwidth\epsffile{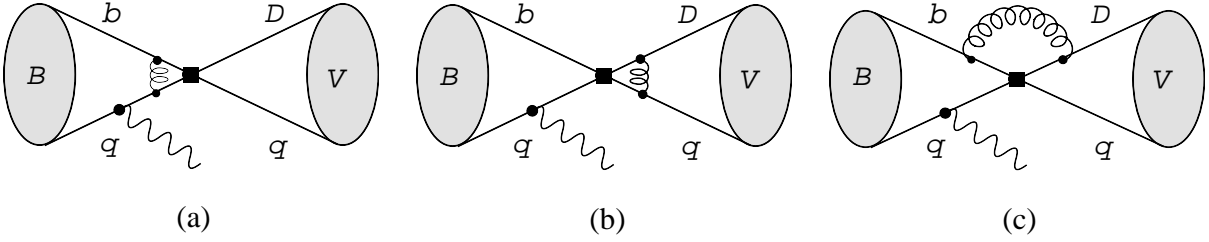}$$
\vspace*{-35pt}
\caption[]{\small Example radiative corrections to 
weak annihilation. The corrections to the $B$ vertex in (a) are known 
\cite{bellnu} and those to the $V$ vertex in (b) are included in $f_V$. 
For the non-factorisable corrections in (c) only preliminary results
are available, see text.}\label{fig2}
\end{figure}

In view of the large size of $a_{7L}^{u,{\rm ann}}(\rho)$ it is
appropriate to have a look at further corrections. The most  obvious ones
are $O(\alpha_s)$ corrections to the QCDF expressions, shown in
Fig.~\ref{fig2}.
As it turns out, the corrections to the
$B$ vertex in Fig.~\ref{fig2}(a)
are known: they also enter the decay $B\to\gamma
\ell\nu$ and were calculated in
Ref.~\cite{bellnu}. Numerically, they are at the level of 10\%. 
Fig.~\ref{fig2}(b) shows the vertex corrections to the $V$ vertex,
which are actually included in the decay constant $f_V$.
For the non-factorisable corrections shown in
Fig.~\ref{fig2}(c) preliminary results have been reported in
Ref.~\cite{chamonix}; according to \cite{chamonix}, these corrections
are of a size similar to the $B$ vertex corrections. 
Another class of corrections is suppressed by
one power of $m_b$ with respect to the QCDF contributions and is 
due to  long-distance photon emission from the
soft $B$ spectator quark. A first calculation of this effect was attempted in
Ref.~\cite{WA} and was corrected and extended in
 Ref.~\cite{emi}. The long-distance photon emission from a soft-quark
 line requires the inclusion of higher-twist terms in the expansion of
 the quark propagator in a photon background field, beyond the
 leading-twist (perturbative) contribution; a comprehensive discussion
 of this topic can be found in Ref.~\cite{kivel}.
The quantity calculated in Ref.~\cite{emi} is
\begin{eqnarray}
\lefteqn{
\langle \rho^-(p)\gamma(q) | (\bar d u)_{V-A} (\bar u b)_{V-A}|
B^-(p+q)\rangle =}\hspace*{3cm}\nonumber\\
& = & e\,\frac{m_\rho f_\rho}{m_B} \eta^*_\mu
\left\{F_V \epsilon^{\mu\nu\rho\sigma} e^*_\nu p_{\rho}
  q_\sigma - i F_A [e^{*\mu} (p q) - q^\mu
    (e^* p)]\right\}\nonumber\\
& = & -e \,\frac{m_\rho f_\rho}{m_B} \left\{ \frac{1}{2}\,
  F_V (S_L+S_R) + \frac{1}{2}\, F_A (S_L-S_R)\right\}\label{problem}
\end{eqnarray}
in terms of the photon-helicity amplitudes $S_{L,R}$. The above
relation differs from the one given in \cite{emi} by an overall sign,
which is due to the different convention used in \cite{emi} (and in
\cite{kivel}) for the covariant derivative: $D_\mu = \partial_\mu
- i e Q_f A_\mu$ instead of $D_\mu = \partial_\mu
+ i e Q_f A_\mu$ as in this paper.
In QCDF, $F_{A,V}$ are given by $Q_u f_B/\lambda_B$
and induce a term $Q_u a_2 b^\rho$ in  $a_{7L}^{u,{\rm
    ann}}(\rho^-)$. The long-distance photon contribution to
$F_{V,A}$ was found to be \cite{emi}
\begin{equation}\label{Fsoft}
F^{\rm soft}_A = -0.07\pm 0.02 \equiv Q_u G_A\,,\qquad 
F^{\rm soft}_V = -0.09\pm
0.02 \equiv Q_u G_V\,.
\end{equation}
with $G_A+G_V = -0.24\pm 0.06$ and $G_V-G_A = -0.030\pm 0.015$.
Again, there is a relative sign with respect to the results in
\cite{emi}. This comes from the fact that the product $e
F_{A,V}^{\rm soft}$ is independent of the sign
convention for $e$, and as we have changed the overall sign of
(\ref{problem}) with respect to \cite{emi}, we also have to change the
sign of $F_{A,V}^{\rm soft}$. Stated differently: the relative sign
between $F_{A,V}^{\rm soft}$ and $F_{A,V}^{\rm hard}$ in \cite{emi} is
wrong because of a mismatch in sign conventions for $e$ in the
covariant derivative. 

In order to obtain concise expressions for $a_{7L(R)}^{U,{\rm ann}}$, it
proves convenient to define one more hadronic quantity:
\begin{equation}
g^\rho_{L,R} = \frac{\pi^2}{T_1^\rho}\,\frac{m_\rho f_\rho}{m_b m_B}\,
(G_V\pm G_A)
\end{equation}
and correspondingly for other mesons. $g_L$ is $O(1/m_b^2)$
as $G_V+G_A$ has the same power scaling in $m_b$ as $T_1$, i.e.\ $\sim
m_b^{-3/2}$, as one can read off from the explicit expressions in
\cite{WA}. The difference $G_V-G_A$, on the other hand, is a twist-3
effect due to three-particle light-cone DAs of the
photon and is suppressed by one more power of $m_b$, i.e.\ $g_R\sim
1/m_b^3$. This quantity will enter the CP asymmetry.
Our final expressions for $a_{7L(R)}^{U,{\rm ann}}$ then read:
\begin{eqnarray}
a_{7L}^{U,{\rm ann}}(V) & = & \left. a_{7L}^{U,{\rm
    ann}}(V)\right|_{\rm QCDF} (b^V\to b^V + g^V_L)\,,\nonumber\\
a_{7R}^{U,{\rm ann}}(V) & = & \left. a_{7L}^{U,{\rm
    ann}}(V)\right|_{\rm QCDF} (b^V\to g^V_R, d^V\to 0)\,.\label{20A}
\end{eqnarray}
Numerically, one has $g^{\rho}_L/b^\rho = -0.3$, so these
corrections, despite being suppressed by one more power in $1/m_b$, are
not small numerically and larger than the known $O(\alpha_s)$
corrections to QCDF from $B\to\gamma\ell\nu$. Based on this, we feel
justified in including these long-distance corrections in our analysis,
while dropping the radiative ones of Figs.~\ref{fig2}(a) and (c).

We conclude this section by listing the numerical values of some of
the annihilation coefficients, for central values of the input
parameters, including in particular those to which
$Q_{1,2}$ contribute (with no Cabibbo suppression):
\begin{eqnarray}
a_{7L}^{c,{\rm ann}}(K^{*0}) & = & -0.013-0.001\, {\rm LD}\,,
\qquad a_{7L}^{c,{\rm ann}}(K^{*-}) =  0.004+0.001\, {\rm
  LD}\,,\nonumber\\
a_{7L}^{u,{\rm ann}}(\rho^0) & = & -0.001-0.004\, {\rm LD}\,,
\qquad ~~ a_{7L}^{u,{\rm ann}}(\rho^-) =  0.149-0.043\, {\rm
  LD}\,,\nonumber\\
a_{7L}^{u,{\rm ann}}(\omega) & = & -0.024+0.003\, {\rm LD}\,.
\end{eqnarray}
The contribution from the long-distance photon emission is labelled
``LD'' (LD$\to 1$ at the end). 
The unexpectedly small $a_{7L}^{u,{\rm ann}}(\rho^0)$ is due
to a numerical cancellation between the charged-current and 
penguin-operator contributions. Comparing these results with those from QCDF,
Eq.~(\ref{10}), it is evident that WA is, as expected, 
largely irrelevant for the
branching ratios, except for $B^\pm\to \rho^\pm\gamma$.

\section{Long-Distance Contributions from Quark Loops}\label{sec:4}

In this section we calculate the soft-gluon emission from quark loops
 shown in Fig.~\ref{fig1}. Again, these contributions are suppressed
by one power of $m_b$ with respect to $a_{7L}^{U,{\rm QCDF}}$, but they
also induce a right-handed photon amplitude which is of the same order
in $1/m_b$ as $a_{7R}^{U,{\rm QCDF}}$. As we shall see in the next
section, this amplitude induces the
time-dependent CP asymmetry in $B\to V\gamma$. 
The asymmetry is expected to be very
small in the SM and $\propto m_D/m_b$ due to chiral suppression of the
leading transition, but could be drastically enhanced by NP
contributions. 
It was noticed in Refs.~\cite{grin04,grin05} 
that the chiral suppression is relaxed by
emission of a gluon from the quark loop, which is the topic of this section.
The task is then not so much to calculate these
contributions to high accuracy, but to exclude the possibility of 
{\em  large}
contributions to the CP asymmetry. For this reason we will be very
generous with the theoretical uncertainties of the results obtained in
this section --- which are currently unavoidable
due to the uncertainties of the relevant hadronic input parameters.

Historically, 
soft-gluon emission from a charm loop was first considered in 
Ref.~\cite{KRSW97} as a potentially relevant long-distance
contribution to the branching ratio of $B\to K^*\gamma$, at about the same
time as similar effects were being discussed for its inclusive
counterpart $B\to X_s\gamma$ \cite{voloshin,ligeti,rey}. 
It was pointed out later, 
in Ref.~\cite{grin05}, that the same diagram also contributes dominantly
to the time-dependent CP asymmetry in $B^0\to K^{*0}\gamma$. The size
of this contribution was calculated only very recently, in
Ref.~\cite{cpas}.
The method used in \cite{cpas} relies on the local operator-product 
expansion of a heavy-quark loop in inverse powers of the
quark mass and hence cannot be used to calculate soft-gluon emission
from light-quark loops, which are doubly Cabibbo suppressed for $b\to
s\gamma$ transitions, but not for $b\to d\gamma$. In Sec.~\ref{4.1} we 
will briefly review the results for heavy-quark loops and in
Sec.~\ref{4.2} we will present a new technique for calculating 
light-quark loops;
however, before we do so, we would like to fix our notation and give explicit
expressions for $a_{7L(R)}^{U,{\rm soft}}$.

Potentially the most important contribution to the soft-gluon emission 
diagram in Fig.~\ref{fig1} 
comes from the charged-current operator $Q_2^U$ with the 
large Wilson coefficient $C_2\sim 1$; it vanishes for $Q_1^U$ 
by gauge invariance. In order to calculate the diagram, it
proves convenient to decompose $Q_2^U$ by a Fierz transformation into
\begin{eqnarray}
Q_2^U &=& \frac13  Q_1^U +  2 \tilde Q_1^U \nonumber\\
\mbox{with}\quad
\tilde Q_1^U &=&  (\bar U    \frac{\lambda_a}{2} U)_{V-A}(\bar D  
\frac{\lambda_a}{2} b)_{V-A}\,.\label{19}
\end{eqnarray}
The contribution of the $U$-quark loop to the $\bar B\to
V\gamma$ amplitude can then be written as
\begin{eqnarray}
\lefteqn{{\cal A}(\bar B \to V \gamma)_{Q_2^U} =  
\Big[ \frac{ G_F}{\sqrt{2}} \lambda_U^{(D)}\Big] \,  C_2 \cdot
 \langle V(p) \gamma(q)| 2 \tilde  Q_1^U | \bar B(p_B)  \rangle} 
\hspace*{0.5cm}\nonumber\\
& = & 
\Big[ \frac{ G_F}{\sqrt{2}} \lambda_U^{(D)}\Big] \,  
C_2  \cdot (-i e) e^{*\al}(q) \sum_q Q_f 
\int d^4 x e^{i q \cdot x} \langle V(p)| T
(\bar q \gamma_\al q \big)  (x)  2 \tilde Q_1^U(0)| \bar B(p_B)
\rangle \hspace*{0.7cm} \label{20} \\
& \equiv & \Big[ \frac{ G_F}{\sqrt{2}} \lambda_U^{(D)}  \Big] \, (-eQ_U) \, 
C_2 \cdot \Gamma^U_{VB} \,,\label{eq:loop_cu}
\end{eqnarray}
where the minus sign comes from the sign convention for $e$
as discussed in Sec.~\ref{sec:2}. We decompose $\Gamma^U_{VB}$ into
contributions from the photon-helicity amplitudes 
$S_{L,R}$, Eq.~(\ref{6}), as
\begin{equation}
\label{eq:general}
\Gamma^U_{VB} =  l_U(V) P  + 
\tilde l_U(V) \tilde P 
\end{equation}
with 
\begin{eqnarray}
\label{eq:projector}
P &\equiv& \epsilon^{\mu\nu\rho\sigma} e_\mu^* \eta_\nu^* p_\rho q_\sigma 
= \frac12(S_L + S_R)\,,\quad \nonumber \\
\tilde P &\equiv&  i \{ (e^* \eta^*) (pq) - (e^*p)(\eta^* q)\} = 
\frac12(S_L - S_R)\,.
\end{eqnarray}
In addition to $Q_2^U$, the penguin operators $Q_{3,4,6}$ give a
non-zero contribution to soft-gluon emission. Including all these
contributions, and comparing
(\ref{eq:loop_cu}) with (\ref{ME}) and (\ref{6}), we obtain the
following expression for $a_7^{U,{\rm soft}}$:
\begin{eqnarray}
a_{7L(R)}^{U,{\rm soft}}(V) & = &  \frac{\pi^2}{m_b T_1^{B\to V}(0)}
\left\{ Q_U C_2 (l_U\pm \tilde l_U)(V) + Q_D C_3 (l_D\pm \tilde
l_D)(V)\right.\nonumber\\
&&\left. + \sum_q Q_q (C_4-C_6) (l_q\pm \tilde l_q)(V)\right\}.\label{24}
\end{eqnarray}
Here the sum over $q$ runs over all five active quarks
$u,d,s,c,b$. $D$ denotes the down-type quark in the $b\to D\gamma$
transition. The minus sign in front of $C_6$ is due to Furry's theorem
according to which only the axial-vector current in the $\bar U U$
term in (\ref{19}) contributes to $\Gamma^U_{VB}$. We do not include
the contribution from $Q_5$ because its Fierz transformation changes the
chirality of the current so that the resulting loop contribution is
proportional to $m_D$ and hence helicity suppressed. In the following
we distinguish between heavy ($b$, $c$) and light ($u$,$d$,$s$) quark
loops. Assuming SU(3)-flavour symmetry, one has $l_u=l_d=l_s$, and
ditto for $\tilde l_{u,d,s}$, which
 causes a cancellation of these contributions in the last term
in (\ref{24}). As to be discussed below, in Sec.~\ref{4.2}, we 
estimate the SU(3)-breaking effects to be around 10\%.

We now turn to the calculation of $l_{b,c},\tilde l_{b,c}$ and $l_u,
\tilde l_u$.

\subsection{Heavy-Quark Loops}\label{4.1}

The calculation of $l_c(K^*)$ and $\tilde l_c(K^*)$ was presented in 
Ref.~\cite{cpas}; here we briefly recapitulate the method and present 
results also for $l_b$, $\tilde l_b$ and for $\rho,\omega,\bar K^*,\phi$.

It was first noticed in Ref.~\cite{KRSW97} that soft-gluon emission
from a charm loop, Fig.~\ref{fig1}(b),
 is suitable for an operator product expansion (OPE)
in $1/m_c$ since the on-shell photon is far away from the partonic
threshold $4 m_c^2$. The OPE reads, to leading order in $1/m_c$ 
\cite{KRSW97}:
\begin{eqnarray}
\label{eq:OPEmc2}
i e^\mu \int d^4 x  e^{i q \cdot x} T \bar c \gamma_\mu c (x) 
2 \tilde Q_1^c =  Q_F + O(1/m_c^4)
\end{eqnarray}
where
$$
Q_F \equiv c_F  \bar D \gamma_\rho(1-\gamma_5) \frac{\lambda_a}{2} 
g \tilde G_{\alpha \beta}^a D^\rho
(F^{\alpha\beta}) b
$$
with 
\begin{equation}\label{cF}
c_F = -1/(48 \pi^2m_c^2)
\end{equation}
 and $F_{\alpha \beta} =
i(q_\alpha e^*_\beta- q_\beta e^*_\alpha)$ for an outgoing photon.
Note that here the sign of $g$ corresponds to the covariant derivative
$D_\mu = \partial - i g T^a A^a_\mu$ which differs from the sign
convention used in Sec.~\ref{sec:2}, but agrees with that used as a
standard in hard-perturbative QCD calculations. Our final results for $l_c$,
however, are independent of the sign of $g$, as the matrix element of
$Q_F$ over mesons will be expressed in terms of three-particle
light-cone meson DAs containing an explicit factor $g$ which
refers to the same convention. 

The matrix element of $Q_F$ can be expressed in terms of $l_c$,
$\tilde l_c$ as 
\begin{equation}
\Gamma^c_{VB} =  \langle V(p)| Q_F | B(p_B) \rangle  =  
 l_c(V) P  +  \tilde l_c(V) \tilde P\,.  
\end{equation}
The parameters $l_c(K^*)$ and $\tilde l_c(K^*)$ were first calculated in
Ref.~\cite{KRSW97} from three-point QCD sum-rules. In
Ref.~\cite{cpas} we calculated them from 
LCSRs, which are more suitable for
the problem
than three-point sum rules, see the discussion in \cite{cpas}. 
The sum rules were obtained for the quantities 
$L$ and $\tilde L$,
which are related to $l_c$ and $\tilde l_c$ by
\begin{equation}
c_F L = \frac{1}{2}\, l_c\,,\qquad c_F \tilde L = \frac{1}{2}\, \tilde
l_c
\end{equation}
with $c_F$ given in Eq.~(\ref{cF}).
The LCSR for $L$ reads \cite{cpas}
\begin{eqnarray}
\frac{m_B^2 f_B}{m_b}\, L\, e^{-m_B^2/M^2} = m_b^4\, 
\int_{u_0}^1 du\, e^{-m_b^2/(u M^2)} 
\left[ f_V \left(\frac{m_V}{m_b}\right)  R_1(u) +  
f_V^\perp \left(\frac{m_V}{m_b}\right)^2  R_2(u)  \right],
\end{eqnarray}
where $R_1$ and $R_2$ are given in terms of three-particle
twist-3 and 4 DAs of the vector meson. 
Explicit expressions are given in \cite{cpas}.
The sum rule for $\tilde L$ is analogous.

In this paper we update the values of $l_c(K^*)$ and $\tilde l_c(K^*)$
as determined in Ref.~\cite{cpas} and also calculate these
parameters for the other vector mesons. The twist-3 and 4 parameters
entering $R_{1,2}$ are given in Tabs.~\ref{tab:kappas} and
\ref{twist4}. The results for $l_c$ and $\tilde l_c$ are given in
Tab.~\ref{tab_lc} \footnote{The values obtained in \cite{KRSW97} with
  local three-point sum rules are
$l_c(K^*) =   -(1374 \pm 250) \, {\rm keV}$ and 
$\tilde l_c(K^*) =    -(1749 \pm 250) \, {\rm keV}$. The
  quoted uncertainty includes solely the variation of the
  Borel parameter and therefore probably underestimates
  the uncertainty. The central values are  
substantially larger than those obtained from LCSRs, Tab.~\ref{tab_lc}. 
It is well known that three-point sum rules are inappropriate for 
$b$ transitions since higher-order condensate contributions grow with 
$m_b$ and destroy the hierarchy of perturbative and non-perturbative
contributions.}.
 Those for $l_b$ and $\tilde l_b$ are obtained as
\begin{equation}
l_b = \frac{m_c^2}{m_b^2}\, l_c\,,\qquad \tilde l_b =
\frac{m_c^2}{m_b^2}\, \tilde l_c\,.
\end{equation}
Although the total uncertainties in these parameters are rather large,
their contribution to $a_{7L}^U$ is of $O(2\%)$ at best and has an
only minor impact on the branching ratios. Soft-gluon
emission is also irrelevant for isospin asymmetries; its main impact
is on the CP asymmetry which is small by itself, so even large
uncertainties are acceptable if the aim is to rule out a numerically
sizable time-dependent CP asymmetry in the SM.
\begin{table}[p]
\renewcommand{\arraystretch}{1.3}
\addtolength{\arraycolsep}{3pt}
$$
\begin{array}{l || c | c |  c}
& \rho,\omega & K^*  &  \phi\\
\hline
\zeta_{3V}^\parallel 
& 0.040(8) & 0.026(6)  & 0.03(1) 
\\
\widetilde\lambda_{3V}^\parallel 
& 0 & 0.08(3)& 0
\\
\widetilde\omega_{3V}^\parallel 
& -0.085(25) & -0.07(2) & -0.035(2)
\\
\kappa_{3V}^\parallel 
& 0 & 0.0005(5) & 0
\\
\omega_{3V}^\parallel 
& 0.20(7) & 0.11(3) & 0.045(3)
\\
\lambda_{3V}^\parallel 
& 0 & -0.020(8) & 0
\\
\kappa_{3V}^\perp 
& 0 & 0.005(2) & 0 
\\
\omega_{3V}^\perp 
& 0.65(25) & 0.35(10) &  0.26(10)
\\
\lambda_{3V}^\perp 
& 0 & -0.05(2) & 0 
\end{array}
$$
\renewcommand{\arraystretch}{1}
\addtolength{\arraycolsep}{-3pt}
\caption[]{\small Three-particle twist-3 hadronic parameters 
at the scale  $\mu= 1\,{\rm GeV}$.  The parameters $\lambda$ and
$\kappa$ are G-odd whereas the parameters $\zeta$ and $\omega$ are
G-even. The results for $K^*$ are updates of those published in
\cite{cpas}, those for $\rho$ updates of \cite{BBKT} and those
for $\phi$ are new. We assume the parameters of $\rho$ and $\omega$ to
be equal. A full derivation of these
results will be published elsewhere. 
Note that the absolute sign of all these parameters
  depends on the sign convention chosen for the strong coupling
  $g$, see text, and that $K^*$ refers to an $(s\bar q)$ bound state.
}\label{tab:kappas}
\renewcommand{\arraystretch}{1.3}
\addtolength{\arraycolsep}{3pt}
$$
\begin{array}{l || c | c | l}
& \rho,\omega,\phi & K^* & \mbox{Remarks}\\\hline
\zeta_{4V}^\perp 
& 0.10\pm 0.05 &   0.10\pm 0.05 
& \mbox{from \cite{BBK}; no SU(3) breaking; to be updated in \cite{BBJL}}
\\
\widetilde\zeta_{4V}^\perp 
& =-\zeta_{4V}^\perp &=-\zeta_{4V}^\perp 
& \mbox{ditto}
\\
\kappa_{4V}^\perp 
& 0 & 0.012(4) 
& \mbox{G-odd; quoted from \cite{BZ06a}}
\end{array}
$$
\caption[]{\small Three-particle twist-4 hadronic parameters 
at the scale  $\mu= 1\,{\rm GeV}$. The same remark about the absolute
sign and the meaning of $K^*$ 
applies as for twist-3 parameters. }\label{twist4}
\renewcommand{\arraystretch}{1.3}
\addtolength{\arraycolsep}{3pt}
$$
\begin{array}{l ||r|r|r|r}
& \multicolumn{1}{c|}{l_c} & \multicolumn{1}{c|}{\tilde l_c} 
& \multicolumn{1}{c|}{l_c-\tilde l_c} & \multicolumn{1}{c}{l_c +
    \tilde l_c} \\
 \hline
 B \to K^*   & -355 \pm 280 &  -596 \pm 520 & 242 \pm 370 & -952 \pm 800  \\
 B \to (\rho,\omega)   & -382 \pm 300 & -502\pm  430 & 120 \pm 390 & 
-884\pm 660  \\
 B_s \to \bar K^*   &  -347\pm 260 & -342\pm 400 & -4\pm 300 & -689\pm 600 \\
 B_s \to \phi   & -312 \pm  240 & -618 \pm 500 & 306 \pm 320 & -930 \pm
750 
\end{array}
$$
\caption[]{\small Soft-gluon contributions from $c$-quark loops in units
keV. The quantities $l_c$ and $\tilde l_c$ are defined in \eqref{eq:loop_cu}
and \eqref{eq:general}. We assume equal parameters for $\rho$ and $\omega$.
$l_b$ is obtained as $l_b = l_c m_c^2/m_b^2$ and correspondingly for
$\tilde l_b$.
}\label{tab_lc}
\end{table}
In obtaining these results we use the one-loop pole mass
$m_b = (4.7 \pm 0.15) \, {\rm GeV}$,
$f_B$ and $f_{B_s}$ as given in Tab.~\ref{tab3}, the Borel 
parameter $M^2 = (12 \pm 3) \,{\rm GeV}^2$, the continuum threshold 
$s_0 = (35 \pm 2) \,{\rm GeV}^2$ and the renormalisation scale 
$\mu^2 = m_B^2-m_b^2\pm 1\,{\rm GeV}^2$ \cite{BZ04b}. 
The uncertainties of the DAs are given in 
Tab.~\ref{tab:kappas}. Within the accuracy of the sum rules for these
parameters it is impossible to distinguish between $\rho^0$ and
$\omega$, so we assume them to be equal. In view of the large
uncertainties 
associated with the parameters of the three-particle DAs we have adopted a
conservative way to estimate the total uncertainty of $l_c$ and $\tilde l_c$
in Tab.~\ref{tab_lc} and added the uncertainties linearly.
The uncertainties are sizable in $m_b$, $f_B$, 
$\zeta_3^\parallel$ and $M^2$. It is worth noting that the differences 
$l_c(V)-\tilde l_c(V)$ hardly depend on the Borel parameter $M^2$.

Let us turn to the issue of the convergence of the $1/m_c$ expansion
which was discussed, for the inclusive case, in
Refs.~\cite{ligeti,rey}. Higher-order terms in the expansion of
(\ref{eq:OPEmc2}) contain operators of type $\bar D (q\cdot D)^n
\tilde G b$; the expansion can be resummed with the result given in
Ref.~\cite{rey}. For inclusive decays, the relevant matrix elements
are  $\matel{B}{\bar b D^n \tilde G b}{B}$, which can be estimated, 
on dimensional grounds, as $\matel{B}{\bar b D^n \tilde G b}{B} \approx
\Lambda_{\rm QCD}^n \matel{B}{\bar b  \tilde G b}{B}$.
The expansion parameter is then $t\equiv (m_b \Lambda_{\rm QCD})/(4 m_c^2) 
\approx 0.2$, which is not power suppressed, but not large numerically.
For $t=0.2$ the effect of  resummation is to enhance the leading-order
matrix element by 15\%, whereas for 
$t = 0.4$ it amounts to a 30\% enhancement. We expect the resummation 
to have a similar effect in exclusive decays. We shall
include the effect of truncating the $1/m_c$ expansion by
doubling the theoretical uncertainty of our final result for the CP
asymmetries, which depend on $l_c-\tilde l_c$; the impact of
$l_c+\tilde l_c$ on the branching ratios is small.
We also would like to mention, as noted in \cite{rey}, that besides 
the derivative expansion in the  gluon field there are further 
higher-twist contributions from e.g.\  two gluon fields.
These contributions, however, are truly power suppressed and of order
$\Lambda_{\rm QCD}^2/m_c^2$, so we feel justified neglecting them.

\subsection{Light-Quark Loops}\label{4.2} 

For light-quark loops the photon is almost at threshold and the 
local OPE does not apply, unlike the case of heavy quarks 
discussed in the previous subsection. In this subsection we develop a
 method for calculating these contributions which starts from
the calculation of $\Gamma_{VB}^u(q^2)$ from LCSRs for an 
off-shell photon momentum
$q^2\neq 0$. We then shall use a dispersion relation to relate
the off-shell matrix element to $\Gamma^u_{VB}(0)$, which in turn can be
expressed  in terms of the wanted quantities
$l_u(V)$ and $\tilde l_u(V)$, Eq.~(\ref{eq:general}). 
The  starting point of the method
developed here is
similar to the one used by Khodjamirian for the calculation of
soft-gluon contributions to $B\to\pi\pi$ \cite{K00}. 
In order to simplify notations, we drop
the superscript $u$ on the correlation function.

\begin{figure}
$$\epsfxsize=0.3\textwidth\epsffile{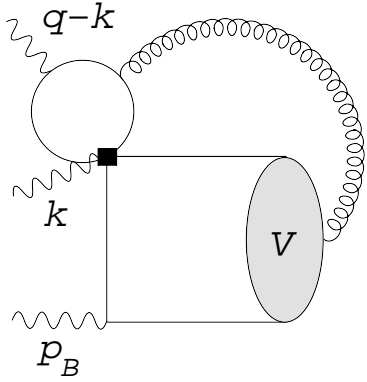}$$
\vspace*{-20pt}
\caption[]{\small Leading contribution to the correlation function $\Gamma_V$
in (\ref{32}). The black square denotes insertion of the operator
$Q_i$ with $i=1,\dots,6$. The $B$ meson momentum is $p_B = p+q $
and the vector meson carries the momentum $p$.}\label{fig3}
\end{figure}

A suitable correlation function for extracting $\Gamma^u_{VB}(0)$ for
the weak interaction operator $Q_2$ is
\begin{equation}\label{32}
\Gamma_V((q- k)^2,p_B^2,P^2) = i^2 e^{*\rho} \int d^4x d^4y 
e^{i(q- k)\cdot x}e^{-ip_B\cdot y}
\langle V(p)| T [\bar u\gamma_\rho u](x) 2 \tilde Q_1^u(0) J_B(y)|0 \rangle
\end{equation}
with $\tilde Q_1^u$ defined as in the previous subsection and
$p_B \equiv p+q$, $P\equiv p_B-k$. 
The current $J_B = m_b \bar b i \gamma_5 q$ is the interpolating field
of the $B$ meson with
\begin{equation}
\langle  B(p_B) |J_B|0\rangle = m_B^2  f_B\,.
\end{equation}
The leading-order contribution to this correlation function, with a
soft gluon, is shown in Fig.~\ref{fig3}.
Following Ref.~\cite{K00}, we have introduced 
an unphysical momentum $k$ at the weak vertex. This additional
momentum serves to avoid unphysical low-lying cuts in $p_B^2$, also 
known as parasitic terms. We will choose the momentum configuration in
such a way that $k$ disappears when extracting
$\Gamma_{VB}^u(0)$. The kinematics of the correlation function 
describes  
a 2--2 scattering process and therefore depends on six independent 
momentum squares. Three of those, namely
\begin{equation}
 P^2  \, , \, (q-k)^2  \,,\, (p_B^2-m_b^2)  
 \ll -\Lambda_{\rm QCD} ^2
\end{equation}
are chosen to lie below their respective thresholds assuring 
that the correlation function is dominated by light-like
distances and therefore suitable for a light-cone expansion.
The other three independent variables are $p^2$, $k^2$ and
$q^2$. Neglecting higher-order corrections in the vector-meson mass,
we set $p^2=0$ and, for simplicity, $k^2=0$. We also set  
$q^2=(q-k)^2$, which will be necessary for avoiding 
a subtraction constant in the dispersion relation in $q^2$ and also 
leaves only one remnant of the presence of the unphysical momentum
$k$: $P^2=(p_B-k)^2\neq p_B^2$.
The rationale for this choice of kinematics 
will become more transparent below. 

Inserting a complete set of hadron states, the correlation function becomes
\begin{equation}
\label{eq:mainsr}
\Gamma_V(q^2,p_B^2,P^2) =  (m_B^2 f_B) \, 
\frac{\Gamma_{VB}^*(q^2,P^2)}{m_B^2-p_B^2} + \dots,
\end{equation}
where the dots stand for higher states and the star 
on $\Gamma_{VB}^*$ is to 
remind one of the presence of the unphysical momentum $k$ in 
$P^2$.
We can decompose the correlation function as
$$
\Gamma_V =  \gamma_V P + 
\tilde \gamma_V \tilde P   + O(k)\,,
$$ 
with the projectors $P$ and $\tilde P$  given in Eq.~\eqref{eq:projector}.
Additional structures in $k$ are unphysical and  can be dropped.
Calculating the $u$-quark loop to twist-3 accuracy, 
we get
\begin{eqnarray}
\label{eq:eq1}
\gamma_V  &=&   \frac{f_V m_b^2 m_V}{48 \pi^2}
\int_{(v,\underline \alpha)} 
\frac{   v\,  (P^2-(q-k)^2) }{ l^2 (p_b^2-m_b^2)}\,
{\cal V}(\underline{\alpha})\,, \\
\label{eq:eq2}
\tilde \gamma_V &=&   \frac{f_V m_b^2 m_V}{48 \pi^2}   
\int_{(v,\underline \alpha)} 
\frac{   v \, ((q-k)^2-P^2)  }{l^2 (p_b^2-m_b^2)}\,
{\cal A}(\underline{\alpha})\, ,
\end{eqnarray}
where $l \equiv q - k + v\alpha_3 p$ and $p_b \equiv q + \bar \alpha_1
p$, $\bar\alpha_1 \equiv 1-\alpha_1$ 
and therefore
\begin{equation}
\label{eq:momenta}
l^2 = v\alpha_3 P^2 + (1-v \alpha_3) (q-k)^2\,, \qquad 
 p_b^2 = \alpha_1 q^2 + \bar \alpha_1 p_B^2\,,
 \end{equation} 
where in our choice of kinematics $(q-k)^2 \to q^2$ in the
sequel. ${\cal V}$
and $\cal A$ are twist-3 three-particle DAs of the vector meson; they
are discussed in detail in Ref.~\cite{BBKT}.
The quantities $\gamma_V$ and
$\tilde \gamma_V$ also receive contributions of higher twist, which we do
not include in this paper. The 
integration measure is defined as
\begin{equation}
\int_{(v,\underline \alpha)} =   \int_0^1  dv  \int_0^1  
d\alpha_1  d\alpha_2  d\alpha_3 \delta ( 1- \alpha_1 - 
\alpha_2 - \alpha_3)\, .
\end{equation}
Eqs.~\eqref{eq:eq1}, \eqref{eq:eq2} and \eqref{eq:momenta} 
clearly show that the introduction of
the unphysical momentum $k$ avoids a low-lying cut (parasite) from the 
$u$-quark loop in the variable $p_B^2$.

The parasitic term in  $q^2$, however, which originates from
the $b$-quark propagator going on-shell,
is not absent  for our choice $q^2 = (q- k)^2$. 
It induces a parasitic term to be added to 
\eqref{eq:mainsr} which is of the form
\begin{equation}\label{40}
\matel{V(p)}{J_B}{B_D(q)}\frac{e^{*\rho}}{m_{B_D}^2-q^2} \int d^4 x e^{i
  (q- k)\cdot x} \matel{B_D(q)}{T 2 \tilde O_1(0) [\bar
  u\gamma_\rho u](x)}{0} + O(k)\,.
\end{equation}
The matrix element on the left-hand side is just the form factor
$A_0(p_B^2)$ for $B\to V$ transitions, which was calculated from LCSRs in 
\cite{BZ04b} and exhibits a pole $\sim 1/(m_B^2-p_B^2)$ in $p_B^2$ 
inducing a parasitic contribution to the first  
term in Eq.~\eqref{eq:mainsr}. Before we can proceed any further, 
we need to determine the size of this parasitic contribution.
If we were dealing with a $c$-quark loop, we could apply a local
  OPE to the integral and calculate  its value from
\eqref{eq:OPEmc2} and  the following  estimate based on dimensional analysis:
\begin{equation}
\matel{B_D}{\bar D \gamma_\rho (1-\gamma_5) \tilde G_{\alpha\beta} b}{0}
\simeq f_B (p_B)_\rho  t_{\alpha \beta} \cdot  \Lambda_{\rm QCD}^2\,,
\end{equation}
where $t_{\alpha_\beta}$ is an antisymmetric dimensionless tensor.
The ratio of the parasitic term (\ref{40}) to the main term in 
(\ref{eq:mainsr}) is then of order 
$\Lambda_{\rm QCD}^2/m_b^2 \sim 1\%$ and  negligibly small. 
For the $u$ loop, on the other hand, the local OPE is not applicable
and one is back to our initial problem of devising a method to
calculate a non-local correlation function, although in this case a
simpler one
than that in Eq.~(\ref{20}). Given, however, the smallness of the
parasitic term for heavy quarks $\sim 1\%$ ,  it is unlikely that this term
is an order of magnitude larger for light quarks, especially in view
of the numerical closeness of the light-quark and heavy-quark loops. 
For our cases of interest even a
contamination at the level of 50\% would not constitute a major
problem, as these contributions to $B\to V\gamma$ are only relevant
for the time-dependent CP asymmetry which is expected to be near zero
in the SM. Our major aim is to confirm that this is indeed the case and 
to exclude large contributions from soft-gluon emission, but not to
give a precise determination of their size. In view of this, even a large
parasitic contamination is perfectly acceptable. 

The next step is to write (\ref{eq:eq1}) and (\ref{eq:eq2}) 
in terms of a dispersion relation in $p_B^2$,   
\begin{eqnarray}
\label{eq:gdis}
\gamma_V(q^2,p_B^2,P^2) = \frac{1}{\pi} \int_{m_b^2}^\infty 
\frac{ds}{s-p_B^2}\, {\rm Im}_s \gamma_V(q^2,s,P^2)\,, 
\end{eqnarray}
in order to match them to the hadronic representation (\ref{eq:mainsr}).
The quantity $l_u^*(q^2,P^2)$ can then be obtained by applying the 
standard QCD sum rule
techniques, namely Borel transformation and continuum subtraction,
which yield
\begin{equation}
\label{eq:lu}
l_u^*(q^2,P^2) = \frac{1}{m_B^2 f_B} \frac{1}{\pi} \int
_{m_b^2}^{s_0^B} ds\,  e^{(m_B^2-s)/M^2} {\rm Im}_s \gamma_V(q^2,s,P^2)\,,
\end{equation}
where the star again indicates the presence of the unphysical momentum 
$k$ in $P^2= (p_B-k)^2$; also note that the photon is still off-shell. 
Once ${\rm Im}_s \gamma_V(q^2,s,P^2)$ is known, $l_u^*(q^2,P^2)$ can be 
analytically continued in $P^2 \to m_B^2 + i 0 $. For the $B$-meson
ground state this removes the last trace of the unphysical momentum
$k$.
The analytic continuation in $P^2$ is justified because it is far above
the other hadronic scales in the corresponding channel.
This yields the physical $u$-quark amplitude, yet still for an off-shell
photon. 

After the conceptual outline given above we will now
outline
how to proceed from the intermediate results \eqref{eq:eq1} and
\eqref{eq:eq2}.
In order to obtain the imaginary parts of $\gamma_V$ and
$\tilde\gamma_V$, it proves convenient to perform some of the 
integrations over 
the variables $v$ and $\alpha_i$ until logarithms appear whose
imaginary parts (cuts) can  easily be identified.
It turns out that the integrals over $dv$ and $d\alpha_1$ with $\alpha_2
=  1- \alpha_1 - \alpha_3$ are elementary 
since the involved variables are space-like which
guarantees the absence of singularities. 
We obtain an expression of the form 
\begin{eqnarray}
\gamma_V\sim
\int_0^1 d\alpha_3 \frac{1}{(P^2-q^2)(p_B^2-q^2)^3}  \Big\{ 
\big(({\rm ln}[m_b^2 - p_B^2]-{\rm ln}[m_b^2 - \alpha_3  p_B^2- \bar 
\alpha_3 q^2])P_1+P_2\big)  \nonumber \\
\big({\rm ln}[ - q^2]-{\rm ln}[- \alpha_3  P^2- \bar
  \alpha_3 q^2]P_3+P_4\big) \Big\} P_5 \,,
\end{eqnarray} 
where $P_i$ stands for polynomials. 
The poles in $q^2$ are integrable, i.e.\  removable.
The dispersion representation in $p_B^2$ is now obtained from the cuts 
of the logarithms. We find
\begin{displaymath}
- \int\limits_{m_b^2}^{\infty} \frac{ds}{s-p_B^2}
\int\limits^{\frac{m_b^2-q^2}{s-q^2}}_0 d \alpha_3  
\Big(({\rm ln}[- q^2]-{\rm ln}[ - \alpha_3  P^2- \bar 
\alpha_3 q^2])P_3+P_4\Big) \frac{P_1 P_5}{(P^2-q^2)(p_B^2-q^2)^3}\,.
\end{displaymath}
The integral over $d \alpha_3$ is elementary and we finally obtain 
the imaginary part for the dispersion relation \eqref{eq:gdis}: 
\begin{eqnarray}
\left.\frac{1}{\pi}{\rm Im}_s \gamma_V(q^2,s,P^2)\right|_{s \geq m_b^2} &=& 
\frac{f_V m_b^2 m_V}{
8 \pi^2(P^2-q^2)^3(s-q^2)^5}\,
\big({\rm ln}[ -q^2]+{\rm ln}[s - q^2]\nonumber\\
&&{}- {\rm ln}[ -m_b^2 P^2 -q^2(s  - 
m_b^2-P^2) ]+P_6\big) P_7\,.\label{eq:lstar}
\end{eqnarray}
This is the expression to be used in Eq.~(\ref{eq:lu}).
As discussed above, the momentum $k$ completely disappears
upon analytic continuation of $P^2 \to m_B^2+i 0 $ 
and we obtain the amplitude for an off-shell photon.
The analytic continuation
is rather straightforward: $l_u^*$ acquires an imaginary part from
those logarithms whose arguments depend on $P^2$. 
The imaginary part is proportional to the mass of the $u$ quark and
originates from the quark going on-shell. After the analytic
continuation of (\ref{eq:lstar}), 
all remnants of the unphysical momentum $k$ have
disappeared and we can drop the star from now on:
\begin{equation}
l_u(q^2) \equiv l_u(q^2,m_B^2+i0) = \frac{1}{m_B^2 f_B} \frac{1}{\pi} \int
_{m_b^2}^{s_0^B} ds\,  e^{(m_B^2-s)/M^2} {\rm Im}_s 
\gamma_V(q^2,s,m_B^2+i0)\,, \label{46}
\end{equation}
for $q^2 \ll -\Lambda_{\rm QCD}^2$.
It is interesting to note that if one does not project
onto the $B$ ground state the analytic continuation leads to
unphysical cuts  in negative $q^2$ which come from the fact that for 
higher states the unphysical momentum $k$ is still present.

There remains only one step to be done, namely to put 
the photon on-shell, i.e.\ $q^2\to 0$. To do so, we follow the method 
used in Ref.~\cite{K97}, where the pion-photon-photon 
 transition form
factor $F_{\gamma^*\gamma \pi}$ was estimated with one on-shell photon.
Since $l_u(q^2)$ is an analytic function in $q^2$, it has the 
standard dispersion representation
\begin{equation}
\label{eq:lu1}
l_{u}(q^2) = \frac{1}{\pi}\int_{{\rm cut}}^\infty dt\, 
\frac{{\rm Im}_t l_{u}(t)}{t-q^2}
\end{equation}
for $q^2$ below the cut. Potential subtraction terms spoiling 
the above representation are absent. This can be seen as follows:
for very large Euclidian values $-q^2\gg \Lambda_{\rm QCD}^2$ one can
perform a local OPE very much the same way as for $c$-quark loops with 
the expansion coefficient $1/m_c^2 \to 1/q^2$. Using this result 
we have explicitly
verified that  $ l_u(q^2),\, \tilde l_u(q^2) \stackrel{q^2 \to \infty}{\sim} 
1/q^2 $.
Another indication, although not sufficient,
is that the explicit calculation of $l_u^*(q^2,P^2)$ does
not contain a constant or polynomial terms in $q^2$.
The imaginary part in $q^2$ comes from the logarithms in 
\eqref{eq:lstar} and
the poles in $1/(q^2-s)$. The poles in $P^2 = m_B^2 +i0$ are again
integrable or removable.

The perturbative or parton representation has a cut starting at $0$,
Eq.~(\ref{eq:lu1}),
and it is therefore impossible to set $q^2 =0$ because
it is right below the perturbative threshold.
The idea is then to cut out this lower part by inserting resonances 
that couple to the $\bar u \gamma_\mu u$ current;
$q^2 = 0$ is then sufficiently below the resonances
and the corresponding continuum threshold.
We shall content ourselves with the two lowest 
resonances $\rho$ and $\omega$. Treating them as equal, we have
\begin{equation}
\label{eq:lu2}
l_{u}(q^2) =   \frac{2 r_{\rho}}{m_{\rho}^2 -q^2}   + 
\frac{1}{\pi}\int_{s_0^\rho}^\infty  \frac{{\rm Im}_t l_{u}(t)}{t-q^2}
\end{equation}
where 
$$
r_\rho  \;  P  + \tilde r_\rho  \;  \tilde P    = e^*_\mu \sum_{\rm
  pol} \matel{0}{\bar u \gamma^\mu
  u}{\rho} \matel{\rho V} { 2 \tilde Q_1} {B}\,,
$$
and the sum runs over the polarisation of the $\rho$.
It remains to  determine  $r_\rho$ (and $\tilde r_\rho$), so 
that Eq.~\eqref{eq:lu2} can be used to extract $l_u(0)$.
This can be achieved by applying a Borel transformation 
in the variable $q^2$ which yields the estimate
\begin{equation}\label{49}
2 r_\rho =  \frac{1}{\pi } \int_{0}^{s_0^\rho} dt \, {\rm Im}\,
l_{u} (t) e^{(m_\rho^2-t)/M^2}
\end{equation}
and  finally
\begin{equation}
\label{eq:lu3}
l_u \equiv l_{u}(0) = \frac{1}{\pi } \int_0^{s_0^\rho}
\frac{dt}{m_\rho^2}e^{
(m_\rho^2-t)/M^2}\; {\rm Im}\, l_{u}(t)  + 
\frac{1}{\pi}\int_{s_0^\rho}^\infty 
\frac{dt}{t}  \; {\rm Im}\, l_{u} (t)\,.
\end{equation}
\begin{table}
\renewcommand{\arraystretch}{1.3}
\addtolength{\arraycolsep}{3pt}
$$
\begin{array}{l ||r|r|r|r}
& \multicolumn{1}{c|}{l_u} & \multicolumn{1}{c|}{\tilde l_u} &   
\multicolumn{1}{c|}{l_u-\tilde l_u} & \multicolumn{1}{c}{l_u + \tilde l_u} \\
 \hline
 B \to K^*   & 536 \pm 70\% &  635 \pm 70\% & -99 \pm 300 & 1172 \pm 70 \% \\
 B \to (\rho,\omega)   & 827 \pm 70\% & 828\pm 70\% & -1\pm 300&
 1655\pm 70\%  \\
 B_s \to \bar K^*   &  454\pm 70\% & 572\pm 70\% & -118\pm 300 & 1025\pm
70\% \\
 B_s \to \phi   & 156\pm 70\% & 737\pm 70\% & -581\pm 300 & 893\pm 70\% \\
\end{array}
$$
\caption[]{\small Soft-gluon contributions from $u$-quark loops in units
keV. The quantities $l_u$ and $\tilde l_u$ are defined in \eqref{eq:loop_cu}
and \eqref{eq:general}. We assume $l_u(\rho)=l_u(\omega)$ and
similarly for $\tilde l_u$.
The uncertainty for $l_u-\tilde l_u$  is given in absolute
numbers because of cancellations.
In the SU(3)-flavour limit assumed in this calculation one has 
$l_u = l_d = l_s 
\equiv l_q$.}\label{tab_lu}
\end{table}
The crucial point here is that for $t$ below the continuum threshold 
the factor $1/t$ gets replaced by
$e^{(m_\rho^2-t)/M^2}/m_\rho^2$.

At this point we would also like to clarify in what respect our
method to calculate soft-gluon emission in $B \to V \gamma$ differs
from that developed in Ref.~\cite{K00} for analogous contributions to the 
non-leptonic $B \to \pi \pi$ decay.
In both cases the problem is a light-quark loop which is almost
on-shell, the corresponding non-perturbative effects are estimated
from light-cone sum rules and, in order to avoid parasitic terms in the
correlation function, 
an auxiliary momentum is introduced into the weak vertex.
The distinction is that, in contrast to the pion, the photon is a perturbative
state and therefore cannot
be represented by an interpolating current, but appears 
directly in the diagram with on-shell momenta. 
In order to set the photon on its mass shell we
use a dispersion representation, Eq.~\eqref{eq:lu1}, and 
estimate the truly non-perturbative part of the spectral
function  from the corresponding sum rule, Eq.~\eqref{49}.
Moreover we have checked, by inspecting the OPE in the deep Euclidian, 
that the dispersion representation has
no subtraction terms, which is implicitly assumed in \eqref{eq:lu1}.
In order to assure the absence of these terms we had to set
the two in principle independent momentum squares $q^2$
and $(q\!-\!k)^2$ equal to each other. This reintroduced 
a parasitic contribution of the form \eqref{40} which we
estimated to be of $O(1\%)$ as compared to the main contribution.

The sum rule (\ref{eq:lu3}) gives the  numerical results collected in
Tab.~\ref{tab_lu}. We use the Borel parameter
$M^2 = (1.2 \pm 0.3)\, {\rm GeV^2}$ and the threshold 
$s_0^\rho = (1.6 \pm 0.1) \, {\rm GeV^2}$.
Comparing these results with those from the $c$ loop, Tab.~\ref{tab_lc}, 
we see that they are roughly of the same size, but come with opposite sign. 
The smallness of 
 $(l_u-\tilde l_u)(\rho)$ is due to an accidental numerical cancellation. The
uncertainties are large, which is no cause of concern, however, because we
are only interested in the approximate size of these contributions
which set the size of the time-dependent CP asymmetry in $B\to V\gamma$.

In the above, we have assumed SU(3)-flavour  symmetry 
which implies $l_u = l_d = l_s \equiv l_q$. 
We can estimate the size of SU(3)-breaking effects  
by taking into account that
$l_s$ couples to the $\bar s\gamma_\mu s$ current via 
the $\phi$ and higher resonances in \eqref{eq:lu2} 
and requires a slightly higher continuum threshold $s_0^\phi$.
This leads to a numerical difference with respect to $l_u$ which 
is around 5\%. The effect of neglecting the quark masses is of order
$m_q/m_b$ and therefore even smaller. 
We conclude that it seems unlikely that $l_s$ differs from 
$l_{u,d}$ by more than 10\%.

\section{Phenomenological Results}\label{sec:5}

In this section we combine the different contributions to the
factorisation coefficients $a_{7L(R)}^U$ calculated in Secs.~\ref{sec:2},
\ref{sec:3} and \ref{sec:4} and give results for the observables in
$B\to V\gamma$ transitions, namely the branching ratio, the isospin
asymmetry and the time-dependent CP asymmetry. 

\subsection{Branching Ratios}\label{5.1}

The (non-CP-averaged) branching ratio of the $b\to D\gamma$ decay 
$\bar B\to V\gamma$ is given by
\begin{eqnarray}
{\cal B}(\bar B\to V\gamma) & = & \frac{\tau_B}{c_V^2}\,
\frac{G_F^2\alpha m_B^3
  m_b^2}{32 \pi^4} \left(1-\frac{m_V^2}{m_B^2}\right)^3
\left[T_1^{B\to V}(0)\right]^2\nonumber\\
&&\times \left\{ \left| \sum_U \lambda_U^{(D)} a_{7L}^U(V)\right|^2 + 
\left| \sum_U \lambda_U^{(D)}
  a_{7R}^U(V)\right|^2\right\}\label{BR}
\end{eqnarray}
with the isospin factors $c_{\rho^\pm,K^*,\phi}=1$ 
and $c_{\rho^0,\omega} = \sqrt{2}$. The branching ratio for the
CP-conjugated channel $B\to \bar V\gamma$ ($\bar b\to \bar D\gamma$ at
parton level) is obtained by replacing
$\lambda_U^{(D)}\to (\lambda_U^{(D)})^*$. Experimental results for $B\to
K^*\gamma$ and $B\to (\rho,\omega)\gamma$ are collected in
Tab.~\ref{tab1}. For $B_s\to\phi\gamma$ there is only an
upper bound ${\cal B}(B_s\to\phi\gamma)<120\times 10^{-6}$
\cite{PDG}. No experimental information is available for $B_s\to \bar
K^*\gamma$. 

With the input parameters from Tab.~\ref{tab3} and the lifetimes given
in Tab.~\ref{tab9} we find the following CP-averaged branching ratios 
for $B\to K^*\gamma$, making explicit various sources of uncertainty:
\begin{eqnarray}
\overline{\cal B}(B^- \to K^{*-}\gamma) & = & (53.3\pm 13.5(T_1)
\pm 4.8(\mu)
\pm 1.8(V_{cb})\pm 1.9(l_{u,c}) \pm 1.3(\mbox{other}))\times
10^{-6}\nonumber\\
& =& (53.3\pm 13.5(T_1)\pm 5.8)\times 10^{-6}\,,
\nonumber\\
\overline{\cal B}(\bar B^0 \to K^{*0}\gamma) & = & (54.2\pm 13.2(T_1)
\pm 6.0(\mu)
\pm 1.8(V_{cb})\pm 1.8(l_{u,c}) \pm 1.4(\mbox{other}))\times
10^{-6}\nonumber\\
& = & (54.2\pm 13.2(T_1)\pm 6.7)\times 10^{-6}\,.\label{50}
\end{eqnarray}
We have added all individual uncertainties in quadrature, except for
that induced by the form factor. The uncertainty in $\mu$ is that
induced by the renormalisation-scale dependence, with 
$\mu= m_b(m_b)\pm 1\,$GeV. The uncertainty in
$l_{u,c}$ refers to the soft-gluon terms calculated in
Sec.~\ref{sec:4}. ``Other'' sources of uncertainty include the
dependence on the parameters in Tab.~\ref{tab3}, on the size of LD WA
contributions and the replacement of NLO by LO Wilson coefficients. 
The above results
agree, within errors, with the experimental ones given in
Tab.~\ref{tab1}, within the large
theoretical uncertainty induced by the form factor.

\begin{table}
\renewcommand{\arraystretch}{1.3}
\addtolength{\arraycolsep}{3pt}
$$
\begin{array}{c|c|c}
\tau_{B^0} & \tau_{B^\pm}/\tau_{B^0} & \tau_{B_s^0}/\tau_{B^0}
\\\hline
1.530(9)\,{\rm ps} & 1.071(9) & 0.958(39)
\end{array}
$$
\vspace*{-10pt}
\caption[]{\small $B$ lifetimes from HFAG \cite{HFAG}.}\label{tab9}
\end{table}
As the uncertainties of all form factors in Tab.~\ref{tab3} are of
roughly the same size, one might conclude that the predictions for all
branching ratios will carry uncertainties similar to those in
(\ref{50}). This is, however, not the case:
the accuracy of the theoretical predictions can be improved by making
use of the fact that the {\em ratio} of form factors is known much
better than the individual form factors themselves. The reason is that
the values given in
Tab.~\ref{tab3}, which were calculated using the same method, LCSRs,
and with a common set of input parameters, 
include common systematic uncertainties (dependence on
$f_B$, $m_b$ etc.) which partially cancel in the ratio. In
Ref.~\cite{BZ06b} we have investigated in detail the ratio of the
$K^*$ and $\rho$ form factor and found
\begin{equation}\label{xirho}
\xi_\rho \equiv \frac{T_1^{B\to K^*}(0)}{T_1^{B\to \rho}(0)} = 1.17\pm
0.09\,.
\end{equation}
The uncertainty is by a factor 2 smaller than if we had calculated $\xi_\rho$
from the entries in Tab.~\ref{tab3}; an analogous calculation for
$\omega$ yields
\begin{equation}\label{xiomega}
\xi_\omega \equiv \frac{T_1^{B\to K^*}(0)}{T_1^{B\to \omega}(0)} = 1.30\pm
0.10\,.
\end{equation}
The difference between $\xi_\rho$ and $\xi_\omega$ is mainly due to
the difference between $f_\omega^\perp$ and $f_\rho^\perp$, see
Tab.~\ref{tab3}. 
For the $B_s$ form factors, we also need the ratio of decay constants
$f_{B_s}/f_{B_d}$. The status of $f_B$ from lattice was reviewed in 
Ref.~\cite{Onogi}; the present state-of-the-art calculations are
unquenched with $N_f=2+1$ active flavours \cite{unquenchedfB}, 
whose average is $f_{B_s}/f_{B_d}=1.23\pm
0.07$. Again, this ratio is fully consistent with that quoted in
Tab.~\ref{tab3}, but has a smaller uncertainty. We then find the
following ratios for $B_s$ form factors:
\begin{equation}\label{xis}
\xi_\phi \equiv \frac{T_1^{B\to K^*}(0)}{T_1^{B_s\to\phi}(0)} = 1.01\pm 0.13
\,,\qquad
\xi_{\bar K^*} \equiv \frac{T_1^{B\to K^*}(0)}{T_1^{B_s\to\bar K^*}(0)} 
= 1.09\pm 0.09\,.
\end{equation}
The uncertainty of $\xi_{\bar K^*}$ is smaller than that of $\xi_\phi$
because the input parameters for $K^*$ and $\bar K^*$ are the same
(except for G-odd parameters like $a_1^\perp$) and cancel in the
ratio; the uncertainty is dominated by that of $f_{B_s}/f_{B_d}$.

To benefit from this reduced
theoretical uncertainty in predicting branching
ratios, one has to calculate ratios of branching ratios,
which mainly depend on $\xi_V$ and only mildly on  $T_1$ 
itself: in addition to the
 overall normalisation, $T_1$ also enters hard-spectator interactions
 and power-suppressed corrections, whose size is set by hadronic
 quantities $\propto 1/T_1$. As these corrections are subleading (in
 $\alpha_s$ or $1/m_b$),
 however, a small shift in $T_1$ has only very minor impact on the 
branching ratios.
The absolute scale for the branching ratios is set by the CP- and
isospin-averaged  branching ratio with the smallest experimental
uncertainty, i.e.\ $B\to K^*\gamma$; from 
Tab.~\ref{tab1}, one finds:
\begin{equation}\label{x}
\overline{\cal B}(B\to K^*\gamma) = \frac{1}{2}\left\{ \overline{\cal
  B}(B^\pm \to K^{*\pm}\gamma) + \frac{\tau_{B^\pm}}{\tau_{B^0}}\, 
  \overline{\cal B}(\bar B^0 \to K^{*0}\gamma)\right\} = (41.6\pm
  1.7)\times 10^{-6}\,.
\end{equation}
That is, we obtain a theoretical prediction for $\overline{\cal B}(B\to
V\gamma)$ as 
\begin{equation}
\left.\overline{\cal B}(B\to V\gamma)\right|_{\rm th}
= \left[ \frac{\overline{\cal B}(B\to V\gamma)}{
\overline{\cal B}(B\to K^*\gamma)}\right]_{{\rm th}} \, \left.
\overline{\cal B}(B\to K^*\gamma)\right|_{\rm exp}\,,
\end{equation}
where $\left[\dots\right]_{\rm th}$ depends mainly on $\xi_V$ and only
in subleading terms on the individual form factors $T_1^{B\to K^*}$
and $T_1^{B\to V}$.
It is obvious that, except for these subleading terms,
this procedure is equivalent to extracting an {\em effective form factor}
$\left.T_1^{B\to  K^*}(0)\right|_{\rm eff}$ from $B\to K^*\gamma$ and using
$\left.T_1^{B\to V}(0)\right|_{\rm eff} = \left.T_1^{B\to
  K^*}(0)\right|_{\rm
  eff}/\xi_{V}$ for calculating the branching ratios for $B\to
V\gamma$. From (\ref{x}) we find
\begin{equation}
\left. T_1^{B\to K^*}(0)\right|_{\rm eff} = 0.267\pm 0.017({\rm th}) \pm
0.006({\rm exp}) = 0.267\pm 0.018\,,
\end{equation}
where the theoretical uncertainty 
follows from the second uncertainty given in (\ref{50}).
Eqs.~(\ref{xirho}), (\ref{xiomega}) and (\ref{xis}) then yield
\begin{eqnarray}
\left. T_1^{B\to \rho}(0)\right|_{\rm eff} &=& 0.228 \pm 0.023\,, \qquad
\left. T_1^{B\to \omega}(0)\right|_{\rm eff} = 0.205 \pm
0.021\,,\nonumber\\
\left. T_1^{B_s\to \bar K^*}(0)\right|_{\rm eff} &=& 0.245 \pm
0.024\,, 
\qquad
\left. T_1^{B_s\to \phi}(0)\right|_{\rm eff} = 0.260 \pm
0.036\,.\label{56}
\end{eqnarray}
Note that all effective form factors agree, within errors, with the
results from LCSRs given in Tab.~\ref{tab3}, which confirms the
results obtained from this method; the crucial point, however, is that
the uncertainties are reduced by a factor of 2 (except for
$T_1^{B_s\to \phi}$). We would like to stress that the motivation for this
procedure is to achieve a reduction of the theoretical uncertainty of the
predicted branching fractions in $B\to (\rho,\omega)\gamma$ and $B_s$
decays. 
The effective form factors do {\em not} constitute a new and
independent theoretical determination, but are derived from the experimental
results for $B\to K^*\gamma$ under the following assumptions:
\begin{itemize}
\item there is no NP in $B\to K^*\gamma$;\footnote{Which is motivated by the
  results from inclusive $B\to X_s \gamma$ decays \cite{misiak}.}
\item QCDF is valid with no systematic uncertainties;
\item LCSRs can reliably predict the ratio of form factors at zero
  momentum transfer.
\end{itemize}
 From (\ref{BR}) and
(\ref{56}), we then predict the following CP-averaged branching ratios:
\begin{eqnarray}
\overline{\cal B}(B^-\to \rho^-\gamma) & = & (1.16\pm 0.22(T_1)\pm 
0.13)\times 10^{-6}\,, \nonumber\\
\overline{\cal B}(B^0\to \rho^0\gamma) & = & (0.55\pm 0.11(T_1)\pm 
0.07)\times 10^{-6}\,, \nonumber\\
\overline{\cal B}(B^0\to \omega\gamma) & = & (0.44\pm 0.09(T_1) \
\pm 0.05)\times 10^{-6}\,,\nonumber\\
\overline{\cal B}(B_s\to \bar K^*\gamma) & = & (1.26\pm 0.25(T_1)\pm 
0.18)\times 10^{-6}\,, \nonumber\\
\overline{\cal B}(B_s\to \phi\gamma) & = & (39.4\pm 10.7(T_1) \
\pm 5.3)\times 10^{-6}\,,\label{57}
\end{eqnarray}
where the first uncertainty is induced by the effective form factors and 
the second includes the variation of all inputs from Tab.~\ref{tab3} 
except for
the angle $\gamma$ of the UT, which is fixed at
$\gamma=53^\circ$. The total uncertainty in each channel is $\sim
20\%$, except for $B_s\to \phi\gamma$, where it is 30\%.
The results for $\rho$ and $\omega$ agree very well with those of
BaBar, Tab.~\ref{tab1},
but less so with the Belle results, although present experimental
and theoretical uncertainties preclude a firm conclusion. Our
prediction for $B_s\to \phi\gamma$ is well below the current
experimental bound $120\times 10^{-6}$ \cite{PDG}. A branching ratio
of the size given in (\ref{57}) implies that $O(10^3)$ $B_s\to\phi\gamma$ 
events will be
seen within the first few years of the LHC.
In Tab.~\ref{tab10}
we detail the contributions of individual terms to the branching
ratios. In all cases ${\cal B}$ is dominated by the QCDF contribution,
with WA most relevant for $B^-\to \rho^-\gamma$. This is expected as
WA enters with the large Wilson coefficient $C_2\sim 1$. The effect is
extenuated by long-distance (LD) photon emission, which itself is
compensated by soft-gluon emission. The other channels follow a similar
pattern, although the size of the effects is smaller.
\begin{table}
\renewcommand{\arraystretch}{1.3}
\addtolength{\arraycolsep}{3pt}
$$
\begin{array}{l||c|c|c|c}
& \mbox{QCDF} & \mbox{+ WA (no LD)} & \mbox{+ WA (incl.\ LD)}& 
\mbox{+ soft gluons}\\\hline
B^-\to \rho^-\gamma & 1.05&1.17 & 1.11&1.16
\\
B^0\to \rho^0\gamma & 0.49&0.53& 0.53&0.55
\\
B^0\to \omega\gamma & 0.40& 0.42&0.42 &0.44
\\
B^-\to K^{*-}\gamma &39.7&38.4& 38.3&39.4
\\
B^0\to K^{*0}\gamma & 37.1&39.7 &39.9 &41.0
\\
B_s^0\to \bar K^{*0}\gamma & 1.12& 1.22& 1.23&1.26
\\
B_s^0\to \phi\gamma &34.6 & 38.2& 38.3&39.4
\end{array}
$$
\vspace*{-10pt}
\caption[]{\small Individual contributions to CP-averaged branching
  ratios, using effective form factors and  central values of all
  other input parameters given in Tab.~\ref{tab3} 
(in particular $\gamma=53^\circ$). LD stands for
  the long-distance photon-emission contribution to WA. Each column
  labelled ``+X'' includes the contributions listed in the previous
  column plus the contribution induced by X. The entries in the last
  column are our total central values.}\label{tab10}
\end{table}

Let us now turn to the determination of CKM parameters from the
branching ratios. In this context, two particularly interesting
observables are 
\begin{equation}\label{58}
R_{\rho/\omega}\equiv\frac{\overline{\cal B}(B\to (\rho,\omega)\gamma)}{
\overline{\cal B}(B\to K^*\gamma)}\,,\qquad
R_{\rho}\equiv\frac{\overline{\cal B}(B\to \rho\gamma)}{
\overline{\cal B}(B\to K^*\gamma)}\,,\qquad
\end{equation}
given in terms of the CP- and isospin-averaged branching ratios of
$B\to(\rho,\omega)\gamma$ and $B\to \rho\gamma$, respectively, 
and $B\to K^*\gamma$ decays, see Tab.~\ref{tab1}. $R_{\rho/\omega}$
has been measured by both BaBar and Belle \cite{Babar,Belle}, a first
value of $R_\rho$ has been given by BaBar \cite{Babar}. 
The experimental determinations
actually assume exact isospin symmetry, i.e.\
$\overline{\Gamma}(B^\pm\to \rho^\pm\gamma) \equiv
2 \overline{\Gamma}(B^0\to \rho^0\gamma)$, and also
$\overline{\Gamma}(B^0\to \rho^0\gamma) \equiv 
\overline{\Gamma}(B^0\to \omega\gamma)$; as we shall discuss in 
the next subsection, these relations are not exact, and the
symmetry-breaking corrections can be calculated. Hence, the present
experimental results for $R_{\rho/\omega}$ are theory-contaminated. 
As the isospin asymmetry between the charged and neutral $\rho$ decay 
rates turns out to be smaller than the asymmetry
 between $\rho^0$ and $\omega$, 
it would actually be preferable, from an experimental point of view,
to drop the $\omega$ channel and
measure $R_\rho$ instead of $R_{\rho/\omega}$, as done in the most
recent BaBar analysis on that topic \cite{Babar}. 
We will give numerical results and
theory uncertainties for both $R_{\rho/\omega}$ and $R_{\rho}$.

One parametrisation of $R_{\rho/\omega}$ often quoted, 
in particular in experimental papers, is
\begin{equation}\label{Brat}
R_{\rho/\omega} = \left|\frac{V_{td}}{V_{ts}}\right|^2
\left(\frac{1-m_{\rho}^2/m_B^2}{1-m_{K^*}^2/m_B^2}\right)^3
\frac{1}{\xi^2_{\rho}} \left [ 1 +  \Delta R\right],
\end{equation}
with $\Delta R=0.1\pm 0.1$ \cite{BVga1} and again assuming isospin
symmetry for $\rho$ and $\omega$. This parametrisation creates the impression
that $\Delta R$ is a quantity completely unrelated to
and with a fixed value independent of
$|V_{td}/V_{ts}|$. We would like to point out here that this 
impression is {\em wrong}: $\Delta R$ contains both QCD
(factorisable and non-factorisable) effects and such from weak
interactions. In Ref.~\cite{BZ06b}, we have expressed $\Delta R$ in
terms of the factorisation coefficients $a_{7L}^U$, assuming isospin
symmetry for $\rho^0$ and $\omega$, as
\begin{eqnarray}
1+\Delta R & = & \left|
  \frac{a_{7L}^c(\rho)}{a_{7L}^c(K^*)}\right|^2 \left( 1 +
  {\rm Re}\,(\delta a_\pm + \delta a_0) \left[\frac{R_b^2 - R_b
  \cos\gamma}{1-2 R_b \cos\gamma + R_b^2}\right]\right.\nonumber\\
& & \left. + \frac{1}{2}\left( |\delta a_\pm|^2 + |\delta a_0|^2\right)
  \left\{ \frac{R_b^2}{1-2 R_b \cos\gamma + R_b^2}\right\} \right)
\label{delR}
\end{eqnarray}
with $\delta a_{0,\pm}=
a_{7L}^u(\rho^{0,\pm})/a_{7L}^c(\rho^{0,\pm})-1$. Here $\gamma$ is
one of the angles of the UT ($\gamma = {\rm arg}\, V_{ub}^*$ in the
standard Wolfenstein parametrisation of the CKM matrix) 
and $R_b$ one of its sides:
$$
R_b =
\left(1-\frac{\lambda^2}{2}\right)\frac{1}{\lambda}
\left|\frac{V_{ub}}{V_{cb}}\right|.
$$
Eq.~(\ref{delR}) shows explicitly that $\Delta R$ depends both on QCD
 ($\delta a_{\pm,0}$) and CKM parameters ($R_b,\gamma$).
The point we would like to make is that the calculation of $\Delta R$
requires input values for $R_b$ and $\gamma$. Once these parameters
(and the Wolfenstein parameter $\lambda$)
are fixed, however, $|V_{td}/V_{ts}|$ is also fixed and given by
\begin{equation}\label{61}
\left|\frac{V_{td}}{V_{ts}}\right| = \lambda \sqrt{1-2 R_b \cos\gamma
  + R_b^2} \left[ 1 + \frac{1}{2}\,( 1 - 2 R_b \cos\gamma) \lambda^2 +
  O(\lambda^4)\right]\,.
\end{equation} 
Hence, as $|V_{td}/V_{ts}|$ and $(R_b,\gamma)$ are not independent
of each other, 
it is {\em impossible} to extract $|V_{td}/V_{ts}|$ from (\ref{Brat})
with a fixed value of $\Delta R$. We hasten to add that our
arguments rely on the unitarity of the CKM matrix, and its well-known
consequence, the existence of the UT. The unitarity of the CKM matrix
is, however, already hard-wired into the effective Hamiltonian
(\ref{heff}); without it, the theory would look quite 
different because of the absence of the GIM mechanism, as mentioned in
Sec.~\ref{sec:2}, Eq.~(\ref{SMunitarity}). Stated
differently: as long as Eq.~(\ref{heff}) is adopted as the relevant
effective Hamiltonian for $b\to D\gamma$ transitions, unitarity of the 
CKM matrix is implied. Obviously, the unitarity of the CKM matrix is 
subject to
experimental scrutiny, but any test of it has to involve the
comparison of {\em different} measurements described within the same
framework (by the same effective Hamiltonian), while a mixture of
different frameworks (unitary vs.\ non-unitary CKM matrix) within {\em
  one} observable, like $R_{\rho/\omega}$, does not make any sense.

\begin{figure}
$$\epsfxsize=0.45\textwidth\epsffile{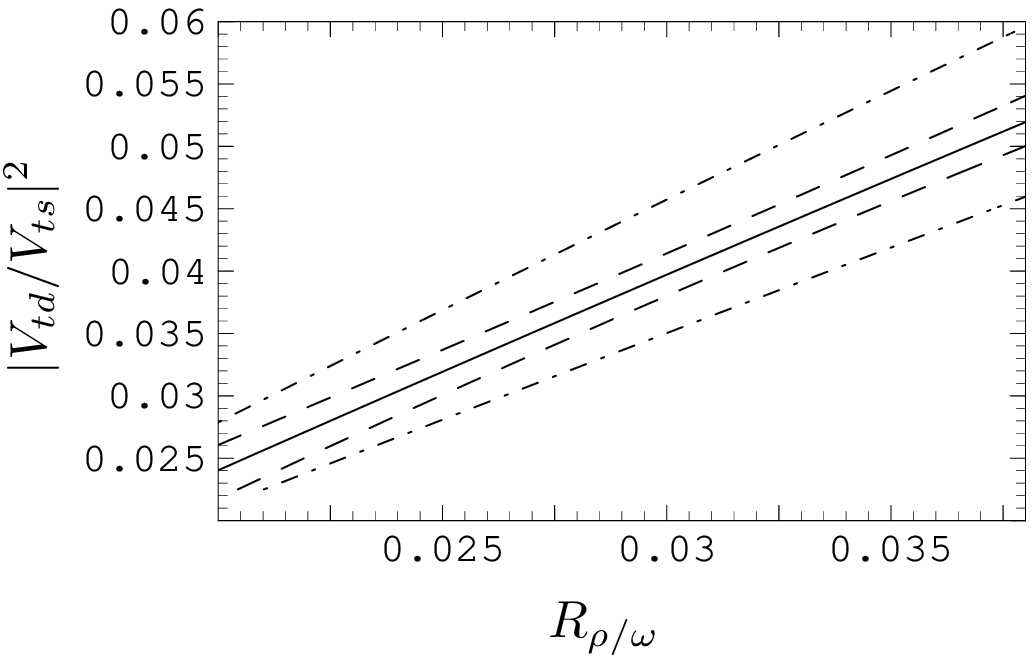}\qquad
\epsfxsize=0.45\textwidth\epsffile{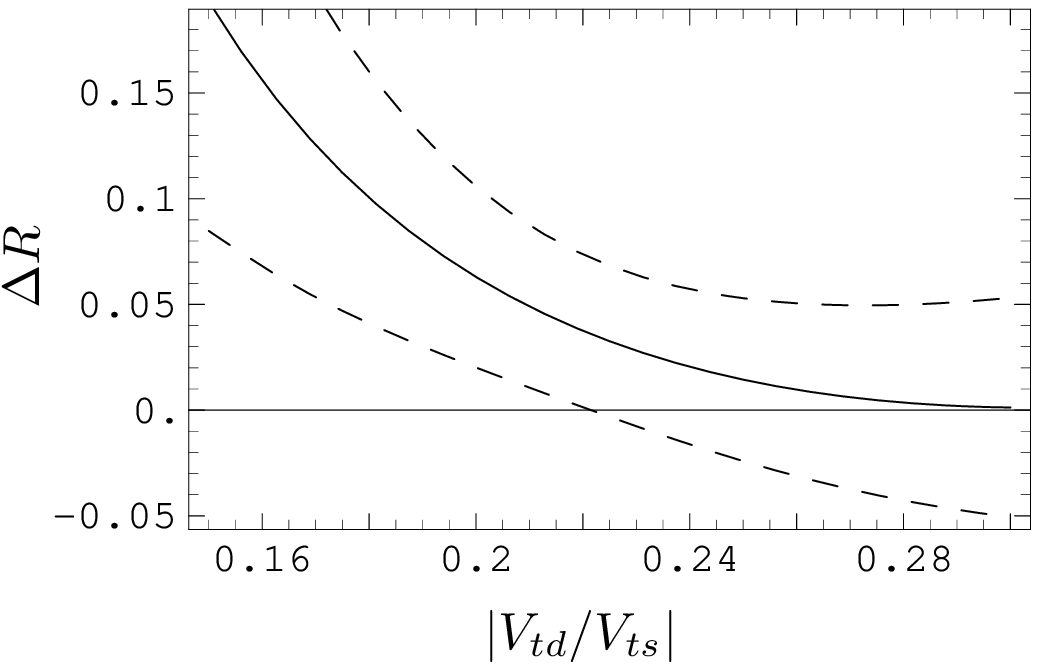}
$$
\vspace*{-25pt}
\caption[]{\small Left panel: $|V_{td}/V_{ts}|^2$ as function of
  $R_{\rho/\omega}$, Eq.~(\ref{58}),
  in the $|V_{tx}|$ basis, see
  text.  Solid line: central values. Dash-dotted lines: theoretical
  uncertainty induced by $\xi_\rho = 1.17\pm 0.09$, (\ref{xirho}). 
Dashed lines: other
  theoretical uncertainties, including those induced
  by $|V_{ub}|$, $|V_{cb}|$ and the hadronic parameters of
  Tab.~\ref{tab3}. Right panel: $\Delta R$ from 
 Eq.~(\ref{delR}) as function of
  $|V_{td}/V_{ts}|$ for the $|V_{tx}|$ set of CKM parameters. Solid
  line: central values. Dashed
  lines: theoretical uncertainty.}\label{fig4}
$$\epsfxsize=0.45\textwidth\epsffile{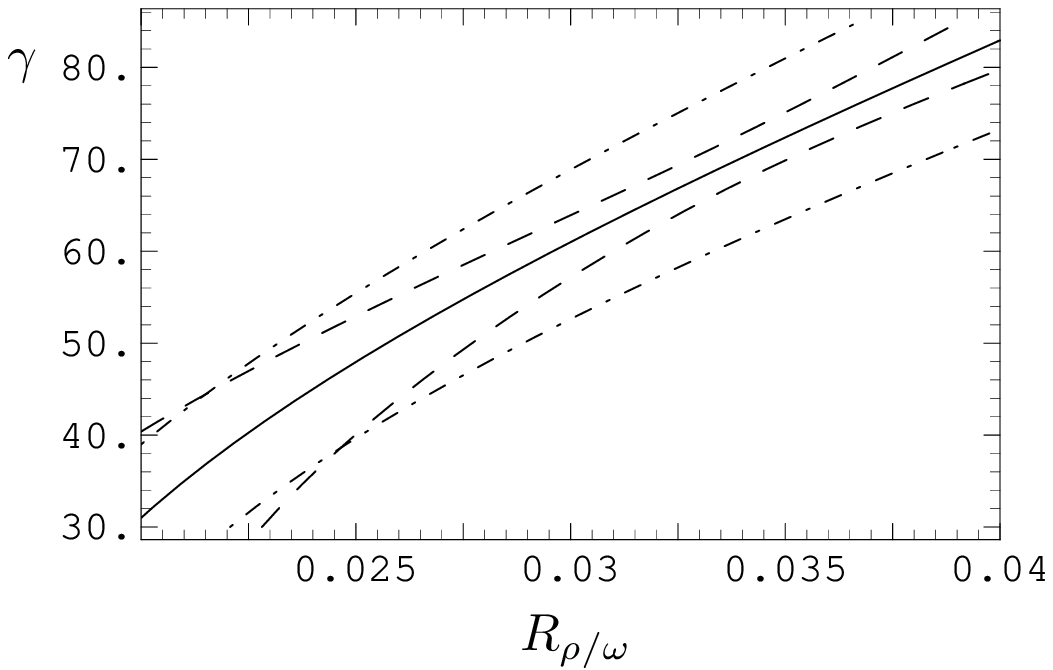}$$
\vspace*{-25pt}
\caption[]{\small The UTangle $\gamma$ as function of 
 $R_{\rho/\omega}$ in the $\gamma$ set of CKM parameters. Solid
  lines: central values of input parameters. Dash-dotted lines: theoretical
  uncertainty induced by $\xi_\rho = 1.17\pm 0.09$. Dashed lines: other
  theoretical uncertainties.}\label{fig5}
$$\epsfxsize=0.45\textwidth\epsffile{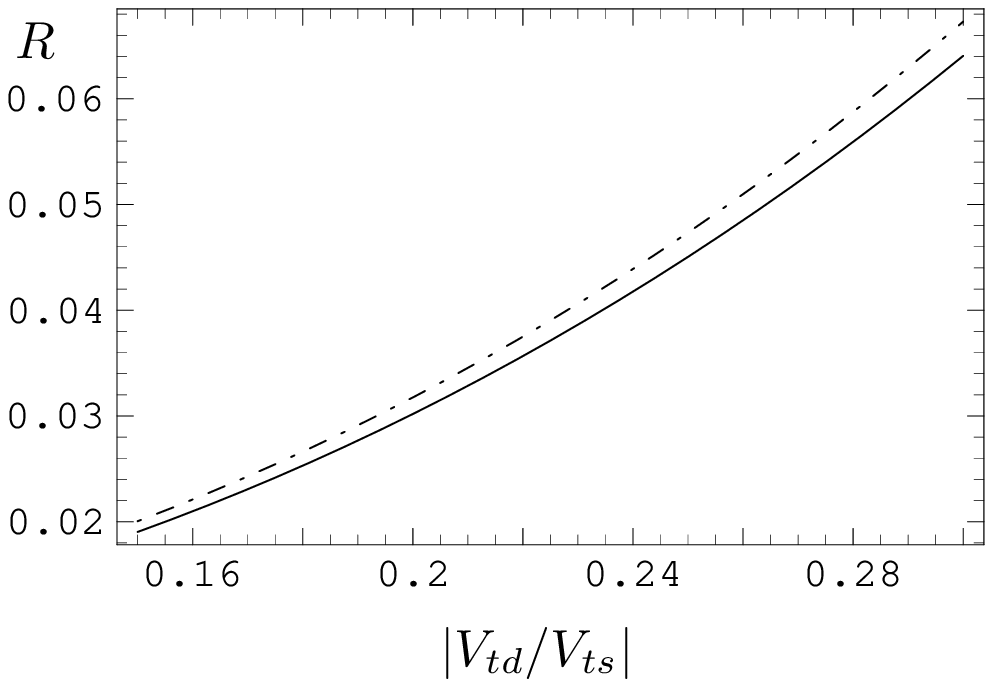}$$
\vspace*{-25pt}
\caption[]{\small Central values of 
$R_{\rho/\omega}$ (solid line) and $R_{\rho}$
  (dash-dotted line) as function of $|V_{td}/V_{ts}|$.}\label{fig6}
\end{figure}

Of course $R_{\rho/\omega}$ and $R_\rho$ of (\ref{58}) 
{\em can} be used in a meaningful way to extract information
about CKM parameters, but in order to do so one has to settle for a set
of truly independent parameters. Based on (\ref{61}), one can
exchange, say, $\gamma$ for
$|V_{td}/V_{ts}|$~.\footnote{Strictly speaking, (\ref{61}) only
  fixes $\cos\gamma$ as function of $|V_{td}/V_{ts}|$, leaving
  a twofold degeneracy of $\gamma$. Eq.~(\ref{delR}), however, only
  depends on $\cos\gamma$, so that indeed one can unambiguously 
replace $\gamma$ by $|V_{td}/V_{ts}|$.
}  So we can either consider $R_V$ as a
function of the CKM parameters $R_b$ and $\gamma$ (let us call this
the $\gamma$ set of parameters) or as a function of $R_b$ and
$|V_{td}/V_{ts}|$ (to be called the $|V_{tx}|$ set). Using the
$\gamma$ set, a measurement of $R_V(\gamma,R_b)$ allows a
determination of $\gamma$, whereas 
$R_V(|V_{td}/V_{ts}|,R_b)$ allows the
determination of $|V_{td}/V_{ts}|$. 
In either case, the simple quadratic relation
(\ref{Brat}) between $R_V$ and $|V_{td}/V_{ts}|$ becomes more
complicated. In Figs.~\ref{fig4} and \ref{fig5} we plot the
resulting values of $|V_{td}/V_{ts}|^2$ and $\gamma$, respectively, 
as a  function of $R_V$.
Although the curve in Fig.~\ref{fig4}(a) looks like a straight line,
as naively expected from (\ref{Brat}),
this is not exactly the case, because of the dependence of $\Delta R$
on $|V_{td}/V_{ts}|$. In  Fig.~\ref{fig4}(b) we plot $\Delta R$ for
the $|V_{tx}|$ set of
parameters. The dependence of $\Delta R$ on $|V_{td}/V_{ts}|$ is
rather strong. Apparently indeed $\Delta R=0.1\pm 0.1$ in the expected
range $0.16<|V_{td}/V_{ts}|<0.24$, but this estimate does not reflect 
the true theoretical uncertainty which is indicated by
the dashed lines in the figure. 
\begin{table}[tb]
\renewcommand{\arraystretch}{1.3}
\addtolength{\arraycolsep}{3pt}
$$
\begin{array}{c||c|c|c||c|c|c}
R_{\rho/\omega} & |V_{td}/V_{ts}| & \Delta_{\xi_\rho} & \Delta_{\rm
  other\,th}
& \gamma & \Delta_{\xi_\rho} & \Delta_{\rm
  other\,th}\\\hline
0.026 & 0.183 & \pm 0.012 & \pm 0.007 & 50.8 & {}^{+7.5}_{-8.2} &
  \pm 5.8 \\
0.028 & 0.191 & {}^{+0.012}_{-0.013} & \pm 0.006 & 56.0 &
  {}^{+7.7}_{-8.3} & \pm 4.7\\
0.030 & 0.199 & \pm 0.013 & \pm 0.006 & 61.0 & {}^{+7.9}_{-8.4} &
  \pm 4.0\\
0.032 & 0.207 & {}^{+0.013}_{-0.014} & \pm 0.006 & 65.7 &
  {}^{+8.1}_{-8.5} & \pm 3.6\\
0.034 & 0.214 & \pm 0.014 & \pm 0.006 & 70.2 & {}^{+8.4}_{-8.8} & 
\pm 3.5\\
0.036 & 0.221 & {}^{+0.014}_{-0.015} & \pm 0.006 & 74.5 &
  {}^{+8.8}_{-9.0} & \pm 3.7
\end{array}
$$
\vspace*{-0pt}
\caption[]{\small Central values and uncertainties of
  $|V_{td}/V_{ts}|$ and $\gamma$ extracted from representative values of 
$R_{\rho/\omega}$, Eq.~(\ref{58}). 
$\Delta_{\xi_\rho}$ is the uncertainty induced by
$\xi_\rho$, Eq.~(\ref{xirho}), and $\Delta_{\rm other\,th}$ that by other input
  parameters, including $\xi_\omega$ and $|V_{ub}|$.}\label{tabx}
$$
\begin{array}{c||c|c|c||c|c|c}
R_{\rho} & |V_{td}/V_{ts}| & \Delta_{\xi_\rho} & \Delta_{\rm
  other\,th}
& \gamma & \Delta_{\xi_\rho} & \Delta_{\rm
  other\,th}\\\hline
0.028 & 0.186 & \pm 0.016 & \pm 0.005 & 52.4 & {}^{+9.9}_{-10.3} & \pm
  5.0\\
0.030 & 0.193 & \pm 0.016 & \pm 0.005 & 57.4 & {}^{+10.2}_{-10.3} &
  \pm 3.9\\
0.032 & 0.201 & \pm 0.017 & \pm 0.005 & 62.0 & \pm 10.5 & \pm 3.1\\
0.034 & 0.208 & \pm 0.017 & \pm 0.004 & 66.4 & {}^{+10.8}_{-10.7} &
  \pm 2.7\\
0.036 & 0.215 & \pm 0.018 & \pm 0.004 & 70.7 & {}^{+11.3}_{-11.0} &
  \pm 2.5
\end{array}
$$
\vspace*{-0pt}
\caption[]{\small Ditto for $R_\rho$. $\Delta_{\xi_\rho}$ is larger
  than in Tab.~\ref{tabx} because of the increased weight of
  $B\to\rho\gamma$ in the isospin average; $\Delta_{\rm
  other\,th}$ is smaller because $\xi_\omega$ does not enter.}\label{taby}
\end{table}

It is now basically a matter of choice whether to use
$R_{\rho/\omega}$ to determine $|V_{td}/V_{ts}|$ or $\gamma$. Once one of
these parameters is known, the other one follows from Eq.~(\ref{61}). In
Fig.~\ref{fig5} we plot $\gamma$ as a function of
$R_{\rho/\omega}$, together with the theoretical uncertainties. In
Fig.~\ref{fig6} we also compare the central values of
$R_{\rho/\omega}$ with those of $R_{\rho}$, as a function of
$|V_{td}/V_{ts}|$. Although the difference is small, $R_{\rho}$ is
expected to be larger than $R_{\rho/\omega}$. In order to facilitate
the extraction of $|V_{td}/V_{ts}|$ (or $\gamma$) from measurements of
$R_{\rho/\omega}$ or $R_{\rho}$, Tabs.~\ref{tabx} and \ref{taby}  
contain explicit values for the theoretical uncertainties for
representative values of $R_{\rho/\omega}$ and $R_{\rho}$. The uncertainty
induced by $\xi_{\rho}$ is dominant. As discussed in
Ref.~\cite{BZ06b}, a reduction of this uncertainty
 would require a reduction of the
uncertainty of the transverse decay constants $f_V^\perp$ of $\rho$
and $K^*$. With the most recent results from BaBar, $R_{\rho/\omega} =
0.030\pm 0.006$ \cite{Babar}, and from Belle, $R_{\rho/\omega} =
0.032\pm 0.008$ \cite{Belle}, we then find
\begin{equation}
\renewcommand{\arraystretch}{1.3}
\begin{array}[b]{l@{\quad}l@{\quad\leftrightarrow\quad}l}
\mbox{BaBar:} & \displaystyle
\left|\frac{V_{td}}{V_{ts}}\right| = 0.199^{+0.022}_{-0.025}({\rm exp})\pm
0.014({\rm th}) &\displaystyle
\gamma = (61.0^{+13.5}_{-16.0}({\rm exp})^{+8.9}_{-9.3}
({\rm th}))^\circ\,,\\[10pt]
\mbox{Belle:} & \displaystyle
\left|\frac{V_{td}}{V_{ts}}\right| = 0.207^{+0.028}_{-0.033}({\rm exp})
^{+0.014}_{-0.015}({\rm th}) &\displaystyle
\gamma = (65.7^{+17.3}_{-20.7}({\rm exp})^{+8.9}_{-9.2}({\rm th}))^\circ\,.
\end{array}
\label{63}
\end{equation}
These numbers  compare well with the Belle result  \cite{Bellegamma} 
from tree-level processes, $\gamma=(53\pm 20)^\circ$, quoted in
Tab.~\ref{tab3}, and results from  global fits
\cite{global}. We also would like to point out that the above
determination of $\gamma$ is actually a determination of
$\cos\gamma$, via Eq.~(\ref{61}), and implies, in principle, a twofold
degeneracy $\gamma\leftrightarrow 2\pi-\gamma$. This is in contrast to the
determination from $B\to D^{(*)} K^{(*)}$ in \cite{Bellegamma}, which
carries a twofold degeneracy
$\gamma \leftrightarrow \pi+\gamma$. Obviously these two
determinations taken together remove the degeneracy and 
select $\gamma\approx 55^\circ<180^\circ$. If 
$\gamma\approx 55^\circ+180^\circ$ instead, one would have 
$|V_{td}/V_{ts}|\approx 0.29$ from
(\ref{61}), which is definitely ruled out by data. Hence, the result
(\ref{63}) confirms the SM interpretation of $\gamma$ from 
the tree-level CP asymmetries in $B\to D^{(*)} K^{(*)}$.

\begin{figure}[tb]
$$\epsfxsize=0.45\textwidth\epsffile{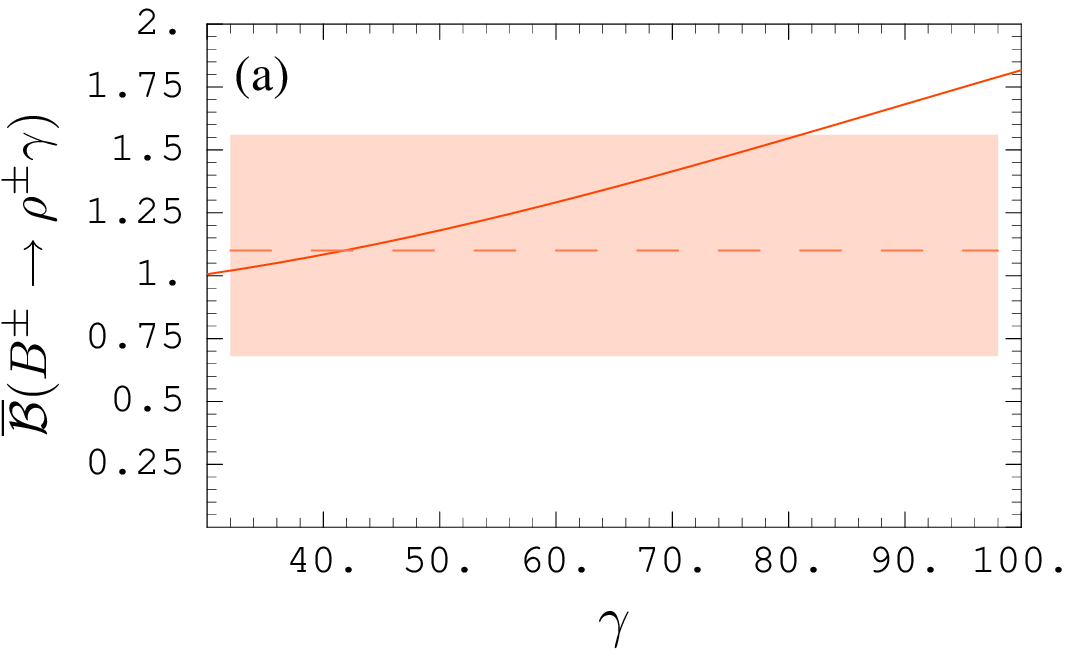}\quad
\epsfxsize=0.45\textwidth\epsffile{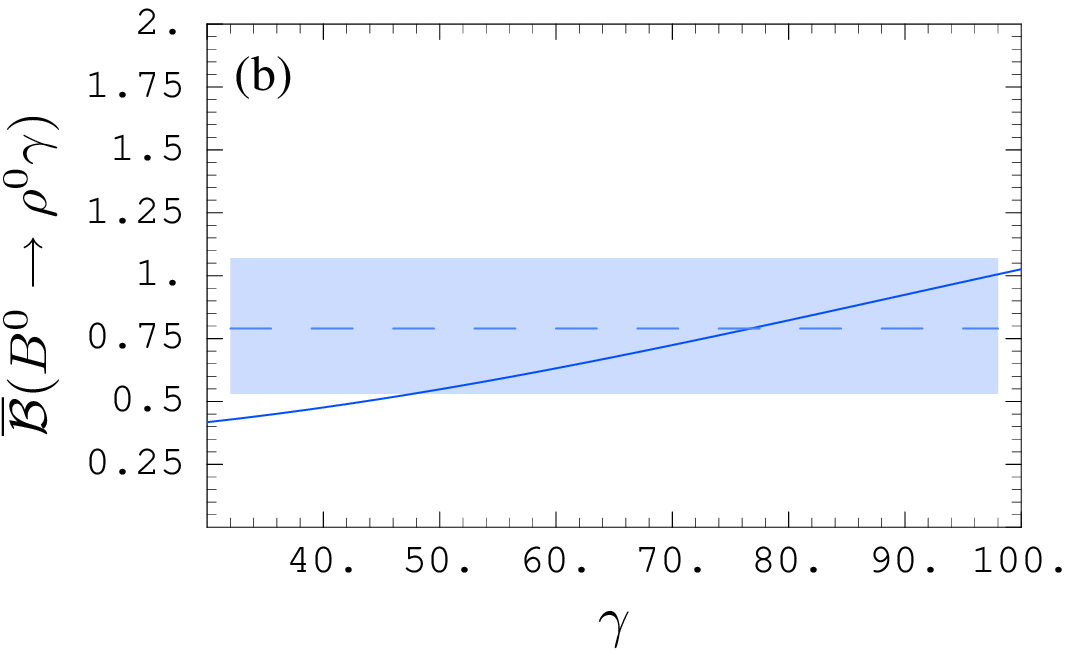}$$
\vspace*{-20pt}
$$\epsfxsize=0.45\textwidth\epsffile{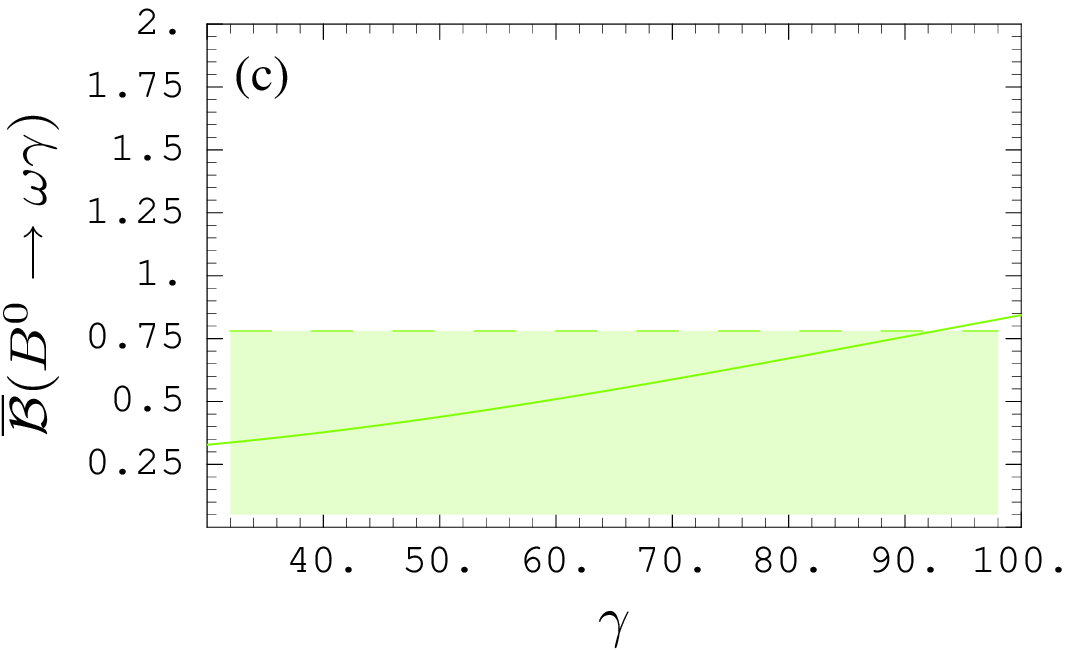}$$
\vspace*{-20pt}
\caption[]{\small CP-averaged branching ratios of 
$B\to(\rho,\omega)\gamma$ as function
  of $\gamma$, using the effective form factors and central values of
  other input parameters. (a): $B^\pm \to \rho^\pm\gamma$, (b):
  $B^0\to \rho^0\gamma$, (c):
  $B^0\to\omega\gamma$. The boxes indicate the 1$\sigma$ experimental results
  from BaBar \cite{Babar}, Tab.~\ref{tab1}. 
Note that the resulting value of $\gamma$
  from the average of all three channels is $\gamma =
  (61.0^{+13.5}_{-16.0}({\rm exp})^{+8.9}_{-9.2})^\circ$, Eq.~(\ref{63}).
}\label{fig7}
\end{figure}

We would like to close this subsection by making explicit the 
dependence of the
three $B\to (\rho,\omega)\gamma$ branching ratios on $\gamma$. 
In Fig.~\ref{fig7} 
we plot these branching ratios, for central values of the input
parameters, as functions of $\gamma$. We also indicate the present
experimental results from BaBar \cite{Babar}, Tab.~\ref{tab1},
 within their 1$\sigma$ uncertainty. 

\subsection{Isospin Asymmetries}\label{5.2}

The  asymmetries are given by
\begin{eqnarray}
A(\rho,\omega) & = &  \frac{\overline{\Gamma}(B^0\to \omega
  \gamma)}{\overline{\Gamma}(B^0\to \rho^0
  \gamma)}-1\,, \label{arw} \\
A_{I}(\rho) & = & \frac{2\overline{\Gamma}(\bar B^0\to 
  \rho^0 \gamma)}{\overline{\Gamma}(\bar B^\pm\to 
  \rho^\pm \gamma)} - 1\,,\label{air}\\
A_{I}(K^*) & = & \frac{\overline{\Gamma}(\bar B^0\to 
  K^{*0} \gamma) - \overline{\Gamma}(B^\pm\to 
  K^{*\pm} \gamma)}{\overline{\Gamma}(\bar B^0\to 
  K^{*0} \gamma) + \overline{\Gamma}(B^\pm\to 
  K^{*\pm} \gamma)}\,;\label{aik}
\end{eqnarray}
the partial decay rates are CP-averaged; $A_I(\rho)$, $A_I(K^*)$ are isospin
asymmetries. 

\begin{figure}[tb]
$$\epsfxsize=0.45\textwidth\epsffile{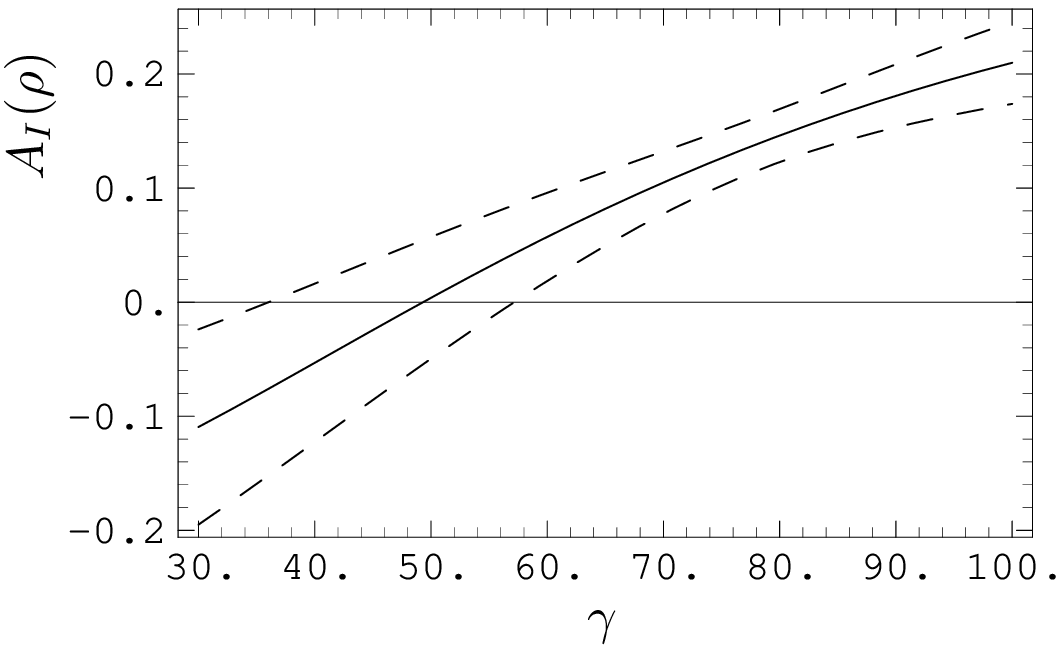}\quad
  \epsfxsize=0.45\textwidth\epsffile{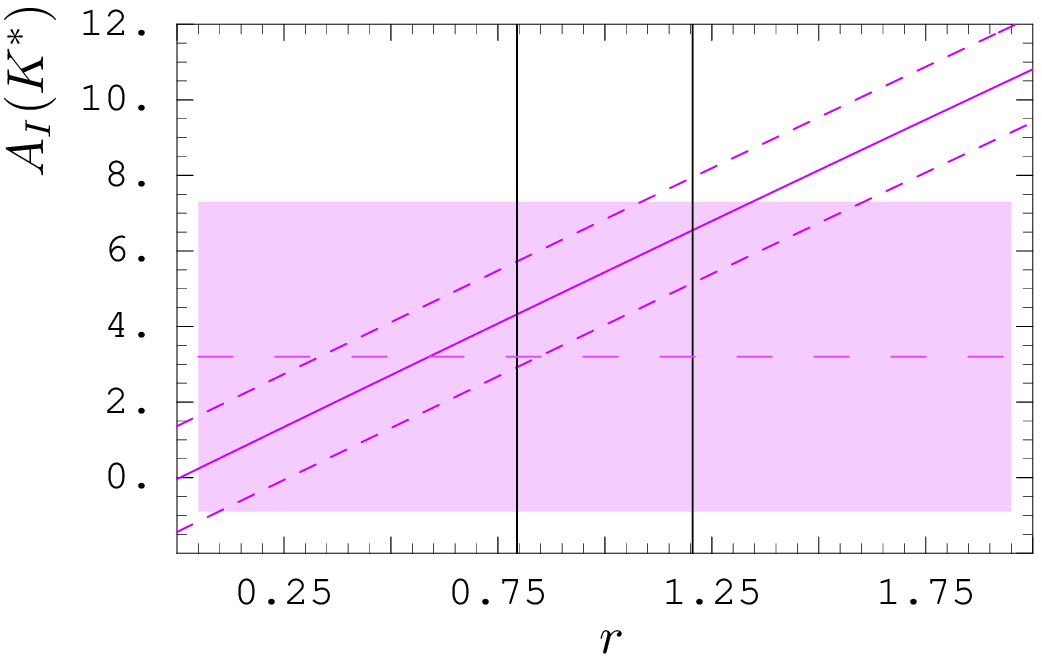}$$
\vspace*{-30pt}
\caption[]{\small Left panel: isospin asymmetry $A_I(\rho)$, Eq.~(\ref{air}), 
as function of the UTangle $\gamma$.
Solid line: central values of input parameters; dashed lines:
theoretical uncertainty. Right panel: $A_I(K^*)$, Eq.~(\ref{aik}), in percent, 
as function of the ratio $r\equiv a_6/a_6^{\rm
    SM}$ of the combination of penguin Wilson coefficients
    $a_6\equiv C_6+C_5/3$. Solid line: central value of input parameters,
    dashed lines: theoretical uncertainty. The box indicates the
    present experimental uncertainty and the straight black lines the
    theory uncertainty in $r$.}\label{fig8}
\end{figure}

Let us first discuss $A(\rho,\omega)$ and $A_{I}(\rho)$ which are 
relevant for the
experimental determination of $\overline{\cal
  B}(B\to(\rho,\omega)\gamma)$, 
which in turn is used for the determination of
$|V_{td}/V_{ts}|$ (or $\gamma$), see Sec.~\ref{5.1}. The present experimental
statistics for $b\to d\gamma$ transitions is rather low, so the
experimental value of $\overline{\cal B}(B\to(\rho,\omega)\gamma)$ is 
obtained
under the explicit assumption of perfect symmetry, i.e.\
$\overline{\Gamma}(B^\pm\to \rho^\pm \gamma) = 2
\overline{\Gamma}(B^0\to \rho^0 \gamma) = 2 \overline{\Gamma}(B^0\to
\omega \gamma)$. 
In reality, the symmetry between $\rho^0$ and $\omega$ is broken by
different values of the form factors, and isospin symmetry between
neutral and charged $\rho$ is broken by photon emission from the spectator
quark, the dominant mechanism of which is WA,
as discussed in Sec.~\ref{sec:3}. From the formulae for
individual branching ratios, Eq.~(\ref{BR}), 
and the various contributions to the
factorisation coefficients $a_{7L(R)}^U$ collected in
  Secs.~\ref{sec:2}, \ref{sec:3} and \ref{sec:4}, we find
\begin{table}
\renewcommand{\arraystretch}{1.3}
\addtolength{\arraycolsep}{3pt}
$$
\begin{array}{c||c|c|c|c}
\gamma & 40^\circ & 50^\circ & 60^\circ & 70^\circ\\\hline
A_I(\rho) & -(5.3\pm 6.9)\% & (0.4\pm 5.3)\% & (5.7\pm 3.9)\% &
(10.5\pm 2.7)\%
\end{array}
$$
\vspace*{-10pt}
\caption[]{\small Isospin asymmetry $A_I(\rho)$, Eq.~(\ref{air}), 
for different values
  of $\gamma$.}\label{tab:iso}
\end{table}
\begin{equation}
\label{eq:AIrw}
A(\rho,\omega) =-0.20\pm 0.09({\rm th})\,.
\end{equation}
The uncertainty is dominated by that of the
 form factor ratio  $T_1^{B\to\omega}(0)/T_1^{B\to\rho}(0)=0.90\pm
0.05$.\footnote{Note that this result is dominated by the ratio of
 decay constants given in Tab.~\ref{tab3} and discussed in the
 appendix. The experimental results entering these averages have a
 large spread which may cast a shadow of doubt on the averaged final
 branching ratios for $(\rho^0,\omega)\to e^+ e^-$ quoted by PDG 
\cite{PDG}.} The
dependence on all other input parameters is marginal. $A_{I}(\rho)$,
on the other hand, is very sensitive to $\gamma$, whereas the form
factors drop out. It is driven by the WA contribution and, in the QCDF
 framework, vanishes if WA is set to zero.
In Fig.~\ref{fig8}(a) we plot $A_{I}(\rho)$ as
 function of $\gamma$, including the theoretical uncertainties. As
 suggested by the findings of Ref.~\cite{chamonix}, these results are
 not expected to change considerably upon inclusion of 
the non-factorisable radiative corrections of
 Fig.~\ref{fig2}(c).
In Tab.~\ref{tab:iso}, we give the corresponding
 results for several values of $\gamma$, together with the theoretical
 uncertainty. Our result agrees very well with that
 obtained by the BaBar collaboration: $A_I(\rho)_{\rm BaBar} = 
0.56\pm 0.66$ \cite{Babar}.

$A_I(K^*)$ was first discussed in Ref.~\cite{kagan}, including
power-suppressed $O(\alpha_s)$ corrections which unfortunately violate
QCDF, i.e.\ are divergent. It is for this reason that we decide to
drop these corrections and include only leading-order terms in
$\alpha_s$. We then find
\begin{eqnarray}
\label{eq:aik}
A_I(K^*) &=& (5.4\pm 1.0(\mu) \pm 0.6({\rm NLO}\leftrightarrow{\rm
  LO}) \pm 0.6 (f_B) \pm 0.6({\rm other}))\%\nonumber\\
&=& (5.4\pm 1.4)\%\,,\label{62}
\end{eqnarray}
where ${\rm NLO}\leftrightarrow{\rm LO}$ denotes the uncertainty
induced by switching from NLO to LO accuracy in the Wilson
coefficients and ``other'' summarises all other sources of theoretical
uncertainty. 
As can be inferred from the entries in Tab.~\ref{tab1}, the 
present experimental result is $A_I(K^*)_{\rm exp}=(3.2\pm 4.1)\%$. 
In Ref.~\cite{kagan},
Kagan and Neubert pointed out that $A_I(K^*)$
is very sensitive to the values of the Wilson coefficients
$C_{5,6}^{\rm BBL}$ in the combination $a_6\equiv C_{5}^{\rm BBL}+C_6^{\rm
  BBL}/3$. In the SM, varying the renormalisation scale as
$\mu=m_b(m_b)\pm 1\,{\GeV}$ and switching between LO and NLO accuracy
for the Wilson coefficients, one has $a_6= -0.039\pm 0.008$, which
actually induces the bulk of the uncertainty in (\ref{62}). In
Fig.~\ref{fig8}(b) we plot $A_I(K^*)$ as function of
$a_6/a_6^{\rm SM}$, with $a_6^{\rm SM}=-0.039$.
The figure clearly indicates that, although there is presently no
discrepancy between theoretical prediction and experimental result,
a reduction of the experimental uncertainty
of $A_I(K^*)$ may well reveal some footprints of NP
in this observable.

\subsection{CP Asymmetries}\label{5.3}

The time-dependent CP asymmetry in $\bar B^0\to V^0\gamma$ is given by
\begin{eqnarray}
A_{CP}(t) &=& \frac{\Gamma(\bar B^0(t)\to V\gamma) -
               \Gamma(     B^0(t)\to   \bar   V\gamma)}{
               \Gamma(\bar B^0(t)\to  V\gamma) +
               \Gamma(     B^0(t)\to \bar     V\gamma)}\nonumber\\
&=& S(V\gamma) \sin(\Delta m_B t ) - C(V\gamma) 
\cos(\Delta m_B t)\,,\label{-1}
\end{eqnarray}
where we have neglected the width difference $\Delta\Gamma$ 
of the two neutral $B$ mesons. This approximation is well justified 
for $B_d$, but less so for $B_s$. Although the above formula can
easily be adapted to
non-zero $\Delta \Gamma_s$, we refrain from doing so: the whole point
in calculating the CP asymmetry is not so much to give precise
predictions for $S$ and $C$, but rather to exclude the
possibility of large corrections to the naive expectation $S\sim
m_D/m_b$. With this is mind, small corrections from a non-zero
$\Delta\Gamma_s$ are irrelevant.

Let us briefly recall the reason for the expected smallness of $S$.
In the process $b\to D\gamma$, in the SM, the emitted photon is 
predominantly 
left-handed in $b$, and right-handed in $\bar b$
decays. This  is due to the fact that the dominant contribution to the 
amplitude comes from the chiral-odd dipole operator $Q_7$,
Eq.~(\ref{eq:WC}). As only left-handed quarks participate in the weak 
interaction, an effective operator of this type necessitates, in the SM, a
helicity flip on one of the external quark lines, which results in a factor
$m_b$ (and a left-handed photon) in $b_R\to D_L\gamma_L$
and a factor $m_D$ (and a right-handed photon) in $b_L\to
D_R\gamma_R$. Hence, the emission of right-handed photons is suppressed by
 a factor $m_D/m_b$, which leads to the QCDF prediction
(\ref{11}) for $a_{7R}^U$. 

The interesting point is not the smallness of the CP asymmetry {\em
  per se}, but
the fact that the helicity suppression can easily be alleviated 
in a large number of NP scenarios where the spin flip occurs on an
internal line, resulting in a factor $m_i/m_b$ instead of $m_D/m_b$.
A prime example is left-right symmetric models
\cite{LRS}, whose impact on the photon polarisation was discussed in
Refs.~\cite{alt, grin04,grin05}. 
These models also come in a supersymmetric version
whose effect on $b\to s\gamma$ was investigated in Ref.~\cite{frank}.
Supersymmetry with no left-right symmetry can also provide large
 contributions
to $b\to D\gamma_R$, see Ref.~\cite{susy} for recent studies. Other
potential sources of large effects 
are warped extra dimensions \cite{warped} or anomalous right-handed top
couplings \cite{anomalous}. 
Unless the amplitude for $b\to D\gamma_R$
is of the same order as the SM prediction for $b\to D \gamma_L$, or the
enhancement of $b\to D \gamma_R$ goes along with a suppression of
$b\to D \gamma_L$, the impact on the branching ratio is small, 
as the two helicity amplitudes add incoherently. This implies there can be a
substantial contribution of NP to $b\to D\gamma$
escaping detection when only branching ratios are measured.

Although the photon helicity is, in principle, an observable, it is
very difficult to measure directly. It can, however, be accessed
indirectly, in the time-dependent CP asymmetry in $\bar B^0\to
V\gamma$, which relies on the interference of both left- and right-helicity
amplitudes and vanishes if one of them is absent. 
In terms of the left- and right-handed photon amplitudes of Eq.~(\ref{4})
one has
\begin{equation}\label{54}
S(V\gamma) 
 =  \frac{2 \,{\rm Im}\,\left(\frac{q}{p}({\cal A}_L^* \bar{\cal A}_L + 
                                       {\cal A}_R^* \bar{\cal A}_R)\right)}{
        |{\cal A}_L|^2 + |{\cal A}_R|^2 + |\bar{\cal A}_L|^2 + |\bar{\cal
                                       A}_R|^2}\,,
\quad
C(V\gamma)  =   \frac{|{\cal A}_L|^2 + |{\cal A}_R|^2 - |\bar{\cal A}_L|^2 - 
               |\bar{\cal A}_R|^2}{
        |{\cal A}_L|^2 + |{\cal A}_R|^2 + |\bar{\cal A}_L|^2 + |\bar{\cal
                                       A}_R|^2}\,.\label{5}
\end{equation}
Here $q/p$ is given in terms of the $B^0_q$-$\bar B^0_q$ 
mixing matrix $M_{12}$, in 
the standard convention for the parametrisation of the CKM matrix, by
$$
\frac{q}{p} = \sqrt{\frac{M_{12}^{*}}{M_{12}}} = e^{-i\phi_q}
$$
with, in the Wolfenstein parametrisation of the CKM matrix,
\begin{equation}\label{73a}
\phi_d \equiv {\rm arg}[(V_{td}^* V_{tb})^2] = 2 \beta\,,\qquad 
\phi_s \equiv {\rm arg}[(V_{ts}^* V_{tb})^2] = -2 \lambda
\left|\frac{V_{ub}}{V_{cb}}\right|  \sin\gamma\,.
\end{equation}

This method of accessing the right-handed photon amplitude via $S(V\gamma)$
was first suggested in Ref.~\cite{alt} and later discussed in more
detail in Refs.~\cite{grin04,grin05}. The direct CP asymmetry
$C(V\gamma)$ is less sensitive to $\bar{\cal A}_R$, but very sensitive to the
   strong phase of $\bar{\cal A}_L$ and vanishes if the radiative
   corrections to $a_{7L}^{U,{\rm QCDF}}$, Eq.~(\ref{10}), are
     neglected. As the accuracy of the prediction of strong phases in
     QCDF is subject to discussion, and in any case $C(V\gamma)$ is
     less sensitive to NP than $S(V\gamma)$, we shall
  not consider direct CP asymmetries in this paper. $S(V\gamma)$
 is rather special in the
sense that usually NP modifies the SM predictions for
time-dependent CP asymmetries by affecting the mixing phase (as in
$B_s\to J/\psi \phi$, see for instance Ref.~\cite{BF06}), introducing
new weak phases or moderately changing the size of the decay 
amplitudes which,
in the absence of precise calculational tools, makes it difficult to
trace its impact. In contrast, the time-dependent CP asymmetry in
$\bar B^0\to V\gamma$ is very small in the SM, irrespective of
hadronic uncertainties, and NP manifests itself by relieving this
suppression. The smallness of the asymmetry in the SM, and the possibility of
large effects from NP, makes it one of the prime candidates
for a so-called ``null test'' of the SM, as recently advertised in 
Ref.~\cite{null}. 

The fly in the ointment, however, is that in addition to the
helicity-suppressed contribution from $Q_7$, $\bar{\cal A}_R$ also 
receives contributions from the parton process $b\to
D\gamma g$, which come without a helicity-suppression factor
\cite{grin04,grin05}. These contributions are dominated by soft-gluon
 and long-distance photon emission in weak annihilation and
are also included in $a_{7R}^U$, Eq.~(\ref{asplit}). In
Ref.~\cite{grin05} it was inferred from a dimensional estimate 
that these contributions could be as
large as $\sim 10\%$, but a recent explicit calculation of 
the contribution of $Q_2^c$ to 
$S(K^*\gamma)$ has shown that their true size is much smaller
\cite{cpas}. In this paper, we extend the calculation of \cite{cpas}
to all $\bar B^0\to V^0\gamma$ channels and include the effects from 
all four-quark operators in the effective Hamiltonian
(\ref{heff}) and also the contribution from weak annihilation.

With ${\cal A}_{L,R}$ and $\bar{\cal A}_{L,R}$ as given in (\ref{ME})
we can calculate $S$ directly from (\ref{54}) and obtain, making
explicit the contributions from  different sources:\footnote{These
  results are obtained using LO Wilson coefficients. The difference
  between LO and NLO results is marginal.}
\begin{equation}
\renewcommand{\arraystretch}{1.3}
\begin{array}[b]{rll}
\label{eq:SVgamma}
S(\rho\gamma) = 
& \phantom{-}(\underbrace{~0.01~}_{m_D/m_b}+\underbrace{~0.02~}_{\rm
  LD~WA}+\underbrace{~0.20~}_{{\rm soft}~g}~\pm~ 1.6)\%
& = \phantom{-}(0.2\pm 1.6)\%\,,\\
S(\omega\gamma) =
& \phantom{-}(0.01-0.08+0.22\pm 1.7)\% 
& = \phantom{-}(0.1\pm 1.7)\%\,,\\
S(K^*\gamma) =
& -(2.9-0+0.6\pm 1.6)\%
& = -(2.3\pm 1.6)\%\,,\\
S(\bar K^*\gamma) =
& \phantom{-}(0.12+0.03+0.11\pm 1.3)\%
& = \phantom{-}(0.3\pm 1.3)\%\,,\\
S(\phi\gamma) =
& \phantom{-}(0+0+5.3\pm8.2)\times 10^{-2}\,\%
& =  \phantom{-}(0.1\pm 0.1)\%\,.
\end{array}
\end{equation}
Including only the helicity-suppressed contribution, one expects, for
$B\to K^*\gamma$, neglecting the doubly Cabibbo suppressed
amplitude in $\lambda_u^{(s)}$, see Eq.~(\ref{heff}),
\begin{equation}
\left.S(K^*\gamma)\right|_{\mbox{\footnotesize no soft gluons}} 
 =  -2\, \frac{m_s}{m_b}\,\sin\,\phi_d\approx -2.7\%\,.\label{75}
\end{equation}
For $B_s\to\phi\gamma$, one expects the CP asymmetry to vanish if the
decay amplitude is proportional to $\lambda_t^{(s)}$, which, at tree
level, precludes
any contributions of type $\sin(\phi_s) m_s/m_b$ and also any
contribution from WA.\footnote{This is
    because the mixing angle $\phi_s$ is given by ${\rm
      arg}[(\lambda_t^{(s)})^2]$, Eq.~(\ref{73a}), 
and the interference of amplitudes in
    (\ref{5}) also yields a factor $(\lambda_t^{(s)})^2$, 
if the individual amplitudes are proportional to  $\lambda_t^{(s)}$ or 
$(\lambda_t^{(s)})^*$, respectively; this is indeed the case for the
helicity-suppressed term $m_s/m_b$ induced by the operator $Q_7$,
Eq.~(\ref{eq:WC}), and the WA contributions to $a_{7R}^U(\phi)$,
Eqs.~(\ref{15}), (\ref{20A}), so that the phases cancel in (\ref{5}).} 

The actual results in (\ref{eq:SVgamma}) disagree with the above
expectations because of the contributions from soft-gluon emission,
which enter $a_{7R}^U$, and, for $S(\phi\gamma)$, because the
soft-gluon emission from quark loops is different for $u$ and $c$ loops,
  see Sec.~\ref{sec:4}, so that $a_{7R}^c\neq a_{7R}^u$ and hence
  $\bar {\cal A}_{R}$ (${\cal A}_{L}$) is not proportional to
  $\lambda_t^{(s)}$ ($(\lambda_t^{(s)})^*$).
Note that a substantial enhancement of $S(\phi\gamma)$ by NP requires
not only an enhancement of $|\bar{\cal A}_R|$ (and $|{\cal
  A}_L|$), but also the presence of a large phase in (\ref{5}); 
this could be  either
a large $B_s$ mixing phase which will also manifest itself in
a sizable CP violation in, for instance, $B_s\to J/\psi \phi$, see
Ref.~\cite{BF06}; or it could be a new weak phase in $\bar{\cal A}_{R}$
(and ${\cal A}_L$); or it could be a non-zero strong phase in
one of the $a_{7R}^{c,u}$ coefficients. Based on the calculation in
Sec.~\ref{4.2} we do not see much scope for
a large phase in $a_{7R}^{u}$ (whose contribution is, in addition,
doubly Cabibbo suppressed), but the situation could be different for
$a_{7R}^{c,{\rm soft}}$, where we only included the
leading-order term in a $1/m_c$ expansion, which does not carry a
complex phase, see Sec.~\ref{4.1}. It is not excluded that a
resummation of higher-order terms in this expansion will generate a
non-negligible strong phase --- which is not really relevant for our
results in Eq.~(\ref{eq:SVgamma}), but could be relevant for the
interpretation of any NP to be found in that observable. For
$S(K^*\gamma)$, on the other hand, no new phases are required, and
any enhancement of $ |\bar{\cal A}_R|$ (and $|{\cal A}_L|$) by NP will
result in a larger value of $S(K^*\gamma)$.

For all $S$ except $S(K^*\gamma)$,
the uncertainty is entirely dominated by that of the soft-gluon emission
terms $l_{u,c}-\tilde l_{u,c}$, whose uncertainties we have doubled with
respect to those given in Sec.~\ref{sec:4}. The smallness of
$S((\rho,\omega)\gamma)$ is due to the fact that the helicity
factor is given by $m_d/m_b$ (we use $m_{u,d}/m_s = 1/24.4$ from
chiral perturbation theory). For $\bar K^*$,
the suppression from the small mixing
angle is relieved by the fact that both weak amplitudes in
$\lambda_U^{(d)}$ contribute, with different strength,
so that the CP asymmetry is comparable
with that of $\rho$ and $\omega$. Despite the generous uncertainties, it is
obvious that none of these CP symmetries is larger than
4\% in the SM, which makes these observables very interesting
for NP searches. The present experimental result from the $B$
factories, $S(K^*\gamma)=-0.28\pm 0.26$ \cite{HFAG}, certainly encourages
the hope that NP may manifest itself in that
observable. While a measurement of the $b\to d$ CP asymmetries is
probably very difficult even at a super-flavour factory,
$S(K^*\gamma)$ is a promising observable for $B$ factories \cite{superB}, but
not for the LHC.\footnote{$K^*$ has to be traced via its decay into a CP
eigenstate, i.e.\ $K_S\pi^0$. Neutrals in the final state are not
really LHC's favourites.} $B_s\to \phi(\to K^+K^-)\gamma$, on the
other hand, will be studied in detail at the LHC, and in particular at
LHCb, and any largely enhanced value of $S(\phi\gamma)$ 
will be measured within the first years of running. 

\section{Summary and Conclusions}\label{sec:6}

In this paper we have presented a comprehensive study of the observables in
$B\to V\gamma$ decays, namely branching ratios, isospin and CP
asymmetries, for all $B_s$ and $B_{u,d}$ transitions,\footnote{We have
  not included pure annihilation decays, for instance
  $B_d\to\phi\gamma$, as their SM branching ratios are tiny,
  $O(10^{-11})$, and sensitive to higher-order effects in the
  electromagnetic interaction, which are not considered in this paper,
  but for instance in  Ref.~\cite{pureann}.} including 
the most recent results on form factors
from QCD sum rules on the light-cone and hadronic parameters
describing twist-2 and 3 two- and three-particle light-cone
distribution amplitudes of  vector mesons. 
Our study is based on QCD factorisation
\cite{BVga1,AP,BVga2,BoBu1,BoschThesis,BoBu2}, but goes beyond it by including
power-suppressed non-factorisable corrections from long-distance photon
emission and soft-gluon emission from quark loops which are also
calculated from light-cone sum rules. In Sec.~\ref{4.2}
we have devised a method for calculating
such soft-gluon emission from a light-quark loop for an on-shell
photon, building on the calculation of related effects in $B\to\pi\pi$
\cite{K00}.  The main idea is to calculate the loop for an off-shell
photon and then use a dispersion representation to relate it to the 
on-shell amplitude. For  phenomenology, light-quark loops are only 
relevant for $b \to d$ transitions, as otherwise they are 
Cabibbo suppressed or come with 
small Wilson coefficients. Our estimates may be of interest 
also for inclusive $b \to d \gamma$ transitions, where
an interplay between exclusive and inclusive effects could take
place similar to that for $b \to s \gamma$ \cite{voloshin,ligeti,rey}. 

Our main results are given in Sec.~\ref{sec:5}. We find that the
theoretical uncertainty of the branching ratios gets reduced by
exploiting the fact that ratios of form factors from QCD sum rules on
the light-cone are known with better accuracy than the form factors
themselves. This allows us to predict the branching ratios of all
$B\to V\gamma$ transitions with $\sim 20\%$ theoretical uncertainty
(except for $B_s\to\phi\gamma$ which comes with a $\sim\,$30\% uncertainty),
based on the experimental input from $B\to K^*\gamma$. The effect of
power corrections beyond QCD factorisation is non-negligible for
all decay channels, although in some channels the net corrections
nearly cancel. We have determined $|V_{td}/V_{ts}|$ and, equivalently, 
$\gamma$, from the most recent BaBar \cite{Babar} and Belle
\cite{Belle} results for $\overline{\cal
  B}(B\to(\rho,\omega)\gamma)/\overline{\cal B}(B\to K^*\gamma)$ as
$$
\renewcommand{\arraystretch}{1.3}
\begin{array}{l@{\quad}l@{\quad\leftrightarrow\quad}l}
\mbox{BaBar:} & \displaystyle
\left|\frac{V_{td}}{V_{ts}}\right| = 0.199^{+0.022}_{-0.025}({\rm exp})\pm
0.014({\rm th}) &\displaystyle
\gamma = (61.0^{+13.5}_{-16.0}({\rm exp})^{+8.9}_{-9.3}
({\rm th}))^\circ\,,\\[10pt]
\mbox{Belle:} & \displaystyle
\left|\frac{V_{td}}{V_{ts}}\right| = 0.207^{+0.028}_{-0.033}({\rm exp})
^{+0.014}_{-0.015}({\rm th}) &\displaystyle
\gamma = (65.7^{+17.3}_{-20.7}({\rm exp})^{+8.9}_{-9.2}({\rm th}))^\circ\,.
\end{array}
$$
As the relation (\ref{61}) between $|V_{td}/V_{ts}|$ and $\gamma$
relies on $\cos\gamma$, these results have a twofold degeneracy
$\gamma\leftrightarrow -\gamma$. Taken together with the 
tree-level CP asymmetries in $B\to D^{(*)} K^{(*)}$, for instance 
$\gamma=(53\pm20)^\circ$ from Belle \cite{Bellegamma}, which comes
with the 
discrete ambiguity $\gamma\leftrightarrow \gamma+\pi$, our result
removes the ambiguity and confirms that $\gamma<180^\circ$ as
predicted in the SM.

As for the isospin asymmetries, we find a non-zero asymmetry for the
$\rho^0$ and $\omega$ channel which is driven by the difference of the
corresponding form factors.
The asymmetry between the neutral and the
charged $\rho$ channel, on the other hand, is very sensitive to
$\gamma$, neglected radiative corrections and hadronic input
parameters, which precludes a precise statement about its size. 
The isospin asymmetry in $B\to K^*\gamma$ depends only mildly on the
input parameters, but is sensitive to the
contribution of the penguin operators $Q_{5,6}$. 
The sign of the asymmetry is predicted unambiguously. Although the present
experimental uncertainty of the asymmetry is too large to allow any
definite conclusion, any reduction could be translated into a
constraint on NP contributions to the Wilson coefficients of
these operators.

The time-dependent CP asymmetry $S(V\gamma)$ 
in $\bar B^0\to V^0\gamma$ is sensitive to
the photon polarisation amplitudes and is power-suppressed in the
SM. The contribution of $Q_7$ is helicity suppressed; the
contributions of other operators enter via the parton process 
$b\to D\gamma g$ with no
helicity suppression, but are also found to be small. The largest CP
asymmetry $\approx -2\%$ is expected for $B\to K^*\gamma$, whereas all
other CP asymmetries are below the 1\% level. 
Any value significantly different from zero, measured either at the LHC 
or a future flavour factory, will constitute an
unequivocal signal for NP with non-standard flavour-changing
interactions.

We also would like to discuss other results
for $B\to V\gamma$ available in the literature. Obviously, there are
earlier results from SCET \cite{BHN} and QCD factorisation, 
Refs.~\cite{BVga2,BoBu1,BoschThesis,BoBu2}, with which we agree apart
from the effects of the new non-factorisable contributions
calculated in this paper and/or updated hadronic input.
A variant of QCD factorisation has been advocated and pursued
by Ali and Parkhomenko (AP) \cite{BVga1,AP}. Another  approach
is that of perturbative QCD factorisation (pQCD), which has been applied to
$B\to V\gamma$ in Ref.~\cite{pQCD}. Most observables discussed in this
paper, branching ratios, isospin and CP asymmetries, have been
calculated in both approaches, for $B\to (K^*,\rho,\omega)\gamma$, and
we shall compare the corresponding results to ours in turn.
As for the branching ratios, it is evident from Eq.~(\ref{BR}) that
the predictions depend primarily on the form factor $T_1$ and only to a
lesser extent on the specific implementation of QCD factorisation. For
this reason, as AP use the same form factors as we, namely our
predictions from QCD sum rules on the light-cone \cite{BZ04b}, their
results in their latest update Ref.~\cite{AP} 
are very close to ours. The branching ratios obtained in
pQCD, on the other hand, are by more than a factor of two larger than ours.
This discrepancy is very likely to be caused by larger values of
their form factors, calculated within the same formalism;
a more detailed comparison is, however, difficult because
Ref.~\cite{pQCD} does not give any explicit numbers for the $T_1$. 
Turning to isospin asymmetries, AP obtain
approximately the same asymmetry $A(\rho,\omega)$,
Eq.~(\ref{eq:AIrw}), between $\rho^0$
and $\omega$ as we do, for the same reason as above.
The isospin asymmetry between the neutral and the charged 
$\rho$, Eq.~\eqref{air}, is more delicate, driven by weak annihilation
contributions and very  sensitive to the angle 
$\gamma$, see Tab.~\ref{tab:iso}. Our value, $(5.7\pm 3.9)\%$ 
for $\gamma = 60^\circ$,  disagrees with that given by AP, $A_I(\rho) = 
-(2.8 \pm 2.0)\%$ for the same angle. 
A likely reason is the smaller size
of the weak annihilation amplitude obtained in Ref.~\cite{WA}, on
which AP rely for that contribution, as compared to the QCD
factorisation result. Indeed, reducing the size of the weak
annihilation contribution in the $B\to V\gamma$ amplitude, 
our results move closer to those of AP. Ref.~\cite{pQCD}, on the other
hand, obtains $A_I(\rho) = 
(5.7\pm6.0)\%$ for $\gamma \approx 60^\circ$,
which coincides with our result, but comes with a larger uncertainty.
As for the isospin asymmetry in
the $K^*$ system, Eq.~(\ref{aik}), we obtain a slightly lower value
 than Kagan and Neubert, Ref.~\cite{kagan}, which is mainly due to their lower
 value $\lambda_{B_d} =0.35 \,{\rm GeV}$, compared to $0.51\,{\rm
   GeV}$ used by us, see Tab.~\ref{tab3}.
In the pQCD approach, the quoted asymmetry is about half of ours,
but comes with a similar relative uncertainty \cite{pQCD}, so that we
agree within errors. 
Concerning, finally, the time-dependent CP-asymmetry $S({K^*\gamma})$ in
\eqref{eq:SVgamma}, we find approximate numerical agreement with
Ref.~\cite{pQCD}, where the quark loops were modeled by
intermediate vector states. As emphasized earlier, the exact size
of the quark-loop contributions in this channel is not crucial since it 
is small compared to the leading term in $m_s/m_b\sin(2\beta)$,
Eq.~(\ref{75}). 
The CP asymmetries
$S(\rho\gamma)$ and $S(\omega\gamma)$ 
were also calculated by AP, but
unfortunately their formulae miss the very crucial
point that in $B\to V\gamma$ one has to deal with two physically 
distinguishable
final states, namely $V_L\gamma_R$ and $V_R\gamma_L$, whose
amplitudes must be added incoherently, not coherently as done in
Refs.~\cite{BVga1,AP}. We therefore refrain from a direct comparison 
with their results.

As for the relevance of our results for NP searches, the
time-dependent CP asymmetries are the 
cleanest observables since they are very small in the SM and constitute ``quasi
null tests'' of the SM \cite{null}, in the sense that any measurement
of a significantly non-zero value of these observables will be an
unambiguous signal of NP. For $K^*$, the asymmetry  has already been measured,
but is compatible with zero within errors. The asymmetry
in $B_s\to\phi\gamma$ is a very promising 
observable for the LHCb.  Also the isospin asymmetry
$A_I(K^*)$, Eq.~(\ref{aik}), is very interesting for NP searches,
and would become even more interesting upon completion of the NLO
calculation started by Kagan and Neubert, Ref.~\cite{kagan}, by including 
in particular the radiative corrections to the annihilation
contribution shown in Fig.~\ref{fig1}, with the photon emitted from
the final-state quark lines.
In contrast, neither the isospin asymmetry between $\rho^0$ and
$\rho^\pm$ nor the asymmetry between $\rho^0$ and $\omega$  
are likely to be sensitive to NP. As for the branching ratios, we
have, motivated by the inclusive $B\to X_s\gamma$ result,
assumed no significant NP effects in $B\to K^*\gamma$, and as long as
there is no breakthrough in the calculation of the absolute values of
form factors, any moderate NP effects in the branching ratios are
likely to be obscured by the uncertainties.

In summary we feel that exclusive $b\to (s,d)\gamma$ transitions 
have a
massive discovery potential for NP and envisage a great future at
the LHC, which may be surpassed only by that of 
$b\to (s,d)\mu^+\mu^-$ decays.

\bigskip

\noindent {\bf Note added.} After completion of the calculations presented
in this paper, Ref.~\cite{damir} appeared which contains a lattice calculation
of $T_1^{B\to K^*}(0)$ and $\xi_\rho$ in the quenched
approximation. The results are $T_1^{B\to K^*}(0)=0.24\pm
0.03^{+0.04}_{-0.01}$ and $\xi_\rho = 1.2\pm 0.1$. The latter agrees
with ours, Eq.~(\ref{xirho}), but comes with a slightly larger central
value and uncertainty, while the former is a bit on the low side of
the LCSR prediction given in Tab.~\ref{tab3}. 

\section*{Acknowledgments}

G.W.J.\ gratefully acknowledges receipt of a UK PPARC studentship.
R.Z. is grateful to Nikolai Uraltsev for discussions on non-perturbative
matrix elements.
This work was supported in part by the EU networks
contract Nos.\ MRTN-CT-2006-035482, {\sc Flavianet}, and
MRTN-CT-2006-035505, {\sc Heptools}.

\appendix
\setcounter{equation}{0}
\renewcommand{\theequation}{A.\arabic{equation}}

\section{Vector Meson Decay Constants Revisited}
There are two types of decay constants for  vector mesons:
the vector coupling $f_V$, for a longitudinally
polarised meson, and the tensor
coupling $f_V^\perp$, for a transversely polarised meson:
\begin{equation}
\matel{0}{ \bar q \gamma_\mu D}{V(p,e)} = 
e_\mu m_{V}  f_{V}\,, \qquad 
\matel{0}{\bar q \sigma_{\mu\nu} D}{V(p,e)}_\mu  =
i(e_\mu p_\nu - e_\nu p_\mu) f_{V}^\perp(\mu)\,.
\end{equation}
Note that $f_V^\perp(\mu)$ depends on the renormalisation scale.

The numerical values of these couplings are essential for our calculations.
Whereas the extraction of the charged mesons' vector couplings from
experimental data is straightforward, that of the neutral mesons'
$\rho^0$, $\omega$ and $\phi$ is complicated by the mixing of these
particles and deserves a more detailed discussion, which we will give
in Sec.~\ref{A.1}. 
The tensor couplings are not accessible experimentally, but
have to be determined by non-perturbative methods, for instance QCD
sum rules and lattice simulations. In Secs.~\ref{A.2} and \ref{A.3}, 
we briefly review the most recent results from these calculations.

\subsection{Longitudinal Decay Constants from Experiment}\label{A.1}

\subsubsection{\boldmath The Charged Decay Constants $f_{\rho^-,K^{*
      \,-}}$ from $\tau$ Decays}

The longitudinal decay constants of charged vector mesons can be 
extracted from $\tau^- \to V^- \nu_\tau$, with the measured branching
ratios \cite{PDG}:
\begin{eqnarray*}
{\cal B}(\tau^- \to \rho^{-} \nu_\tau) = (25.2 \pm 0.4)\times
10^{-2}\,, \qquad
{\cal B}(\tau^- \to K^{*\,-} \nu_\tau) =  (1.29 \pm 0.05)\times 10^{-2}\, .
\end{eqnarray*}
The decay rate is given by
$$
\Gamma(\tau^- \to V^- \nu_\tau) = \frac{m_\tau^3}{16 \pi} G_F^2 \
|V_{uD}|^2 f_{V^-}^2 \left(1-\frac{m_{V^-}^2}{m_\tau^2}\right)^2 \left(1+ 
2 \frac{m_{V^-}^2}{m_\tau^2}\right).
$$
With $|V_{ud}| = 0.9738 \pm 0.0002$ and 
$|V_{us}| = 0.227 \pm 0.001$ \cite{PDG}, we get
\begin{equation}
\label{eq:fLtau}
f_{\rho^-} = (210 \pm 2_{\cal B}\pm 1_{\Gamma_{\tau}} 
)\,{\rm MeV}\,, \quad \quad  
f_{K^{* \, -}} = (220 \pm 4_{\cal B} \pm 1_{\Gamma_{\tau}} \pm 1_{|V_{\rm
    us}|}) \, {\rm MeV} \, ,
\end{equation}
where we have taken into account the uncertainties in the branching
ratios, total decay rates and CKM matrix elements. The uncertainties of other
input parameters are irrelevant and the size of neglected 
corrections to the decay rate in $\alpha$ and higher powers in
$1/m_W^2$ is expected to be smaller than the total
uncertainty.

\subsubsection{\boldmath The Neutral Decay Constants
  $f_{\rho^0,\omega,\phi}$ from $V^0 \to e^+e^-$}

The decay constants of  $\rho^0$, $\omega$ and $\phi$
can be extracted from the electromagnetic annihilation
process $V^0 \to e^+ e^-$, which is, however, complicated by the mixing 
of these mesons. The states of definite isospin are given by
\begin{equation}
\label{eq:isostates}
\state{\rho^0_I} = \frac{1}{\sqrt{2}}(\state{\bar u u} - \state{\bar d
  d})\,, \qquad
\state{\omega_I} =  \frac{1}{\sqrt{2}}(\state{\bar u u} + \state{\bar
  d d})\,, \qquad
\state{\phi_I} = \state{\bar s s} \,,
\end{equation}
where $\rho$ has isospin 1 and $\omega$ and $\phi$ have 
isospin 0. In view of the subtleties of mixing, let us state clearly
that the neutral decay constants 
$f_{\rho^0}$, $f_\omega$ and $f_\phi$ shall
denote the coupling of the real particles to their isospin currents,
e.g. $ \matel{0} {\bar s \gamma_\mu s } {\phi}  \equiv e_\mu 
m_\phi f_\phi$.
$\rho$-$\omega$ mixing  violates isospin and hence is
a purely electromagnetic effect which can be parametrized as 
\begin{equation}
\state{\rho} \sim \state{\rho_I} - \epsilon_{\rho\omega} \state{\omega_I}\,,
\qquad
\state{\omega} \sim  \state{\omega_I} + \epsilon_{\rho\omega} \state{\rho_I}
\end{equation}
with $\epsilon_{\rho\omega} = \delta_{\rho\omega}/\big((m_\omega-
i \Gamma_\omega/2)^2 - (m_\rho-i \Gamma_\rho/2)^2\big)$ and 
$\delta_{\rho\omega} = -(0.004 \pm 0.002)\, {\rm GeV}^2$
\cite{rw-mixing}, which results
in $\epsilon_{\rho\omega} = (0.036\pm 0.018)i e^{0.15i}$;
since $m_\rho\approx m_\omega$, 
$\epsilon_{\rho\omega}$ is almost purely imaginary. This parameter was
also determined  experimentally \cite{exprhoomega}.
The mixing of $\omega$ and $\phi$, on the other hand, is due to strong
interactions:
\begin{equation}
\state{\omega} \sim \state{\omega_I} - \epsilon_{\omega\phi} 
\state{\phi_I}\,,
\qquad
\state{\phi} \sim \state{\phi_I} + \epsilon_{\omega\phi} \state{\omega_I}\,.
\end{equation}
The mixing parameter has been determined to be $\epsilon_{\omega\phi} =
0.045 \pm 0.01$  \cite{longtimeago} by parametrising the SU(3)
breaking in order to match the Gell-Mann--Okubo mass relation for
light mesons with the observed masses. 
A direct measurement of this quantity was reported in $\omega \to e^+e^-$
decays \cite{expomegaphi}.
Evidently a full description of the mixing would involve all three
states, but $\rho$--$\phi$ mixing is expected to be very small because it 
is a second-order effect that requires both electromagnetic and strong
interactions to be at work.

As the $V^0 \to e^+ e^-$ transition is an electromagnetic decay process, 
one needs the relevant electromagnetic currents of  light quarks: 
\begin{equation}
\label{eq:jem}
j^{\rm em}_\mu = Q_u \bar u \gamma_\mu u + 
 Q_d \bar d \gamma_\mu d +
 Q_s \bar s \gamma_\mu s = \frac{1}{3\sqrt{2}} \,(j_\mu^{I=0} + 
3 \, j_\mu^{I=1}) - \frac{1}{3} \bar s \gamma_\mu s\,; 
\end{equation}
the isospin currents are defined as 
$j^{I = 0/1}_\mu = \frac{1}{\sqrt{2}}(\bar u \gamma_\mu u \pm 
\bar d \gamma_\mu d$).
The experimental rates are \cite{PDG}:
\begin{eqnarray}
\label{eq:Bee}
& & {\cal B}(\rho^0 \to e^+ e^- ) = (4.7 \pm 0.08) \times 10^{-5}\,, \quad
{\cal B}(\omega  \to e^+ e^-) = (7.18 \pm 0.12) \times 10^{-5}\,,  
\nonumber \\
& & {\cal B}(\phi \to e^+ e^- ) = (2.97 \pm 0.04) \times 10^{-5}\,.
\end{eqnarray}
The theoretical expression for the decay rate is given by
\begin{equation}
\label{eq:Vee}
\Gamma(V^0 \to e^+ e^-) = \frac{4 \pi}{3} \frac{\alpha^2}{m_V} f_V^2
c_V\,,
\end{equation}
where the coefficients $c_V$ in the limit of no mixing can be read-off
from 
\eqref{eq:isostates} and \eqref{eq:jem}: 
$c_{\rho_I^0} = (Q_u-Q_d)^2/2 = 1/2$,  $c_{\omega_I} = (Q_u+Q_d)^2/2=
1/18$ and $c_{\phi_I} = Q_s^2 = 1/9$. The effect of, for instance,
$\rho$--$\omega$ mixing is to change $c_{\rho^0}$ to 
$c_{\rho^0} = |\sqrt{c_{\rho_I^0}} - \epsilon_{\rho\omega}  
\sqrt{c_{\omega_I}}|^2/|1+\epsilon_{\rho\omega}|^2$ and
correspondingly for $c_{\omega}$. Including the mixing effects we
finally get
\begin{eqnarray}
f_{\rho^0} &=&  (222 \pm  2_{\rm Br} \pm 1_{\Gamma_{\rho}})\, 
{\rm MeV}\,,\nonumber\\ \label{eq:fLee}
f_{\omega}  &=&  (187 \pm  2_{\rm Br} \pm 1_{\Gamma_\omega}  \pm
4_{\omega\phi} \pm 1_{\rho\omega}) \, {\rm MeV}\,, \nonumber \\
f_{\phi} &=&  (215 \pm  2_{\rm Br} \pm 1_{\Gamma_\phi}  \pm 4_{\omega\phi}
) \, {\rm MeV}\,,
\end{eqnarray}
where again the uncertainties in the other input parameters are irrelevant 
and the corrections to \eqref{eq:Vee} are expected to be 
smaller than the total uncertainty.
$\rho$--$\omega$ mixing has a negligible effect on $f_{\rho^0}$, 
but raises $f_\omega$ by 2 to 3$\,{\rm MeV}$. $\omega$--$\phi$ mixing
is much more relevant and lowers $f_\omega$ by about 
$10 \, {\rm MeV}$ and $f_\phi$ by about $13 \, {\rm MeV}$ . 


\subsection{Decay Constants from QCD Sum Rules}\label{A.2}

The calculation of decay constants from QCD sum rules was one of the
earliest applications of this method \cite{SVZ2}.
More recent determinations include more (radiative and mass)
corrections and updated values of input parameters.
The most recent results were obtained in Refs.~\cite{BZ06b,BZ05b} and read
\begin{equation}
 \label{eq:fSR}
\renewcommand{\arraystretch}{1.3}
\begin{array}[b]{lcl@{\qquad}lcl}
  f_\rho &=&(206 \pm 7)\, {\rm MeV}\,, 
& f_\rho^\perp(1\, {\rm GeV}) &=& (165 \pm 9)\, {\rm MeV}\,,\\
  f_{K^*} &=& (222 \pm 8)\, {\rm MeV}\,,
& f_{K^*}^\perp(1\, {\rm GeV}) &=& (185 \pm 10)\, {\rm MeV}\,.
\end{array}
\end{equation}
Note that the determination of $f_{K^*}^\perp$ is more complicated
than that of the other couplings and requires the inclusion of higher
resonances in the hadronic dispersion relation \cite{BZ05b}.
The above results refer to charged mesons;  isospin
breaking and meson mixing are not included. 
For comparison with lattice results, it proves convenient
to also quote the results for the ratio of 
couplings:\footnote{Including the 
NLO scaling factor $f^\perp_V(2 \, {\rm GeV})/f^\perp_V(1 \, {\rm GeV}) =
0.876$.}
\begin{equation}
\label{eq:sumruleratio}
\left( \frac{f_{\rho}^\perp}{f_\rho^\parallel}\right)_{\rm SR} 
(2\,{\rm GeV}) = 0.70\pm
0.04\,, \qquad 
\left( \frac{f_{K^*}^\perp}{f_{K^*}^\parallel}\right)_{\rm SR} 
(2\,{\rm GeV}) = 0.73\pm 0.04\,.
\end{equation}

\subsection{Decay Constants from Lattice QCD}\label{A.3}

The ratio of decay constants $f_V^\perp/f_V$ has been calculated by
two  lattice collaborations, in the quenched approximation. 
Ref.~\cite{fperplatt1} obtains
\begin{eqnarray}
\left( \frac{f_{\rho}^\perp}{f_\rho^\parallel}\right)_{\rm latt} 
(2\,{\rm GeV}) &=& 0.72\pm 0.02\,,  \qquad 
\left( \frac{f_{K^*}^\perp}{f_{K^*}^\parallel}\right)_{\rm latt} 
(2\,{\rm GeV}) = 0.74\pm 0.02 \,,\nonumber \\
\left( \frac{f_{\phi}^\perp}{f_\phi^\parallel}\right)_{\rm latt} 
(2\,{\rm GeV}) &=& 0.76\pm 0.01\,,
\label{eq:flattice}
\end{eqnarray}
in the continuum limit, whereas Ref.~\cite{Braunlatt} quotes:
\begin{equation}
\left( \frac{f_{\rho}^\perp}{f_\rho^\parallel}\right)_{\rm latt} 
(2\,{\rm GeV}) = 0.742\pm 0.014\,,\qquad 
\left( \frac{f_{\phi}^\perp}{f_\phi^\parallel}\right)_{\rm latt} 
(2\,{\rm GeV}) = 0.780\pm 0.008\,,
\end{equation}
at the finite lattice spacing $a=0.10\,$fm. 

The results from both lattice collaborations are roughly in agreement.
It is evident that the ratios depend only weakly on the quark masses. 
 
\subsection{Discussion, Conclusions and Results} 

In this paper we use the experimental results \eqref{eq:fLtau}
and \eqref{eq:fLee} for the longitudinal decay constants,
averaging the two results for the $\rho$ meson. For the tensor
couplings of $\rho$ and $K^*$
we use the sum rule results \eqref{eq:fSR}. For $\omega$, we assume
isospin symmetry of the ratio of decay constants and use
$f^\perp_\omega(2\,{\rm GeV})/f_\omega = 0.71 \pm 0.03$,
which is the average of QCD sum rule and lattice results, 
to obtain a value for $f_\omega^\perp$ from the measured $f_\omega$.
Finally, for $\phi$ we use the lattice ratio
\eqref{eq:flattice}, with the more conservative uncertainty 
$\pm 0.03$, and the experimental value for $f_\phi$. Our final
results which enter Tab.~\ref{tab3} are:
\begin{equation}
\renewcommand{\arraystretch}{1.3}
\begin{array}[b]{lcl@{\qquad}lcl}
  f_\rho &=& (216 \pm 3) \,{\rm MeV}\,, 
& f_\rho^\perp(1\,{\rm GeV}) &=&  (165 \pm 9) \,{\rm MeV}\,,\\
  f_\omega &=& (187 \pm 5) \,{\rm MeV}\,, 
& f_\omega^\perp(1\,{\rm GeV}) &=& (151 \pm 9) \,{\rm MeV}\,,\\
 f_{K^*} &=& (220 \pm 5) \,{\rm MeV}\,, 
& f_{K^*}^\perp(1\,{\rm GeV}) &=& (185 \pm 10) \,{\rm MeV}\,, \\
 f_{\phi} &=& (215 \pm 5) \,{\rm MeV}\,, 
& f_{\phi}^\perp(1\,{\rm GeV}) &=& (186 \pm 9) \,{\rm MeV}\,.  
\label{eq:deacy_final}
\end{array}
\end{equation}

The experimental vector couplings, 
\eqref{eq:fLtau} and \eqref{eq:fLee}, come with rather small
uncertainties and indicate an isospin breaking in $f_{\rho^{0,\pm}}$
of $\approx 5\%$. Is it really justified to average both results into
only one decay constant for the $\rho$? In order to answer this
question, let us have a look at the experimental information on
isospin breaking in other light-meson decay constants. For the $\pi$, PDG
gives $f_{\pi^\pm} = (130.7\pm 0.4)\,$MeV, whereas $f_{\pi^0}$ is
extracted from $\pi^0\to\gamma\gamma$ as $(130\pm5)\,$MeV which is
perfectly compatible with $f_{\pi^\pm}$; the
uncertainty is dominated by that of the $\pi^0$ lifetime
\cite{PDG2}. 
A further confirmation of the smallness of isospin breaking comes from
the CLEO measurements
of $D^0 \to (K^-,\pi^-) e^+ \nu$ and $D^+ \to  (\bar K^0,\pi^0) e^+
\nu$ \cite{CLEO}. These decays are sensitive to the $\pi$ and $K$
meson decay constants via the form factors, which, at least in the
LCSR approach, are directly proportional to $f_{\pi,K}$, see e.g.\
Ref.~\cite{me_again}. The data for
$D\to K e\nu$ indicate that the isospin breaking of the form factor is
$(3\pm 2)\%$. Although this result also includes potential isospin
breaking of both $f_{D}$ and the dynamical part of the form factor, the
corresponding effects are neither expected to be sizable nor to cancel
each other, so that indeed for both $\pi$ and $K$ decay
constants isospin breaking is smaller than 5\% at 1$\sigma$.
Taking this as an indication for the generic size of isospin breaking
in light mesons, we conclude that a 5\% difference between
$f_{\rho^0}$ and $f_{\rho^\pm}$ is not excluded, but on the large
side. This implies that it is indeed appropriate to average
the experimental results for neutral and charged $\rho$ as done in 
\eqref{eq:deacy_final}.

Note that the QCD sum rule results for the vector couplings,
\eqref{eq:fSR}, agree rather well with the experimental results, 
\eqref{eq:fLtau} and 
\eqref{eq:fLee}, which increases confidence in the corresponding
results for the tensor couplings,  particularly as the sum rule
results for the ratios, \eqref{eq:sumruleratio}, also agree with those
from lattice, \eqref{eq:flattice}.
Nevertheless there is one fact which remains somewhat puzzling,
namely the difference of nearly  20\%  between the $\omega$ and
$\rho^0$ couplings from $V^0 \to e^+e^-$, which is larger than the 
expectation
from QCD sum rules. This difference could be caused by 
electromagnetic corrections, different values of the up and down quark 
condensate and differences in the continuum thresholds and Borel windows. 
The latter effects should not exceed the typical accuracy of sum rules
themselves, which is
about 10\%, and the former effects are expected to be very small.
On the other hand, the individual experimental results entering the
PDG averages \eqref{eq:Bee} are spread over a wide range, 
particularly for $\omega$, which indicates that the uncertainties 
quoted in \eqref{eq:Bee} may be on the optimistic side.

There remains one subtle point to be discussed, namely that for 
LCSRs for $b \to d \gamma$ transitions and 
$\rho^0$ or $\omega$ in the final state,
one needs the decay constant
$$
\sqrt{2} \matel{0}{ \bar d \gamma_\mu d}{V^0(e)} = 
e_\mu \, m_{V^0} \,  f_{V^0}^{(d)}\,,
$$
rather than $f_{V^0}$. The quantity $f_{V^0}^{(d)}$ could differ from 
$f_{V^0}$ 
through mixing with the other neutral mesons. Fortunately,
$\omega$-$\phi$ mixing is irrelevant because the coupling of $\phi$ 
to the down-quark current is highly suppressed, and  $\rho$-$\omega$ 
mixing has a small effect because the mixing parameter
is almost imaginary. The total impact of mixing is hence below the
theoretical uncertainty and can safely be neglected.

\end{document}